# Hot Streaks in Artistic, Cultural, and Scientific Careers

Lu Liu,[1,2,3] Yang Wang,[1,2] Roberta Sinatra,[4] C. Lee Giles,[3,5] Chaoming Song,[6] Dashun Wang[1,2]

[1]*Northwestern Institute on Complex Systems, Northwestern University, Evanston, IL, USA*

[2]*Kellogg School of Management, Northwestern University, Evanston, IL, USA*

[3]*College of Information Sciences and Technology, Pennsylvania State University, State College, PA, USA*

[4]*Center for Network Science and Mathematics Department, Central European University, Budapest, Hungary*

[5]*Department of Computer Science and Engineering, Pennsylvania State University, State College, PA, USA*

[6]*Department of Physics, University of Miami, Coral Gables, FL, USA*

**The hot streak, loosely defined as winning begets more winnings, highlights a specific period during which an individual's performance is substantially higher than her typical performance. While widely debated in sports[1–3], gambling[4–7], and financial markets[8–10] over the past several decades, little is known if hot streaks apply to individual careers. Here, building on rich literature on lifecycle of creativity[11–23], we collected large-scale career histories of individual artists, movie directors and scientists, tracing the artworks, movies, and scientific publications they produced. We find that, across all three domains, hit works within a career show a high degree of temporal regularity, each career being characterized by bursts of high-impact works occurring in sequence. We demonstrate that these observations can be**

**explained by a simple hot-streak model we developed, allowing us to probe quantitatively the hot streak phenomenon governing individual careers, which we find to be remarkably universal across diverse domains we analyzed: The hot streaks are ubiquitous yet unique across different careers. While the vast majority of individuals have at least one hot streak, hot streaks are most likely to occur only once. The hot streak emerges randomly within an individual's sequence of works, is temporally localized, and is unassociated with any detectable change in productivity. We show that, since works produced during hot streaks garner significantly more impact, the uncovered ho t streaks fundamentally drives the collective impact of an individual, ignoring which leads us to systematically over- or under-estimate the future impact of a career. These results not only deepen our quantitative understanding of patterns governing individual ingenuity and success, they may also have implications for decisions and policies involving predicting and nurturing individuals with lasting impact.**

A creative career is often defined by the sequence of works an individual produces at various stages[14,15,20,22–25]. According to the Matthew effect[12,26–28], victories bring reputation and recognition that can translate into tangible assets which in turn help bring future victories. This school of thought supports the existence of hot streaks in a career, which is also consistent with the innovation literature showing that peak performance clusters in time, typically occurring around mid career[11,14,24]. Yet, on the other hand, the random impact rule uncovered in arts[13,24] and sciences[13,23] predicts the opposite: The best works occur randomly within a career, primarily driven by productivity. Following this school of thought, works after a major breakthrough are not affected by what preceded them, supporting the viewpoint of regression toward the mean. The two divergent schools of thought, together with the consequential nature of this question to individual ingenuity and success, raise a fundamental question: Do hot streaks exist in creative careers?

To answer this question, we collected systematically large-scale datasets recording career histories of individual artists, movie directors and scientists (Supplementary Information S1), allowing us to explore across three major domains involving human creativity. The first dataset ($D_1$) consists of auction records curated from online auction databases, allowing us to reconstruct career histories of 3,480 artists through the sequence of works they each produced, together with impacts of the artworks, approximated by hammer prices in auctions[20]. $D_2$ contains profiles of 6,233 movie directors recorded in the IMDB database, each career being represented by the sequence of movies he/she directed. Since metrics that quantify impacts of a movie correlate closely with each other[29], here we use the IMDB ratings to measure the goodness of a movie. Finally, our third dataset ($D_3$) includes publication records of 20,040 individual scientists through a large-scale name disambiguation effort that combined the Web of Science and Google Scholar datasets (Supplementary Information S1.3). The impact of each paper is measured by citations garnered after 10 years of its publication[17,21,23,30], $C_{10}$. Since hammer price ($D_1$) and $C_{10}$ ($D_3$) both follow fat-tailed distributions (Fig. S3), here we take the logarithmic of these measures, i.e., $\log$(price) and $\log(C_{10})$ to approximate the goodness of an artwork and scientific publication.

Motivated by Merton's theory of multiples[12,31], we start by investigating the timing of the three most impactful works produced in each career. In a sequence of $N$ works by an individual, we denote with $N^*$ the position of the highest impact work within a career, $N^{**}$ the second high-est and $N^{***}$ the third. We find each of the three highest impact works occurs randomly within a career (Fig. S4). That is, when it comes to any of the three most expensive artworks by an artist, three highest rated movies by a director, and three highest impact publications by a scientist, each of them is randomly distributed among all the works one produces. These results offer strong endorsement for the random impact rule[13,23,24], hence supporting an unpredictable view of individual creativity.

Yet, as we show next, the random impact rule observed across creative careers is only apparent, because the timing between creative works follows highly predictable patterns. Indeed, we measure the correlation between the timing of the two biggest hits within a career (e.g., $N^*$ and $N^{**}$) by calculating the joint probability $P(N^*, N^{**})$, and compare it with a null hypothesis in which $N^*$ and $N^{**}$ each occurs at random. We find that, the normalized joint probability, $\phi(N^*, N^{**}) = P(N^*, N^{**})/(P(N^*)P(N^{**}))$, is significantly overrepresented along the diagonal elements of matrices (Figs. 1a–c), demonstrating that $N^*$ and $N^{**}$ are much more likely to colocate with each other than what we would expect from the random impact model. Moreover, the colocation pattern is universal across a wide range of careers we studied, including artists (Fig. 1a), movie directors (Fig. 1b) and scientists (Fig. 1c). The diagonal pattern disappears if we shuffle the order of works within each career, thereby breaking the temporal correlations between highest impact works while preserving the random impact rule (Figs. 1d–f, Fig. S6).

To quantify the temporal colocation of hits observed in Figs. 1a–c, we calculate the distance between two highest impact works for every individual, measured by the number of works produced in between, $\Delta N = N^* - N^{**}$. We compare $P(\frac{\Delta N}{N})$ of real careers with $P_S(\frac{\Delta N}{N})$ of shuffled careers by defining $R(\frac{\Delta N}{N}) = P(\frac{\Delta N}{N})/P_S(\frac{\Delta N}{N})$. For artists, movie directors, and scientists, $R(\frac{\Delta N}{N})$ all exhibits a clear peak centering around zero and decays quickly as $\Delta N$ deviates from zero (Figs. 1j–l). Indeed, the two most important works of an artist is $1.48$ times more likely to occur back-to-back than expected by chance (Fig. 1j). The same is true for movie directors (Fig. 1k) and scientists (Fig. 1l), where such colocation is $1.42$ and $1.57$ times more likely than their baseline occurrence rate, respectively. Also important to note is the interesting fact that $R(\frac{\Delta N}{N})$ is mostly symmetric around zero (Figs. 1j–l), indicating a comparable likelihood for the biggest hit to arrive before or after the second biggest for all three types of careers. This symmetry was also captured by Figs. 1a–c, where $\phi$ features a roughly even split across the diagonal. The colocation patterns documented in Figs. 1a–c and 1j–l are not limited to the two highest impact works within a career. Indeed, we repeated our analyses for other pairs of hit works, such as $N^*$ vs. $N^{***}$ and $N^{**}$ vs. $N^{***}$, uncovering the same colocation patterns (Figs. 1j–l and Fig. S5).

Do high impact works come in streaks within a career? To answer this question, we count the number of consecutive works whose goodness exceeds a certain threshold across various careers (Figs. 1m–o). Here we choose the median goodness of all works within a career as the threshold. We calculate the length of the longest streak $L$ for each career, and measure the distribution of $L$ across our user base in each of the three domains. We then shuffle the order of works within each career, and measure again their longest streaks $L_s$. We find $P(L)$ is characterized by a much longer tail, compared with $P(L_s)$ (Figs. 1p–r), indicating real careers are characterized by long streaks of excellent works clustered together in sequence. Note that for the three types of careers, the tail part of both $P(L)$ and $P(L_s)$ follows approximately an exponential function, meaning that the likelihood to observe a longer streak diminishes rather rapidly. Hence, the deviations observed between $P(L)$ and $P(L_s)$ are rather significant (Fig. S11). To test the robustness of these results, we repeated our analyses by controlling for individual career length, and also varying our threshold used to calculate $L$, finding our conclusions remain the same (Supplementary Information S2.4 and S2.5).

Taken together, results presented in Fig. 1 paint a rather unexpected portrait of individual careers. Indeed, while the timing of high impact works each appears at random by itself, their relative timing, however, follows highly reproducible yet previously unknown patterns. As such, individual careers are far from being random, but characterized by bursts of high-impact works occurring in sequence. These fascinating empirical findings raise an important question: What are the mechanisms responsible for the temporal regularities observed across diverse career histories?

To unearth the fundamental mechanisms governing the patterns documented in Fig. 1, let us first consider a null model in which the goodness of works produced in a career (i.e.. $\log$(price) for artists, ratings for directors, and $\log(C_{10})$ for scientists) is drawn from a normal distribution $\mathcal{N}(\Gamma_i,\, \sigma_i^2)$, fixed for an individual. The average $\Gamma_i$ characterizes typical impact of works produced by the individual, and $\sigma_i$ captures the variance. This null model reproduces the fact that each hits occur randomly within a career[13,23] and the differences in typical impact between careers (Supplementary Information S3.2, Figs. S25–S27). Yet it fails to capture any of the temporal correlations observed in Fig. 1. The main reason is illustrated in Figs. 2a–c, where we selected for illustration purposes one individual from each of the three datasets and measure the dynamics of $\Gamma_i$ during his/her career. We find $\Gamma_i$ is not constant throughout a career. Rather, it features deviations from a baseline performance ($\Gamma_0$) at a certain point of a career ($t_\uparrow$), elevating to a higher value $\Gamma_H$ ($\Gamma_H > \Gamma_0$), which is then sustained for some time before falling back to a similar level as $\Gamma_0$ (Figs. 2a–c):

$$\Gamma(t) = \begin{cases} \Gamma_H & t_\uparrow \leq t \leq t_\downarrow \\ \\ \Gamma_0 & otherwise \end{cases}, \tag{1}$$

This observation, combined with the shortcomings of the null model, raises an intriguing hypothesis: Could a simple model based on (1) explain the temporal anomalies documented in Fig. 1?

To test this hypothesis, we apply (1) to real productivity patterns of an individual, allowing us to generatively simulate impacts of the works produced by an individual (Supplementary Information S3.3). As an individual's baseline performance is captured by $\Gamma_0$, during the period in which $\Gamma_H$ operates ($t_\uparrow \leq t \leq t_\downarrow$), the individual seemingly performs at a higher level than her typical performance ($\Gamma_0$), prompting us to call this model, the hot-streak model, and correspondingly, the $\Gamma_H$ period as the hot streak. Hence, in the hot-streak model, the goodness of works produced by an individual is drawn from two distributions $\mathcal{N}(\Gamma_0, \sigma^2)$ and $\mathcal{N}(\Gamma_H, \sigma^2)$, depending on whether the individual is within the hot streak. We introduce to each career one ho t streak that occurs at random with a fixed duration and magnitude, and repeat our measurements in Fig. 1 on careers generated by the model. We find, while equation (1) only introduces a simple temporal variation, surprisingly the hot-streak model is sufficient in reproducing all empirical observations that existing modeling frameworks fail to account for from the temporal colocations among top hit works within a career (Figs. 1g–i), to their temporal distances (Figs. 1j–l), to the occurrences of long streaks of excellent works (Figs. 1p–r). For full comparisons with existing models, see Supplementary Information S4. Equation (1) assumes implicitly each individual has one hot streak, a hypothesis that we test later. Given the myriad factors that can affect career impacts[12–16,23,25,32,33], and the obvious diversity of careers we studied, the level of universality and accuracy demonstrated by the simple hot-streak model is rather unexpected.

The real value of the model arises, however, when we fit the model to real careers to obtain the individual specific $\Gamma_0$, $\Gamma_H$, $t_\uparrow$ and $t_\downarrow$ parameters (Supplementary Information S3.4), allowing us to probe quantitatively the hot streak phenomenon underlying artistic, cultural, and scientific careers, and helping us reveal several fundamental patterns governing individual careers:

1. The hot streak is ubiquitous *across* careers, yet at the same time rather unique *within* a career. We find, the vast majority of artists (91%, Fig. 2d), movie directors (82%, Fig. 2e) and scientists (90%, Fig. 2f) have at least one hot streak throughout their careers, documenting the practical relevance of the uncovered hot streak phenomenon. Yet, despite its ubiquity, the hot streak is most likely to be unique within a career. Indeed, we relax our fitting algorithm to allow for multiple hot streaks (up to three) with different values of $\Gamma_H$, finding that, among those who have a hot streak, 64% of artists, 80% of directors, and 68% of scientists are best captured by one hot streak only (Figs. 2d–f), documenting the precious nature of hot streaks. Second acts may occur but less likely, particularly for movie directors. About 30% of artists and scientists have two hot streaks, but only 11% for directors. Occurrences of more than two hot streaks are rare across all careers. We also find that, between those who have one or two hot streaks, there is no detectable difference in terms of typical performance metrics, including impact, productivity and career length (Fig. S22), suggesting that hot streaks capture an orthogonal dimension to current metrics characterizing individual careers.

2. The hot streak occurs randomly within a career. We estimate the beginning of hot streaks for artistic, cultural, and scientific careers. Denoting with $N_{\uparrow}$, the position of work produced when the hot streak starts ($t_{\uparrow}$), we find that $N_{\uparrow}$ is randomly distributed in the sequence of $N$ works within a career (Figs. 2g–i). This finding reconciles two seemingly divergent schools of thought[12,13,23], providing a further explanation for the random impact rule: If the hot streak occurs randomly within a career, and the highest impact works are statistically more likely to appear within the hot streak, then the timing of the highest impact works would also appear random.

3. Across different domains, a hot streak lasts for a considerably shorter period comparing with the typical career length recorded in our database. We measure the duration distribution of hot streaks ($\tau_H = t_{\uparrow} - t_{\downarrow}$), finding $P(\tau_H)$ peaks around $5.7$ years for artists, $5.2$ years for directors, and $3.7$ years for scientists (Figs. 2j–l). Interestingly, the duration of a hot streak is independent of when it occurs within a career (early, mid or late career, Figs. 2j–l). The temporally localized nature of hot streaks is also captured by its proportion over career length $\tau_H/T$ (Figs. 2j–l, insets), whose median hovers around $20\%$ ($0.17$ for artists, $0.23$ for directors, and $0.20$ for scientists).

4. How much does an individual deviate from her typical performance during hot streaks?

Do people with higher $\Gamma_0$ also experience more performance gain from hot streaks? To answer these questions, we explore correlations between $\Gamma_0$ and $\Gamma_H$, finding them to be well approximated by a linear relationship across three kinds of careers (Figs. 2m–o). Hence the better typical performance, the better individuals perform during their hot streaks. It is interesting to note that the coefficients are slightly less than 1 ($0.99$ for artists, $0.85$ for directors, and $0.9$ for scientists, Figs. 2m–o). Hence $\Delta\Gamma \equiv \Gamma_H - \Gamma_0$ decreases with $\Gamma_0$ (Figs. 2m–o insets), suggesting individuals with smaller $\Gamma_0$ benefit more from hot streaks. These results are again independent of when hot streaks occur along a career (Figs. 2m–o).

5. Are individuals more productive during hot streaks? Surprisingly, the answer is no. We measure the distribution of the total number of works produced during hot streaks $P(N_H)$. We then construct a null distribution, by randomly picking one work out of a career and designating its production year to be the start of the hot streak. We find $P(N_H)$ measured in real careers well aligns with the null model's predictions for all three kinds of careers (Figs. 2p–r). Therefore, individuals show no detectable change in productivity during hot streaks, despite the fact that their outputs during the period are significantly better than typical, suggesting an endogenous shift in individual creativity when a hot streak occurs.

What is the impact of hot streaks on individual careers? To answer this question, we focus on scientific careers ($D_3$), and measure the collective impact of a scientist, $g(t)$, defined as the total number of citations over time collected by all the papers one published. Brought to spotlight by popular websites such as Google Scholar (Fig. 3a), $g(t)$ is playing an increasingly important role in driving many critical decisions, from hiring, promotion and tenure to awarding of grants and rewards. Next we show the collective impact of an individual is fundamentally governed by the uncovered hot streak phenomenon, ignoring which would lead us to systematically under- or over-estimate the future impact of a scientist.

Many factors are known to influence the collective impact of a career, ranging from productivity[15,22,34] to citation disparity and dynamics[17,18,21,25,27,35] to temporal inhomogeneities along a career[14,22–25,36]. Since our goal is to understand impact, here we bypass the need to evaluate the inhomogeneous nature of productivity[22,23] by rearranging publication time of each paper, such that an individual produces a constant number of papers each year, denoted by $n$ (Figs. 3b–c). To calculate $g(t)$ analytically, we need to incorporate papers' citation patterns into our hot-streak model (1). A recent study[21] suggests that citation dynamics of a paper published at time $t_0$ can be approximated by

$$C(t, t_0) = m\left(e^{\lambda\Phi\left(\frac{\ln(t-t_0)-\mu}{\sigma}\right)} - 1\right) \equiv m\left(e^{\Gamma(t_0)\Phi\left(\frac{\ln(t-t_0)-\mu}{\sigma}\right)} - 1\right), \tag{2}$$

where $m$ is a global parameter describing the typical number of references a paper contains, and $\Phi(\cdot)$ is the cumulative normal function, characterized by $\mu$ and $\sigma$, which capture the typical citation life cycle of a paper. The paper's impact is ultimately determined by its fitness[21], $\lambda$. To adapt this model into our framework, we replace $\lambda$ with $\Gamma(t_0)$, and for simplicity assume $\mu$ and $\sigma$ are fixed for different papers published by an individual. The resulting model, combining (1) and (2), can be solved analytically (Supplementary Information S5.1–S5.4), allowing us to express $g(t)$ in terms of hot-streak parameters $(\Gamma_0, \Gamma_H, t_\uparrow, t_\downarrow)$:

$$g(t) = \underbrace{nm(e^{\Gamma_0 \Phi\left(\frac{\ln(t)-\mu}{\sigma}\right)} - 1)}_{g_0(t)} + \underbrace{\begin{cases} 0 & t < t_\uparrow \\ nm(\Gamma_H - \Gamma_0)\Phi\left(\frac{\ln(t-t_\uparrow)-\mu}{\sigma}\right) C(t, t_\uparrow) & t_\uparrow \leq t < t_\downarrow \\ nm(\Gamma_H - \Gamma_0)[\Phi\left(\frac{\ln(t-t_\uparrow)-\mu}{\sigma}\right) C(t, t_\uparrow) - \\ \qquad \Phi\left(\frac{\ln(t-t_\downarrow)-\mu}{\sigma}\right) C(t, t_\downarrow)] & t \geq t_\downarrow \end{cases}}_{\Delta g(t)} . \quad (3)$$

Equation (3) consists of two terms. $g_0(t)$ captures a career's collective impact in the absence of hot streaks (i.e. $\Gamma(t) = \Gamma_0$). Contributions from hot streaks are encoded in $\Delta g(t)$, driven by both the timing and magnitude of hot streaks ($t_\uparrow$, $t_\downarrow$, $\Gamma_H$, and $\Gamma_H - \Gamma_0$). Varying hot streak parameters significantly alters the collective impact of a career (Fig. 3d).

We adopt two measures to quantify the accuracy of our model (3). To account for the inherently noisy career trajectories, we first assign an impact envelope to each individual, explicitly quantifying the uncertainty of model predictions (Fig. 3e, Supplementary Information S5.6). We measure the fraction of $g(t)$ that fall within the envelope, finding the distribution across individuals peaks close to 1 (Fig. 3f), indicating most career trajectories are well encapsulated within the predicted envelopes. The superior accuracy of our model is also captured by the Mean Absolute Percentage Error (*MAPE*) (Fig. 3g), with improvement being most pronounced for an early onset of a hot streak (Fig. 3g), which is also correctly predicted by our model. Hence the hot-streak model captures a wide range of trajectories that collective impacts of scientific careers follow (Fig. 3h).

The observed accuracy prompts us to ask whether the hot-streak model is unique in its ability to capture the impact of individual careers across diverse domains. There are several alternative hypotheses capturing different hot streak dynamics (Supplementary Information S6), each associated with possible origins of the uncovered hot streak phenomena: (*A*) A right trapezoid (Fig. S32b) captures a sudden onset of the hot streak with a more gradual decline, as innovators may stumble upon a groundbreaking idea, which manifests itself in the forms of multiple artworks, movies, and publications. Hence from an evolutionary perspective, the duration of hot streaks may characterize time it takes for the temporary competitive advantage to dissipate. (*B*) An isosceles trapezoid model (Fig. S32c) captures a hot streak that evolves and dissolves gradually over time, which may approximate social tie dynamics, as one individual's hot streak could be the result of a fruitful, repeated collaboration[32,37]. (*C*) Furthermore, individual performance may peak at a certain point of a career, prompting us to test inverted-U shape (Fig. S32d) and tent functions (Fig. S32e). Lastly (*D*) a left trapezoid function (Fig. S32f) captures a gradual startup period with a sharp cutoff, corresponding to career opportunities that can augment impact but last for a fixed duration.

We tested hypotheses *A*–*D* systematically to describe real careers (Supplementary Information S6). Of all hypotheses considered, the proposed hot-streak model is the simplest and least flexible. Yet, surprisingly, it is the only model whose predictions are consistent with real careers (Fig. S32). The fact that none of the alternative hypotheses alone can fully account for empirical observations demonstrates the hot streak phenomena uncovered in creative careers may not be driven by one particular factor but a combination of multiple factors. Identifying its true origin requires additional experimentation and goes beyond the scope of this work. As such, hot streaks uncovered in this paper should be treated in a metaphorical sense, highlighting an intriguing period of outstanding performance tracing individual careers without implying any associated drivers for the phenomena. Yet, crucially, the findings presented in this paper hold the same, regardless of the underlying drivers.

The analytical framework presented here not only offers a new theoretical basis for our quantitative understanding of dynamical patterns governing individual career impact, it also has policy implications for comparing and evaluating scientists (Fig. S30). Indeed, for individuals whose hot streaks are yet to come, ignoring hot streaks may lead to underestimating their impacts (Figs. S30a–b), especially given the ubiquitous nature of hot streaks (Fig. 2f). On the other hand, an early onset of hot streaks leads to a high impact that peaks early but may not sustain unless a second act occurs (Fig. S30c). Given that individuals improve substantially during hot streaks, the uncovered hot streak phenomenon can be particularly crucial for decisions and policies concerning long-term impact of a career.

**Acknowledgements** The authors thank A.-L. Barabasi, B. Uzzi, W. Ocasio, J. Chown, C. Jin, Y. Yin, and all members of Northwestern Institute on Complex Systems (NICO) for invaluable comments. This work is supported by the Air Force Office of Scientific Research under award number FA9550-15-1-0162 and FA9550-17-1-0089, and Northwestern University's Data Science Initiative.

**Correspondence** Correspondence and requests for materials should be addressed to D.W. (email: dashun.wang@northwestern.edu).

## Figure captions

**Figure 1: The hot streak phenomenon in artistic, cultural and scientific careers**. **a–c,** $\phi(N^{*}, N^{**})$, color coded, measures the joint probability of the top two highest impact works within a career for **a** artists, **b** directors, and **c** scientists. $\phi(N^{*}, N^{**}) > 1$ indicates two hits are more likely to colocate than random. **d–f,** we shuffle the order of each work in a career while keeping their impact intact, allowing us to measure the null hypothesis of $\phi(N^{*}, N^{**})$ across three domains, where $N^{*}$ and $N^{**}$ each occurs at random. The diagonal patterns in **a–c** disappear for shuffled careers. **g–i,** $\phi(N^{*}, N^{**})$ predicted by the hot-streak model successfully recovers the diagonal patterns observed in **a–c**. **j–l,** $R(\frac{\Delta N}{N})$ measures the temporal distance between highest impact works relative to null model's prediction. Red dots denote measurements from data, showing a clear peak around 0. Solid lines in red are predictions by the hot-streak model. Different shades of red correspond to different pairs of hit works. Blue dots denote the same measurement but on shuffled careers, and blue lines are predictions from shuffled careers generated by our model. **m–o,** Definitions of the longest streak $L$ within a career for **m** artists, **n** directors and **o** scientists. $L$ measures as the maximum number of consecutive works whose impacts are above the median impact of a career (horizontal dashed line denoting the threshold). Above the threshold, dots are colored in orange, and blue for below the threshold. $L$ in the lower panel highlights the longest streak in a career. **p–r,** The distribution of the length of streaks $P(L)$ for real careers and $P(L_s)$ for shuffled careers, for **p** artists, **q** directors and **r** scientists. Red dots capture empirical observations, whereas blue dots correspond to shuffled careers. Our hot-streak model (red lines) closely reproduces $P(L)$ observed in data, demonstrating the model's validity to capture the impact beyond top three highest impact

works across different domains. The shuffled version of our model (blue lines) also well captures shuffled careers.

**Figure 2: The Hot-streak Model. a–c,** $\Gamma(N)$ for one **a** artist, **b** movie director and **c** scientist, selected from our data for illustration purposes. $\Gamma(N)$ is calculated by the moving average of impact with a window length $= 0.1 \times N$. **d–f,** The histogram of the number of hot streaks for **d** artists, **e** directors, and **f** scientists. **g–i,** $N_{\uparrow}/N$ measures the position of the work when a hot streak occurs, among $N$ works in a career. Their cumulative distributions $P(\geq \frac{N_{\uparrow}}{N})$ for **g** artists, **h** directors and **i** scientists are shown in blue dots. The red line captures the cumulative distribution when the start of a hot streak $N_{\uparrow}$ is distributed randomly among $N$ works. **j–l,** The duration distribution of the hot streak $P(\tau_H)$ for **j** artists, **k** directors and **l** scientists. Dots are measurements from data. Red lines are log-normal fits as guide to the eye. The median $\tau_H$ are $7.3$ years for artists, $7.0$ years for directors, and $4.8$ years for scientists, respectively. Inset, relative duration distribution $P(\tau_H/T)$ for individuals in three domains, where $T$ is the career length of each individual. Solid lines are lognormal fits as guide to the eye. **m–o,** The relationship between $\Gamma_H$ and $\Gamma_0$ for **m** artists, **n** directors and **o** scientists, where the blue background denotes kernel density of data, dots represent binning results of data, and the red line depicts the linear fit. Within each domain, $\Gamma_H$ and $\Gamma_0$ for individuals with early, middle, and late hot streaks can be well approximated by a linear relationship. Inset, the relationship between $\Delta\Gamma$ $(= \Gamma_H - \Gamma_0)$ and $\Gamma_0$ for each domain. **p–r,** The distribution of the number of works produced during hot streaks $P(N_H)$, compared with a null distribution, where we randomly pick one work as the start of the hot streak (**p** artists, **q** directors, and **r** scientists.) We use the Kolmogorov-Smirnov (KS) measure to compare $P(N_H)$ of data with the null distribution, finding that we cannot reject the hypothesis that the two distributions are drawn from the same distribution ($p > 0.05$).

**Figure 3: Hot streaks govern the collective impact of scientific careers.** **a,** Screenshot of Albert Einstein's Google Scholar profile. **b,** Collective impact of a randomly selected scientist in our dataset $D_3$. The publication dates are rearranged such as one produces a constant number of papers each year (lower panel). Vertical lines in the lower panel depict when each paper is published after the rearrangement. The color indicates the order of publications, showing that the sequence of papers published in each career remains intact. The solid line indicates that the paper has been published for at least 10 years (dashed line, otherwise) **c,** For the same scientist as (b), citation patterns of each papers are shown with corresponding colors denoted in the lower panel. The collective impact of a career represents the sum of citation dynamics of all papers published by the individual. **d,** $g(t)$ modelled by (3) given different hot streak parameters (red lines). Here we use $\mu = 7.0$, $\sigma = 1.0$, $\Gamma_0 = 1.0$, and $\tau_H = 3$ years but vary $t_\uparrow$ and $t_\downarrow$. Varying hot-streak parameters of $g(t)$ allows us to reproduce a wide variety of career dynamics that cannot be captured by the null model (blue line). Inset decomposes contributions to $g(t)$ in terms of $g_0(t)$ and $\Delta g(t)$. **e,** The uncertainty envelope of $g(t)$ for an individual in our dataset, where blue dots denote data, the red line is the fitting result of equation (3), and the shaded area illustrates the predicted uncertainty measured in one standard deviation. **f** The fraction of $g(t)$ falling within the envelop $P$(fraction) for the null model (blue area) and our hot-streak model (red area). Fraction = 1.0 indi-

cates the whole $g(t)$ trajectory falls within the envelope. Our model outperforms the null model in capturing individual collective impact as $P$(fraction) peaks close to 1.0. **g,** The average $\langle MAPE \rangle$ of our hot-streak model and the null model for individuals with early, mid and late onset of hot streaks. The difference between the two models is the largest for individuals with early hot streak and smallest for late ones. **h,** $g(t)$ of 50 randomly selected individuals in our dataset whose careers started between 1960 and 1995. Color corresponds to the year when a career started, dots denote collective impact of real careers, and solid lines capture the predictions from the hot-streak model.

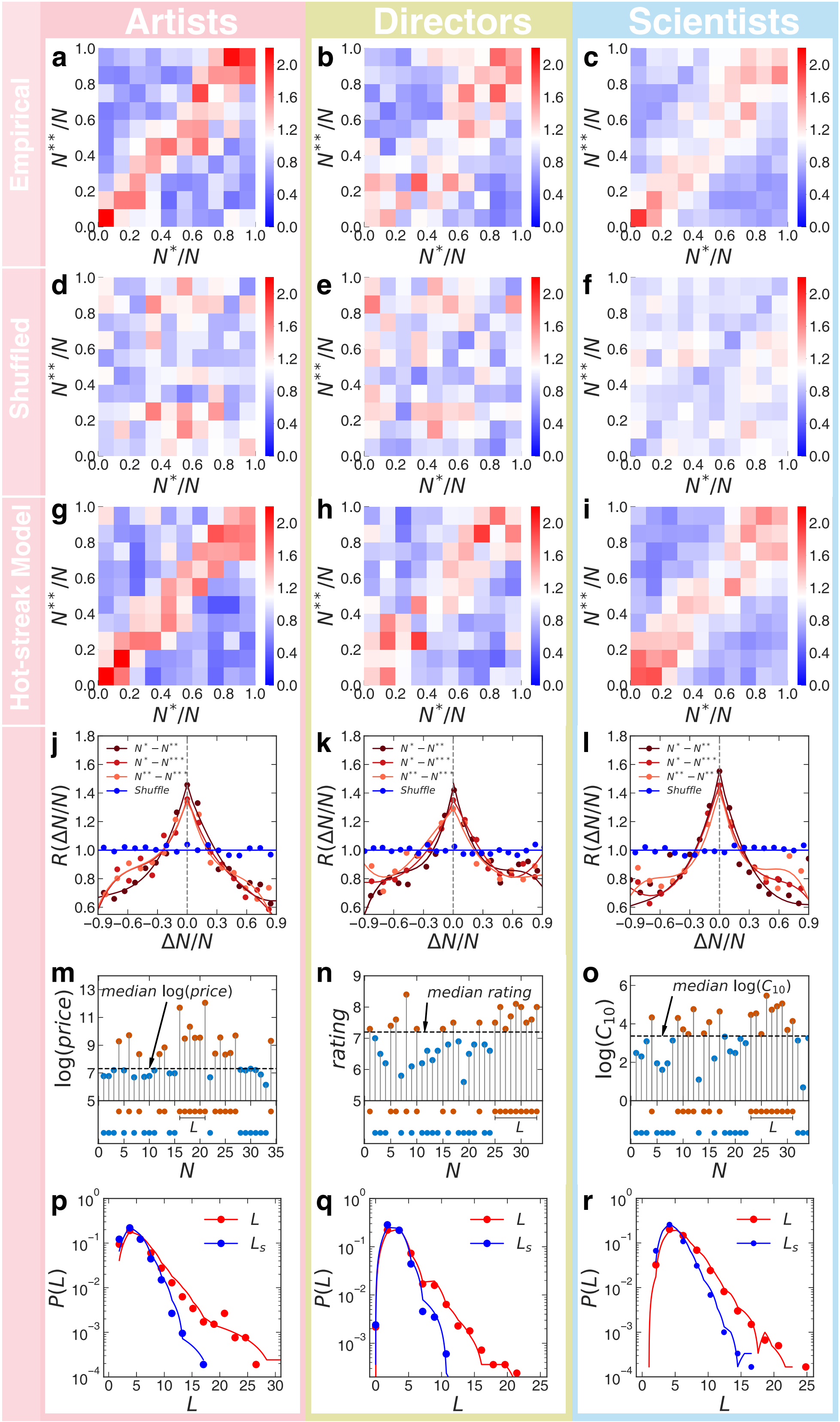

Figure 1:

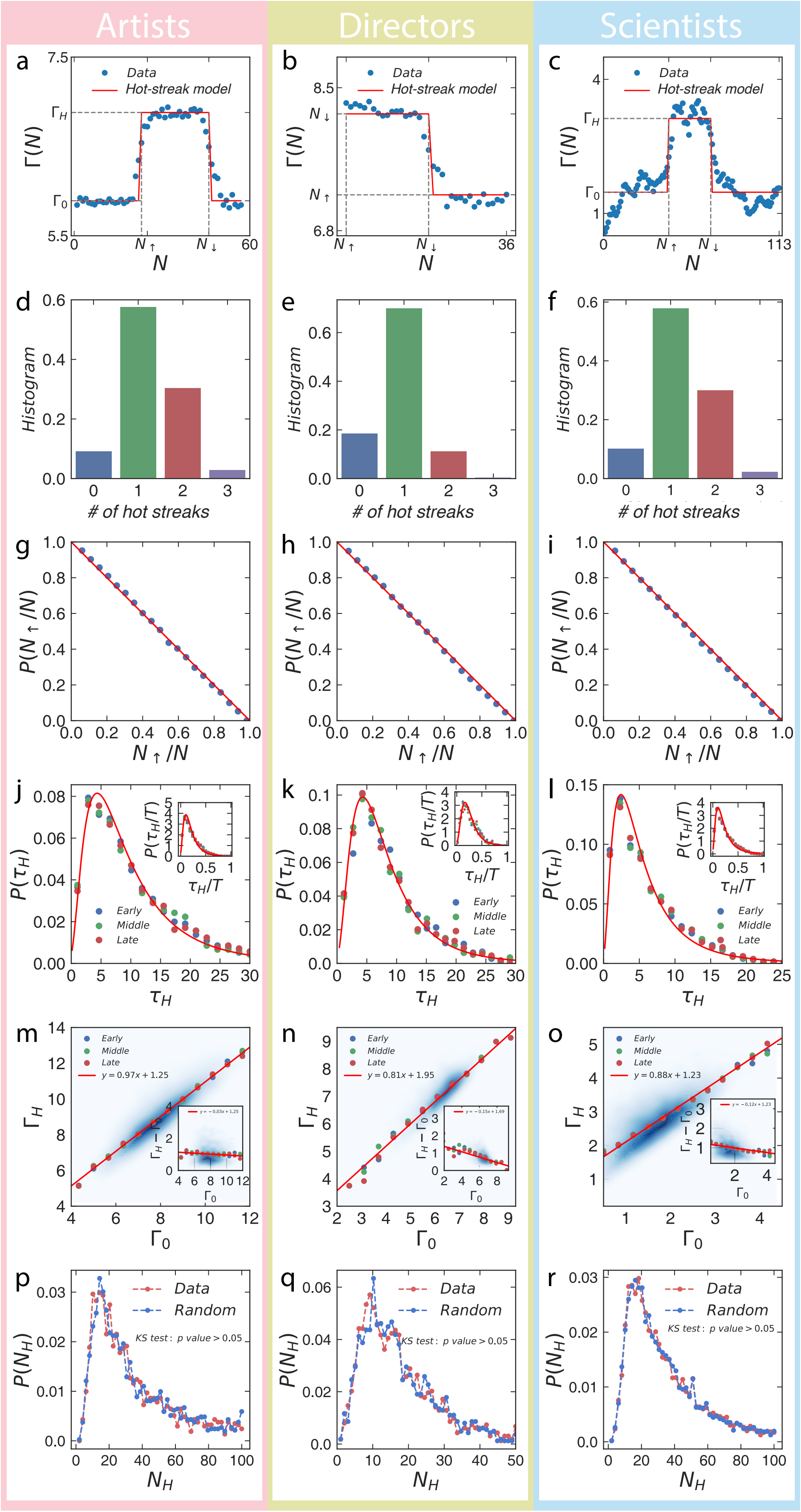

Figure 2:

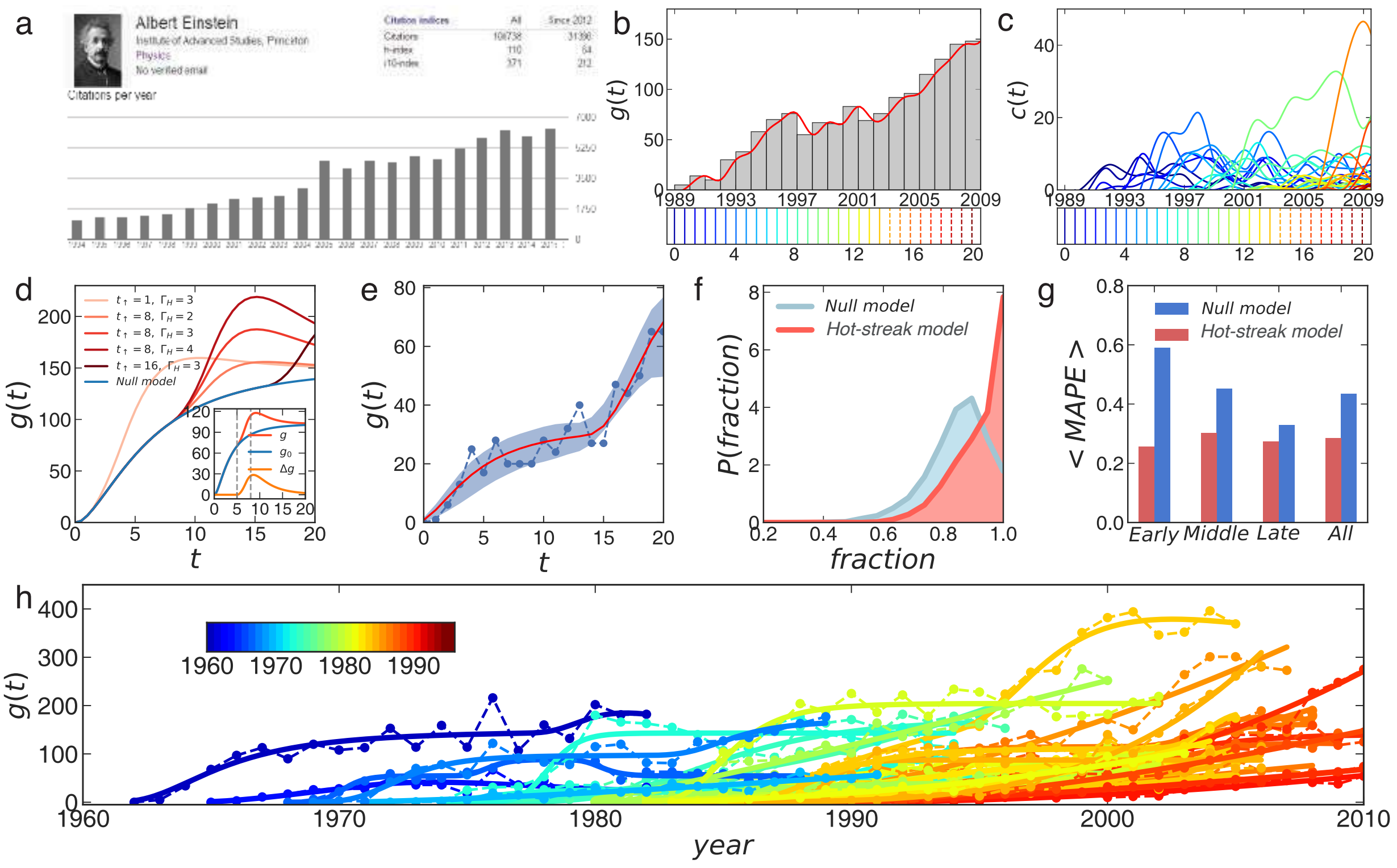

Figure 3:

# Supplementary Information for Hot Streaks in Artistic, Cultural, and Scientific Careers

Lu Liu,[1,2,3] Yang Wang,[1,2] Roberta Sinatra,[4] C. Lee Giles,[3,5] Chaoming Song,[6] Dashun Wang[1,2*]

[1]*Northwestern Institute on Complex Systems, Northwestern University, Evanston, IL, USA*

[2]*Kellogg School of Management, Northwestern University, Evanston, IL, USA*

[3]*College of Information Sciences and Technology, Pennsylvania State University, State College, PA, USA*

[4]*Center for Network Science and Mathematics Department, Central European University, Budapest, Hungary*

[5]*Department of Computer Science and Engineering, Pennsylvania State University, State College, PA, USA*

[6]*Department of Physics, University of Miami, Coral Gables, FL, USA*

**Correspondence should be addressed to D.W. (dashun.wang@northwestern.edu)*

# Contents

## S1 Data Description

In this project, we compiled a comprehensive database consisting of three large-scale datasets of individual careers across three different domains: Dataset $D_1$ contains profiles of artists obtained from online auction databases. Dataset $D_2$ contains profiles of movie directors recorded in the IMDB database. Dataset $D_3$ contains the publication and citation records of individual scientists, obtained by combining Google Scholar and Web of Science. In this section we describe in detail how we collected and reconstructed individual career histories for the three datasets.

**S1.1 Artists $D_1$** Among the three domains we analyzed, the success of artists is probably the most difficult to quantify, hence unsurprisingly the least studied. Indeed, apriori, it may seem that the success of an artist is inherently unquantifiable. Yet more recent developments start to suggest that several measures have the potential to be systematically collected and can be used to measure and compare the impact of artworks[20]. The various measures include auction hammer prices, being selected as illustrations in art textbooks, retrospective exhibitions, museum collections and exhibitions. Just like citations—which have by now been commonly adopted to quantify success of a paper and scientist, despite the fact that they offer at best an incomplete measure of impact that is inherently multi-faceted—these measures of artistic success each capture at best a singular dimension that is to certain degree correlated with the overall "goodness" of an artwork, with their associated limitations (Sec. S1.4). Among all measures, hammer prices are the most commonly used to quantify artistic success[20], perhaps because they reflect the values of artworks judged by art professionals and art markets, serving as a proxy for the impact of artworks. While there are studies showing the career trajectories of artists are rather robust against different measurement choices[20,38], here to ensure our findings are consistent with the state of art, we seek to collect systematically hammer prices of artworks from different sources to reconstruct artistic careers.

We collected information on artistic careers from online art market databases, Artprice[1] and Findartinfo[2]. Both websites offer a comprehensive list of auction records for each artists, with complementary information on various kinds of fine arts ranging from old masters to contemporary art. Indeed, auction records on Artprice offer several useful information that can help quantify artistic careers, including, for each artwork, its auction date, title, year of production, medium, and the price rank among all hammer prices of artworks produced by an artist. Findartinfo contains information on auction date, title, medium, and the actual hammer price for each auction. In this study, we combined the two data sources, allowing us to extract the most comprehensive information tracing individual artistic careers.

Artworks can go through multiple auctions. Artprice website contains auction records from 1983 up to now, helping us ensure that we analyze the latest sales of each artwork. Findartinfo also offers auction records from 2001 till 2015, including the actual hammer price for each record, allowing for cross check and references. Both databases are excellent in their longitudinality, containing artworks produced dating back to the Middle Ages.

In total, we collected 31,101 individual profiles from Artprice and 283,839 profiles from

[1] www.artprice.com

[2] www.findartinfo.com

Findartinfo. Artprice indexed a large number of artists, but here we focused on the top ones with more than 50 records. We conducted a comprehensive entity linking process between the two databases, aiming to match each artist in Artprice with the corresponding profile in Findartinfo. We ignore profiles from the two databases noted as “attributed to” or “attrib.”, since these notions suggest sellers are not clear about the authorship of these artworks. We then cluster remaining profiles with the same last name together in each database, allowing us to match artists within a small subset for computational efficiency. We compare each artist’s profile in Artprice with each profile in Findartinfo with the same last name. Two artists are considered to be the same if they satisfy the following criteria: 1) Initials of the first names are the same. If full names are available for both artists, they have to be identical; 2) They have at least one artwork with an identical title; 3) If more than one artists meet criteria 1) and 2), we pick the one with most matched titles. By applying this entity resolution procedure, we end up with a total of 5,352 matched artists. To evaluate the accuracy of our algorithm in linking the two databases, we compare the number of works for all matched artists in Findartinfo and Artprice in the same period (2001–2015), finding the number of works in the two databases to be similar (Fig. S1).

After linking the two databases, we reconstruct the career histories of artists based on the production year of each work and its hammer price during most recent auction. If an artwork is included in the original Findartinfo dataset, the actual hammer price is used to measure the impact of the artwork. If an artwork is only included in the Artprice dataset, we know the price rank of the artwork among all artworks by the artist. Hence if we need to compute the actual hammer price, we can convert sales rank using rank-to-score conversion given the price distribution measured in Findartinfo[39]. Note that, the uncovered hot streaks documented in the main text (Fig. 1) are independent of whether we use price or rank because the choice does not affect the measurements of relative positions between hit works within each individual. When an artwork was produced during several years, we use the last year as its year of production, corresponding to the year in which the work was finalized. For our final dataset, we selected artists with at least 15 works and 10 years of career length, resulting in 3,480 artists, with careers dating back to as far as 1460.

**S1.2 Movie directors** $D_2$ The Internet Movie Database (IMDB)[3] is the largest movie database around the world, containing information about over one million movies, spanning over 20 genres from 1874 to present. Each movie in the database includes detailed information such as title, release time, casts, crews, an average user rating and the number of votes. IMDB also contains over eight million personal profiles with unique identifiers, each containing personal information, a list of works she was involved in, and specific roles in these works, such as being a director, editor, writer, actress, etc. Although a movie is inherently a product of complex collaborative efforts often involving a large number of individuals, the director of the movie is commonly considered to play a prominent role in cinematic creativity[40,41].

To this end, we gathered 513,306 movie records and profiles of 20,592 directors from IMDB, including those who serve as an assistant director and art director. IMDB provides a rating system for each movie that ranges from one to ten, reflecting the "crowd wisdom" of users, adjusted by the weighting algorithms developed by IMDB to avoid vote stuffing. Previous work has found that metrics quantifying the impact of a movie largely correlate with each other[30]. Here, we use the IMDB rating to approximate the impact of a movie, and construct the sequence of works with their impacts for each director in our dataset. We focused on movies released before 2017 with more than 5 votes. To select directors with long enough career histories, we focused on those who have at least 15 movies and 10 years of career length, resulting in a total of 6,233 directors, whose careers dating back to as far as 1890 ($D_2$).

[3]http://www.imdb.com/interfaces/

**S1.3 Scientists** $D_3$ For studies on scientific careers, automated name disambiguation in large-scale scholarly datasets remains a challenging problem[42,43]. In this study we perform a large-scale disambiguation effort by combining two large datasets, Google Scholar (GS) and Web of Science (WoS). Google offers scholar profile services for individuals scientists to create, maintain, and update their own publication records, assisted by its disambiguation algorithms. Users can adjust the publication records recommended by Google, further ensuring the accuracy and reliability of each profile. Hence, GS offers a comprehensive dataset of individual scientific profiles across different domains that should be at least as accurate as the state of the art, with additional two levels of assurance. As such it has the potential to catalyze more and more research on scientific careers[23,44,45].

To this end, we crawled over 240,000 public profiles from GS in summer 2015. Each GS profile contains the publication records of a scientist, including for each paper its title, publication year, journal, author(s) and cumulative citation count within GS database calculated up to the point when the data was collected. GS encourages users to enter other information including affiliation, research interest, homepage, and collaborators. Users with verified academic email address are noted in their GS pages. Here, we choose those with verified email address and co-authors as they are considered well-maintained. We further remove individuals with more than 1,000 publications in GS, finding most of them have Asian names which are the most difficult to disambiguate. Here again the goal is to follow closely the same procedures used by existing studies in this area[23,44,45], so that every finding in our paper is immediately comparable with previous literature to ensure our results are consistent hence new findings reliable given the existing literature.

Since GS only provides the cumulative citation count of each paper by the time of data collection, to study the dynamical impact of each scientist we linked GS with the WoS dataset, which provides comprehensive citation records of around 46 million journal papers published after 1900. For each scientist, we match each publication in her GS profile with the corresponding paper in WoS. We conducted a comprehensive linkage process that takes into account not only the author name and paper title, but also the metadata available for each publication. For each scientist in GS we first consider a pool of WoS papers having an author with the same last name. Then for each publication in her GS profile, we calculated the cosine similarity between the title of each GS record and each WoS paper in the subset. We then consider a record in GS profile matched with a WoS publication if the following criteria are met: 1) The WoS paper was published within $\pm 2$ years of the GS publication year; 2) There is at least one co-author sharing the same last name in GS and WoS if the paper has more than one authors; 3) The cosine similarity between titles is higher than 0.5 after removing stop words. If there are multiple papers, we choose the one with the most similar title (highest cosine similarity). If, using these criteria, we could not find a corresponding paper in WoS, we consider the GS record is not matched to ensure the quality of the resulting dataset. We choose scientists with at least 15 papers and 20 years of career length, resulting in 20,040 profiles for our analyses ($D_3$).

To approximate the impact of each paper, we follow recent studies[17,21,23] to calculate the number of citations the paper received after 10 years of publication $C_{10}$. Previous studies have shown that for WoS dataset the average citation counts of a paper increase over time[18,23,46], which we verified to be the case for our dataset (Fig. S2). To make sure our results are not affected by this temporal effect, we follow previous studies[18,23] and use a rescaled $C_{10}$, defined as $C_{10}/\langle C_{10}\rangle_y$, to gauge the impact of a paper, where $\langle C_{10}\rangle_y$ is the average impact of all papers in WoS published in the same year $y$. We report in S2 and Fig. 1 the results based on the rescaled $C_{10}$. We find the use of raw or rescaled $C_{10}$ does not alter our conclusions.

**S1.4 Data Limitations** Although the datasets curated in our study are among the largest collections of individual careers, our data are not without limitations. While auction data are most commonly used to quantify impact of an artwork, it is important to understand a potential bias involved in using auction data to measure and compare the relative value of an artwork[20]. This issue is of particular relevance for famous artists, whose work is eagerly sought after by museums. While museums sometimes take part in auctions, they are in general less likely to sell a collection. The "museum bias" may also affect works differently. For example, if we consider the probability that an oil painting may be owned by a museum, late paintings tend to have a higher probability than early paintings. Also museums may not take paintings randomly from a career, but are rather biased toward the best works. Hence the hits we observed in some careers may represent a lower bound of the real number of hits. But thankfully, we find our results are rather robust when we vary the threshold for hits. Despite these limitations, one possible reason why auction prices remain the best measures to quantify the importance of artworks is rooted in the fact that they are significantly better than alternatives. For example, other measures like museum exhibitions mainly focus on masterpieces of famous artists, limiting the scope and data availability for quantitative analysis.

Datasets about movies and scientific publications are more frequently studied quantitatively than artworks, hence the metrics used in those cases are better defined. But it is still important to remember that similar biases also exist. For example, the IMDB rating could be biased toward the judgement of general audience, whereas professional film critics may have different opinions. Papers could be highly cited for many reasons. There are many scientific discoveries that are ground-breaking but have few citations.

Lastly, for any study regarding individual careers, name disambiguation is always an issue that should not be overlooked. For the three domains analyzed here, name disambiguation issues may potentially affect results regarding artists and scientists, but less so for movie directors, as they are associated with unique identifiers. Another issue when studying individual careers is that, as we need enough records to study each individual, the insights obtained are inherently biased towards productive individuals who have produced much with long enough careers.

Amid all the limitations described in this section, it should be clear, however, that these limitations are by no means unique to our study; Rather, they apply to most, if not all studies in this domain. In many ways, research progress on individual careers is hindered by the inherent difficulties to collect and curate high-resolution individual career histories from diverse domains. Hence another contribution of our paper is to make available to broad research communities the comprehensive datasets we collected in this paper. By doing so, the hope is that these datasets could be an important asset for researchers from many different disciplines as they significantly improve the data-scarce situation, allowing researchers to build and develop new findings. To this end, we created a dedicated website to host datasets and descriptions, with the hope to update them as they develop. The website can be access here http://personal.psu.edu/ lpl5107/data/data/data/.

## S2 Empirical Measurements

**S2.1 Impact distribution** We use auction hammer price, IMDB rating, and paper citation in 10 years $C_{10}$, to approximate the impact of works for artists, movie directors, and scientists, respectively.. To study the work impact across three domains, we measure the distribution of hammer prices, IMDB ratings and paper citation count in our datasets, finding they follow quite different distributions. The hammer price for all artworks in our dataset can be approximated by a log-normal distribution (Fig. S3a) following

$$P(x) = \frac{1}{x\sqrt{2\pi}\sigma} e^{-\frac{(\log x - \mu)^2}{2\sigma^2}} \tag{S1}$$

We fit Eq. S1 to $P(\text{price})$ measured from data, and obtain the estimated parameters $\mu = 7.91$ and $\sigma = 1.55$. Similarly, we find the distribution for both raw and rescaled $C_{10}$ in our dataset also follows a log-normal distribution (Figs. S3c–d), which is consistent with previous studies[17,18,23]. In contrast, the IMDB rating follows a normal distribution ranging between 1 and 10 (Fig. S3b).

Hence, to study the impact across three domains, we define a goodness parameter $\Gamma$ for artists, directors and scientists as $\log$(price), IMDB ratings and $\log(C_{10})$, respectively.

**S2.2 Random impact rule** One school of thought on the lifecycle of creative careers suggests a hit work within individual career is largely driven by chance, having a constant probability to appear in unit of works within each career[13,23,47,48]. To verify this hypothesis in our dataset, we study the position $N^i$ of each top $i$ highest-impact work in the sequence of $N$ works within each career, and measure the complementary cumulative distribution function $P(\geq N^i/N)$ within the sequence of works produced by individuals in different domains. We find for each of the top 3 hit works in an artistic career, $P(\geq N^i/N)$ decreases linearly as $(N^i/N)^{-1}$, corresponding to a uniform $P(N^i/N)$ (Fig. S4a), suggesting the most expensive works of an artists is randomly distributed within her career. Similarly, we find the same pattern of $P(\geq N^i/N)$ for directors and scientists (Figs. S4b–c), suggesting the highest rated movies and the most cited papers also appear randomly within each career. Hence, random impact rule applies across three domains.

**S2.3 $\Phi$ for other pairs of hit works** In the main text, we observed the normalized joint probability $\Phi(N^*, N^{**})$ is overrepresented along the diagonal for artists, directors and scientists, suggesting the colocation of the top two hit works within a career. To study if the colocation applies to other pairs of hit works, we calculate the normalized joint probability $\Phi(N^*, N^{***})$ for the highest and third highest impact works, and $\Phi(N^{**}, N^{***})$ for the second and third highest impact works within each career across three domains. We first measure $\Phi(N^*, N^{***})$ and $\Phi(N^{**}, N^{***})$ in real careers (Fig. S5), finding a similar diagonal pattern for $\Phi(N^*, N^{***})$ and $\Phi(N^{**}, N^{***})$, indicating the temporal collocation applies to other pairs of hit works. $\Phi(N^*, N^{***})$ and $\Phi(N^{**}, N^{***})$ both feature an even split across the diagonal, suggesting the equal probability of each hit to appear before or after another hit. We find the colocation of other pairs of hit works is less significant than the top two hits (lighter color along the diagonal), which is also consistent with the lower peak value of $R(\Delta N/N)$ in Figs. 1j–l. However, if we shuffle the order of works within each career while keeping their impact intact, the diagonal pattern of $\Phi(N^*, N^{***})$ and $\Phi(N^{**}, N^{***})$ disappears across three domains (Fig. S6), suggesting the colocation of any pair of hit works observed in data cannot be explained by the random impact rule.

**S2.4 Measurement under different career length** The empirical observations in Fig. 1 are based on artists and directors with at least 10 years of career length, and scientists with at least 20 years of career length. To study if the random impact rule and the temporal correlation of hit works is influenced by the career length we measured, we selected two groups of individuals with longer career length in each domain: artists with at least 20 and 30 years of career length, directors with at least 20 and 30 years of career length, and scientists with at least 30 and 40 years of career length, and repeated the empirical measurements on the random impact rule and the temporal colocation for the two groups of individuals in each domain. First, we find the $P(\geq N^i/N)$ still follows a uniform distribution given different career length for artists, directors and scientists (Fig. S7), suggesting the random impact rule is independent of the career length we measured. Second, we calculate the temporal distance $R(\Delta N^*/N)$ for the hit and second hit within a career, where $\Delta N^*$ is defined as $N^* - N^{**}$. We find $R(\Delta N^*/N)$ peaks around zero featuring a symmetric split along the x-axis (Fig. S8), suggesting the colocation of hit works is not influenced by career length.

Third, we find the tail of both $P(L)$ and $P(L_S)$ follows an exponential distribution. $P(L)$ features a much wider than $P(L_S)$ (Fig. S9), suggesting the long streaks sustain in careers of different length.

**S2.5** $P(L)$ **under different threshold** We have studied the streak of works above the median impact within a career in Fig. 1, finding $P(L)$ observed in real careers has a longer tail than that of shuffled careers. To test if the conclusion is robust to different choices of impact threshold, we changed the threshold to the mean impact and geometric mean of impact within each career for artists, directors, and scientists. We find similar results for $P(L)$ and $P(L_S)$ given different impact threshold.

**S2.6 Difference between** $P(L)$ **and** $P(L_S)$ To quantify the different probabilities to observe streaks for real and shuffled careers (Figs. 1p–r), we measured $P(L)/P(L_S)$, capturing how more likely it is to observe streaks than random at different streak length. Since we care about the probability of long streaks, we focus on the tail of $P(L)$ and $P(L_S)$, and fit their tails to an exponential distribution. We then compare the tail difference of $P(L)$ and $P(L_S)$ based on the fitting results across three domains (Fig. S11). We find for artists and scientists, the probability for an individual to produce consecutive 20 high impact works above the median impact within a career, is around 20 times more likely than shuffled careers. The probability is over 100 times more likely for directors.

## S3 Hot-streak Model

**S3.1 Hot streak studies in the literature** The debate on hot streaks dates back to 1985, when Gilovich *et al.* presented that the hot-hand belief on basketball players is merely a cognitive bias of random process[1] . Since then, hot hand has been widely studied in sports, financial markets and gambling. These studies tried to answer two fundamental questions: 1) the statistical evidence on whether the hot hand exists[1–3], and 2) the psychological origin of the hot hand belief[1,49–52]. Here we conduct a comprehensive review of the existing literature related to the first question. Table S1 reviews selected empirical studies of whether hot hand exists. Table S2 reviews different mathematical models to detect hot hand. Hot hand has been measured and reported by independent research groups, each using different datasets and statistical models mainly in sports, financial markets and gambling. In this study, we extend the analyses on hot hand to different domains, focusing on individual creative careers.

**S3.2 Null model** To uncover the regularities behind the empirical observations, we first introduce a null model motivated by the random impact rule. That is, we assume each individual $i$ produces a sequence of $N$ works whose impact (i.e.. $\log(\text{price})$ for artists, ratings for directors, and $\log(C_{10})$ for scientists) is randomly chosen from an impact distribution $P(\Gamma) = \mathcal{N}(\Gamma_i, \sigma_i^2)$, where $\Gamma_i$ is a constant goodness parameter specific to each individual, and $\sigma_i$ reflects the impact fluctuation within each career. For simplicity, we assume $\sigma_i$ to be the same for each individual in a domain. The null model allows us to simulate impacts of the works produced by an individual. For each individual, we use real productivity $N$ as input, and assume $\Gamma_i$ as the average impact measured from each individual, and $\sigma_i = 1.0$ is a constant for all the individuals. We repeated the measurements of $P(\geq N^i/N)$ and $\Phi$ to verify its prediction on the random impact rule and the temporal correlation, finding the null model can reproduce the random impact rule of the top three hit works across three domains (Fig. S12), while it fails to capture any temporal clustering among hits (Fig. S13), suggesting that there are other factors affecting individual careers.

**S3.3 Generative hot-streak model** The failure of the null model prompts us to abandon our hypothesis that $\Gamma_i$ is constant with each career. Each hit work is random while their relative position is not, indicating the presence of a period of outstanding performance ($\Gamma_H$) that appears randomly with a career. Using real productivity $N$ as input, the hot-streak model allows us to generatively simulate the impacts of the works produced by an individual. For each individual $i$, the impact of a work is randomly drawn from a normal distribution $\mathcal{N}(\Gamma_H, \sigma_i^2)$ if it is produced during hot streaks, or $\mathcal{N}(\Gamma_0, \sigma_i^2)$ otherwise. To define a random hot streak in each career, we randomly pick one work out of the sequence of $N$ works she produced, and denote its year of production as $t_\uparrow$, marking the start of the hot streak. For simplicity, we assume $\Gamma_0$, $\Gamma_H$, $\sigma_i$, and $\tau_H = t_\uparrow - t_\downarrow$ to be the same for each individual in a domain. The result reported in Fig. 1 is based on the following parameters: $\Gamma_0 = 6.9$, $\sigma = 1.1$ and $\tau_H = 6$ years for artists; $\Gamma_0 = 6.5$, $\sigma = 1.1$ and $\tau_H = 6$ years for directors; and $\Gamma_0 = 3.0$, $\sigma = 1.3$ and $\tau_H = 4$ years for scientists, with $\Gamma_H = \Gamma_0 + 1.0$ for individuals in all three domains. Although it is a simple generative model with four parameters, our hot-streak model can reproduce all the empirical findings measured from a variety of individual careers.

**S3.4 Estimation of hot streaks** In order to test how well our hot-streak model matches empirical data, we need to estimate the model parameters for each individual. We show in this section that the impact of works within a career can be captured by a time-dependent variable $\Gamma(N)$, defined as the average of impact calculated by rolling a window of size $\Delta N = \max(5, 0.1N_T)$ over an individuals sequence of works, where $N_T$ is the total number of works produced by an individual. In order to capture the average performance during a period, we assume the window size to account for 10% of all the works one produced, and set the lower bound of $\Delta N$ as 5 in order to calculate $\Gamma(N)$ based on enough works. In order to remove any potential boundary effect, we calculate $\Gamma(N)$ from $\Delta N/2$, whose value is defined by the average $\Gamma$ between 0 and $\Delta N$, and then move the sliding window one work per step until $N_T - \Delta N/2$. $\Gamma(N)$ reflects the average impact of works between $N-\Delta N/2$ and $N+\Delta N/2$. $\Gamma(N)$ is calculate from $\log(\text{price})$ for artists, IMDB ratings for directors, and the raw $\log(C_{10})$ sequence for scientists. We find for scientists calculating the raw or the rescaled $C_{10}$ does not alter the trend of $\Gamma(N)$ (Fig. S14a). Indeed, we measure the Pearson correlation between the $\Gamma(N)$ sequence calculated from raw $\log(C_{10})$ and rescaled $\log(C_{10})$ for each individual, finding the distribution of correlation coefficient $P(\rho)$ peaks around 1 with mean value 0.93 (Fig. S14b), suggesting the two sequences are highly correlated and the rescale does not change the trend of $\Gamma(N)$ in general. Since the raw $\log(C_{10})$ is easier to interpret, we report results related to $\Gamma(N)$ based on the raw $\log(C_{10})$.

For each individual in our dataset, we use a piecewise function to fit the sequence of $\Gamma(N)$ measured from real careers. Specifically, we relax the number of hot streaks to at most three, and allow the $\Gamma_H$ of each hot streak to be different. We used ordinary least square (OLS), and adopted

scipy.optimize package to fit the piecewise function to data. Besides, to overcome the over-fitting problem, we added the $L1$ regularization term to the cost function that penalizes the number of hot streaks. For each individual, we repeated the fitting procedures for 20 realizations, and selected the results with the smallest cost. We show more fitting results for individuals across three domains in Figs. S15–S17. To define the hot streak in each career, we assume the smallest fitted $\Gamma(N)$ as $\Gamma_0$ for an individual. To make sure the fitted hot streak reflects a substantial period of improved performance, we set a threshold for both the duration and the intent of each hot streak. Specifically, for any $\Gamma$ larger than $\Gamma_0$, if the difference between the two is larger than the inherent noise within a career (standard deviation of $\Gamma(N)$), we then define it as a hot streak. Besides, we assume $\Gamma_H$ should at least last for 5 works, otherwise we think the duration is too short, and regard it an outlier of normal performance.

**S3.5 Model validation: Fitting performance** To systematically evaluate the goodness of fitting for the procedure proposed above, we measure the difference between fitted and real $\Gamma(N)$ by calculating the coefficient of determination $R^2$. To study if $R^2$ between fitted and real $\Gamma(N)$ can be explained by the inherent noise of $\Gamma(N)$ sequence, we calculate the expected value of $R^2$ generated by the inherent noise of the impact sequence. To do so, we use fitted $\Gamma(N)$ as input, and simulate the impacts of works within a career for each individual, by assuming the impact of each work is randomly selected from a normal distributions $\mathcal{N}(\Gamma_H,\, \sigma_s^2)$ if it was produced during hot streaks, or from $\mathcal{N}(\Gamma_0,\, \sigma_s^2)$ otherwise. We assume $\sigma_s$ to be the same for all individuals in a domain.

To determine the value of $\sigma_s$, we calculate the difference $\sigma$ between the real $\Gamma(N)$ and fitted

$\Gamma(N)$, and measure the distribution $P(\sigma)$ for all individuals in each domain. We find $P(\sigma)$ follows a normal distribution that peaks around zero for artists, directors and scientists (Fig. S18). The standard deviation for $P(\sigma)$ is $0.186$ for artists, $0.229$ for directors, and $0.189$ for scientists. We approximate the standard deviation for $P(\sigma)$ as the noise in each domain, respectively. Using these numbers as input, we simulated the impacts of works within a career for 1000 realizations, allowing us to calculate a distribution $P(R^2)$ for each individual. To study if $R^2$ of real and fitted $\Gamma(N)$ can be explained by noise, we define a baseline $R^2$ that is the lowest 5% of all simulated $R^2$ (p-value $= 0.05$). If the $R^2$ of data and fitted $\Gamma(N)$ is larger than the baseline $R^2$, we assume the error is mainly generated by noise, and $\Gamma(N)$ is well captured by our hot-streak model (Fig. S19). We find for individuals in our dataset, over 69% artists, 80% directors and 75% scientists have $R^2$ larger than the baseline $R^2$, suggesting our hot-streak model captures the majority of individuals in our dataset.

We also compared the fitting performance of our hot-streak model with the null model by calculating adjusted $R^2$ and Bayesian information criterion (BIC), both penalizing the number of parameters in the model. Compared with the null model, we find our hot-streak model has systematically larger adjusted $R^2$ and smaller BIC for individuals across three domains (Fig. **??**), suggesting out hot-streak model better captures the dynamics of $\Gamma(N)$ than the null model.

**S3.6 Model validation: Impact before and after a hot streak** Our hot-streak model assumes that after a hot streak, the individual performance returns back to her normal performance. To test this assumption, we measure for each individual the average impact of all works produced during the normal performance before and after each hot streak, where the normal performance and the hot streak is defined by the fitted $\Gamma(N)$. We calculate the distribution of the difference between average impact before and after the hot streak $P(\Delta\langle\Gamma\rangle)$ that $\Delta\langle\Gamma\rangle = \langle\Gamma\rangle_{after} - \langle\Gamma\rangle_{before}$, finding $P(\Delta\langle\Gamma\rangle)$ follows a normal distribution peaks around zero (Fig. S20) for individuals across three domains, suggesting there is no systematic different between the average impact before and after a hot streak.

**S3.7 Individuals with different number of hot streaks** In this section, we discuss if there is any difference among individuals with zero, one and more than one hot streaks across three domains. We first compare the distribution of average impact $P(\Gamma)$, the number of works $P(N)$, and the career length $P(T)$ for individuals with and without hot streaks, finding there is no significant difference in terms of impact, productivity and career length between the two groups of people (Fig. S21). However, we observe differences in the distribution of $\Gamma_0$ and $\Gamma_H$ for individuals with different number of hot streaks. Indeed, for scientists and directors, $P(\Gamma_0)$ is smaller for individuals with one hot streak than without hot streaks (Figs. S22b–c), and $P(\Gamma_H)$ is higher for individuals with more than one hot streaks (Figs. S22e–f), reflecting the inherent variance of impact within a career for directors and scientists. While we find for artists the distribution $P(\Gamma_0)$ and $P(\Gamma_H)$ is the same for those who different number of hot streaks. The difference between artists and the other two domains is probably due to the relation between $\Gamma_0$ and $\Gamma_H$, that it is more difficult for individuals with high $\Gamma_0$ to gain better performance during hot streaks for directors and scientists. Hence, the number of hot streaks with a career captures an orthogonal dimension to current metrics characterizing individual careers.

**S3.8 The correlation between $\Gamma_0$ and $\Gamma_H$** To avoid potential bias that our results in Figs. 3M–O are affected by the underlying distribution of $\Gamma_0$ and $\Gamma_H$, we repeated our analysis on the relation of $\Gamma_0$ and $\Gamma_H$ by using the their percentile, finding the linear relationship between $\Gamma_0$ and $\Gamma_H$ remains the same for individuals across three domains (Fig. S23), suggesting our conclusion is insensitive to the distribution of $\Gamma_0$ and $\Gamma_H$.

To test the significance of the correlation between $\Gamma_0$ and $\Gamma_H$ observed in data, we propose a null model that assumes they are uncorrelated. Specifically, we assume for the null model that $\Gamma_H$ and $\Gamma_0$ follow the same distribution. For a given $\Gamma_0$, $\Gamma_H$ is calculated as the expectation of $P(\Gamma)$ ranging from $\Gamma_0$ to infinity under the null model. We find the null model prediction of $\Gamma_H$ for a given $\Gamma_0$ is much higher than the real $\Gamma_H$, suggesting the impact during hot streaks of each individual cannot be explained by chance, but largely depends on her normal performance.

**S3.9 Discussion on scientific disciplines** To compare the hot-streak properties for scientists from different disciplines, we identify each scientist her discipline by using the scientific journal categories provided by WoS[18]. The publications collected in $D_3$ belong to journals from 145 different categories in WoS ranging from Acoustic to Zoology. For each scientist, we count the number of papers published in each of the 145 categories. Since a journal may belong to multiple categories, and each category includes a list of journals, we count each paper multiple times in all related categories. We consider the category with the most publications within a scientific career as her research discipline. Hence, each scientist in our dataset can be only assigned to one discipline. Here we study the top 30 disciplines with the most scientists in our datasets. We find the ratio of scientists with one hot streak to be stable across disciplines, accounting for roughly 60% within each discipline (Fig. S24a), where the ratio is highest (up to 70%) for scientists from oncology, and the lowest (around 50%) for scientists from ophthalmology. We also calculate the average duration of the hot streak $\langle \tau_H \rangle$ for scientists from 30 disciplines (Fig. S24b), finding scientists from geophysics have on average the longest hot streak (around 7.5 years), whereas scientists from chemistry have the shortest hot streak on average (around 3 years). The robustness across different disciplines indicates the generalizability of the hot-streak model.

## S4 Relationship with Existing Models

In this section, we discuss the relationship between the proposed hot-streak model with previous studies on the individual impact. Here we will focus on the most recent model, namely the $Q$-model[23]. Different from the assumption that scientists with similar productivity have indistinguishable impact, the $Q$-model suggests the existence of a hidden parameter $Q$ unique to each individual. Although the hot-streak model builds upon a similar conceptual basis to the $Q$-model, the main findings of our model demonstrate that the $Q$-model is not sufficient to capture the impact dynamics within each career. The reason is simple: the $Q$-model is mainly designed to capture the impact differences *across* individuals, focusing on the overall performance of an individual rather than the how impact changes *within* a career, which is the focus of our paper. Next we show in this section the mathematical consistency between our hot-streak model and the $Q$-model. As such, the hot-streak model not only helps us explain the temporal regularities documented in the main text not anticipated by the $Q$-model, but is also able to reproduce all the predictions the $Q$-model makes on individual careers.

**S4.1 Compare $\Gamma$ and $Q$ parameter** The $Q$-model models the citation after 10 years $C_{10}$ of a paper $\alpha$ for a scientist $i$ as the multiplication of two parts[23]:

$$C_{10,i\alpha} = Q_i p_\alpha, \tag{S2}$$

where $Q_i$ is a individual-specific parameter, and $p_\alpha$ is the luck component that is same for every individual. The $Q$-model assumes $Q_i$ to be a constant over time for each individual, and $p_\alpha$ follows a log-normal distribution that $P(\log p) = \mathcal{N}(\mu_p, \sigma_p)$ same to all scientists. Hence, we can take the logarithms of Eq. S2 that $\log(C_{10,i\alpha}) = \log(Q_i) + \log(p_\alpha)$. Noting that $C_{10}$, $p_\alpha$ and $Q_i$ all follow a log-normal distribution, the impact of a paper produced by a scientist $i$ is randomly drawn from a normal distribution that

$$P(\log C_{10,i\alpha}) = \mathcal{N}(\mu_p + \log Q_i, \sigma_p), \tag{S3}$$

whose mean value is modulated by the individual-specific $Q_i$ parameter. Comparing Eq. S3 with the definition of $\Gamma$ parameter in S2.1, we find our $\Gamma$ parameter can be expressed as

$$\Gamma_{i\alpha} = \log(Q_i) + \mu_\alpha. \tag{S4}$$

Hence, the $\Gamma$ parameter combines the two parts in the $Q$-model, reflecting both the individual differences and the luck component.

The hot-streak model contains all the ingredients of the $Q$-model, while it also captures the temporal colocation of hit works within a career by introducing a temporal variation of $\Gamma$, allowing an individual to improve her typical performance ($\Gamma_0$) to a higher level ($\Gamma_H$) for a while. With this simple modification, we show, for the first time, that the dynamics of impact within an individual career is characterized by a remarkable degree of regularity. The idea behind changing $\Gamma_0$ to $\Gamma_H$ is the improvement of individual skill from $Q_i$ to a higher value. It is also worth noting that the duration of a hot streak is relatively short compared to the career length we measured, usually lasting for 5 years or accounting for 20% works produced by a scientist. Hence, our hot-streak model can be approximated as the $Q$-model in the long run. In sum, our hot-streak model is consistent with the $Q$-model, in terms of the definition of $\Gamma$ parameter, and the generative process of impacts within a career.

**S4.2 Q-model validation for scientists** In this section, we validate the $Q$-model for scientists in our dataset. We also demonstrate the hot-streak model successfully reproduces all the $Q$-model predictions on the impact across scientists. To validate the $Q$-model in our dataset, we first calculated the distribution of impact and productivity, finding $P(\log(C_{10}))$ and $P(N)$ both follow a log-normal distribution that meets the assumption of the $Q$-model (Fig. S3c, Fig. S25a). We then used the maximum likelihood method to estimate the parameters $\mu$ and $\Sigma$ of the trivariate normal distribution $P(\log p, \log Q, \log N) \sim \mathcal{N}(\mu, \Sigma)$ in the $Q$-model. The estimated parameters for the joint probability $P(\log p, \log Q, \log N)$ is

$$\mu = (\mu_p, \mu_Q, \mu_N) = (1.39, 0.72, 3.43)$$
$$\Sigma = \begin{pmatrix} \sigma_p^2 & \sigma_{p,Q} & \sigma_{p,N} \\ \sigma_{p,Q} & \sigma_Q^2 & \sigma_{Q,N} \\ \sigma_{p,N} & \sigma_{Q,N} & \sigma_N^2 \end{pmatrix} = \begin{pmatrix} 1.67 & 0.009 & 0.002 \\ 0.009 & 0.56 & 0.11 \\ 0.002 & 0.11 & 0.77 \end{pmatrix} \tag{S5}$$

The matrix Eq. S5 is consistent with the $Q$-model that $\sigma_{p,Q} = \sigma_{p,N} \approx 0$. The estimated $Q$ parameter also follows a log-normal distribution (Fig. S25b).

The $Q$-model successfully explains the impact differences across individual scientists by capturing the scaling of $\langle C_{10}^{*} \rangle$ with productivity $N$ and with the logarithm of the average impact without the hit $\langle C_{10}^{-*} \rangle$[23]. To validate the $Q$-model in our dataset, we calculated the predictions of the $Q$-model by using Eq. S5, and compared them with the empirical observations, finding the results of the $Q$-model are aligned with data for the relationship between $\langle C_{10}^{*} \rangle$ and $N$, $\langle C_{10}^{*} \rangle$ and $\langle C_{10}^{-*} \rangle$ (Figs. S25c–d). Besides, to test if our hot-streak model can capture the predictions made by the $Q$-model, we simulated the impacts of each scientists using the generative hot-streak model described in Sec. S3.3, and repeated the analyses for $\langle C_{10}^{*} \rangle$ between $N$ and $\langle C_{10}^{-*} \rangle$. We find our hot-streak model can also capture these the empirical observation on the impact differences across scientists (Figs. S25c–d), suggesting our hot-streak model is consistent with the Q-model when comparing impact across different scientists.

**S4.3 Q-model validation for artists and directors** To study if the $Q$-model can be used to model the impacts of works for artists and movie directors, we repeated the procedures to estimate the $Q$-model parameters similar to scientists. Assume the auction price of each work is generated from the joint probability of $p(p, N, Q)$, where $p$ is the global distribution of price. To obtain the parameters for $p(p, N, Q)$, we first measured the distribution of auction price $P(\text{price})$ for all works, and productivity $P(N)$ for all artists in our datasets, finding they both follow a log-normal distribution (Fig. S3a, Fig. S26a). The maximum-likelihood estimation allows us to calculate the parameters for the joint probability of $p(\hat{p}, \hat{N}, \hat{Q})$, where $\hat{p} = \log p$, $\hat{N} = \log N$, and $\hat{Q} = \log Q$. Similarly, we find $p(\hat{p}, \hat{N}, \hat{Q})$ follows a trivariate normal distribution whose

$$\mu = (\mu_p, \mu_Q, \mu_N) = (6.5, 1.8, 4.03),$$
$$\Sigma = \begin{pmatrix} \sigma_p^2 & \sigma_{p,Q} & \sigma_{p,N} \\ \sigma_{p,Q} & \sigma_Q^2 & \sigma_{Q,N} \\ \sigma_{p,N} & \sigma_{Q,N} & \sigma_N^2 \end{pmatrix} = \begin{pmatrix} 1.43 & 0.28 & 0.52 \\ 0.28 & 1.26 & 0.25 \\ 0.52 & 0.25 & 1.41 \end{pmatrix} \tag{S6}$$

Notice that different from scientists, $Q$ and $N$, $N$ and $p$ have positive correlations. The estimated $Q$ parameter follows a log-normal distribution as well (Fig. S26b), suggesting the validity of the $Q$-model for artists. Besides, the $Q$-model predictions for artists are in agreement with data in terms of the correlation between $N$ and $\log$ price*, and the correlation between $\langle \log \text{price}^{-*} \rangle$ and $\log$ price* (Figs. S26c–d). Hence for artists, the $Q$ parameter can be calculated as $Q = e^{<\log price> - \mu_{price}}$, where $u_{price} = 6.5$.

Similarly, we repeated the same procedures for movie directors. We find $P(\text{rating})$ follows a normal distribution and $P(N)$ follows a log-normal distribution (Fig. S3b, Fig. S27a), allowing us to calculate the parameters for the joint probability $p(p, \hat{N}, \hat{Q})$, where $\hat{N} = \log N$, and $\hat{Q} = \log Q$, where $p$ is the global distribution for ratings. $p(p, \hat{N}, \hat{Q})$ follows a trivariate normal distribution whose

$$\mu = (\mu_p, \mu_Q, \mu_N) = (5.9, 0.9, 2.8)$$
$$\Sigma = \begin{pmatrix} \sigma_p^2 & \sigma_{p,Q} & \sigma_{p,N} \\ \sigma_{p,Q} & \sigma_Q^2 & \sigma_{Q,N} \\ \sigma_{p,N} & \sigma_{Q,N} & \sigma_N^2 \end{pmatrix} = \begin{pmatrix} 1.3 & 0.25 & 0.38 \\ 0.25 & 0.8 & 0.22 \\ 0.38 & 0.22 & 1.0 \end{pmatrix} \tag{S7}$$

The estimated $Q$ parameter follows a log-normal distribution as well (Fig. S27b). The $Q$-model predictions for directors slightly deviate from data in terms of the correlation between Rating$^*$ and $N$, and the correlation between $\langle \text{Rating}^{-*} \rangle$ and Rating$^*$ (Figs. S27c–d), which overestimates Rating$^*$ given the same $N$ and $\langle \text{Rating}^{-*} \rangle$. We find it is probably due to the upper bound of the IMDB rating system.

## S5 Model for Individual Collective Impact

**S5.1 $g(t)$ definition** Individual collective impact, $g(t)$, is defined as the sums of citations to all her publications at each year that

$$g(t) = \sum_{i=1}^{N(t)} c_i(t - t_i), \tag{S8}$$

where $t$ measures the career stage, defined by the number of years after her first publication, $N(t)$ is the total number of papers published up to time $t$, and $t_i$ is the publication time of her $i^{th}$ paper, whose yearly citations are denoted as $c_i$. $g(t)$ is driven by the interplay of three factors: productivity $N(t)$, the citation dynamics of each publication $c_i$, and how publications and their impacts are distributed within a career. Due to the non-uniform nature of productivity, we bypass the need to control for $N(t)$ to measure the career stage in unit of publications. That is, we keep the total number of publications in 20 years $N_T$ ($T = 20$) of each scientist, but rearrange the publishing process, such that a scientist produces the same number of papers every year. After this procedure, a scientist publishes constant $n$ papers per year, and $g(t)$ is defined as

$$g(t) = \sum_{i=1}^{N_T} c_i(t - \lfloor \frac{i}{N_T} T \rfloor), \tag{S9}$$

where $\lfloor\rfloor$ means rounding down and $\lfloor \frac{i}{N_T} T \rfloor$ denotes the publication year of the $i^{th}$ paper. As a result, papers published in the late career may be shifted to early career stage, whose citation records are not long enough to cover the rest of the time. Hence, to remove the boundary effect, we only consider $g(t)$ until the career stage that is not influenced by the boundary effect. The shortest career length after adjusting the boundary effect is 16 years.

**S5.2 Modulate paper citation with $\Gamma(t)$** Next, we build a mathematical model for $g(t)$ by incorporating our hot-streak model with paper citation dynamics. We combine $\Gamma(t)$ with a previous model describing the citation dynamics, namely the WSB model[21], allowing us to model how publications and their impacts are distributed within a career. The WSB model captures the cumulative citation of a paper $C_i(t)$ by three parameters: fitness $\lambda_i$, immediacy $\sigma_i$, and longevity $\mu_i$ that

$$C(t, t_i) = m \left( e^{\lambda_i \Phi\left(\frac{\ln(t-t_i)-\mu_i}{\sigma_i}\right)} - 1 \right), \tag{S10}$$

where $t$ and $t_i$ are defined as Eq. S8. $\Phi(\xi)$ is the cumulative normal distribution that follows

$$\Phi(\xi) = \frac{1}{\sqrt{2\pi}} \int_{-\infty}^{\xi} \phi(x) \mathrm{dx}, \tag{S11}$$

which illustrates the aging effect of a paper characterized by $\mu_i$ and $\sigma_i$. The constant $m$ denotes the average number of references a paper contains. Here we assume $m = 30$ to be consistent with previous studies[21]. By replacing fitness $\lambda_i$ with $\Gamma(t_i)$, Eq S10 can be expressed as

$$C(t, t_i) = m \left( e^{\Gamma(t_i) \Phi\left(\frac{\ln(t-t_i)-\mu_i}{\sigma_i}\right)} - 1 \right), \tag{S12}$$

where $\Gamma(t_i)$ represents the goodness parameter when a paper was published that can be measured from $\langle \log(C_{10}) \rangle_{t_i}$, where $\langle \cdot \rangle_{t_i}$ means the average around $t_i$. The replacement is consistent with the WSB model that $\Gamma(t_i)$ in Eq. S12 reflect the fitness $\lambda_i$ in the WSB model: The ultimate citation of a paper follows $C^{\infty} = m(e^{\lambda_i} - 1)$[21] that can be approximated by $C_{10}$[23], allowing us to get $\Gamma(t_i) = \langle \log(C_{10}) \rangle_{t_i} \approx \langle \log(C^{\infty}) \rangle_{t_i} \approx \log(m) + \langle \lambda \rangle_{t_i}$, where $\log(m)$ is a constant whose value does not change the results of WSB model[21]. Hence, Eq. S12 captures both the paper impact and how impact distributes within career by incorporating $\Gamma(t_i)$ into the WSB model.

**S5.3 $g(t)$ under the null model** We first discuss the null model prediction, $g_0(t)$, that assumes $\Gamma$ remains constant across a career ($\Gamma(t) \equiv \Gamma_0$). The integral of Eq. S12 over $t$ allows us to get the cumulative individual citation $G_0(t)$ that

$$
\begin{aligned}
G_0(t) &= \sum_{i=1}^{N_T} C_i(t, t_i) \\
&\approx N_T \int_0^t \int_{\mu', \sigma'} C(t, t') P(\mu', \sigma') \mathrm{d}t' \mathrm{d}\mu' \mathrm{d}\sigma' \\
&= \int_0^t \int_{\mu', \sigma'} n C(t, t') \delta(\mu') \delta(\sigma') \mathrm{d}t' \mathrm{d}\mu' \mathrm{d}\sigma', \\
&= \int_0^t m \left( e^{\Gamma_0 \Phi\left(\frac{\ln(t-t')-\mu}{\sigma}\right)} - 1 \right) \mathrm{dt}'
\end{aligned}
\tag{S13}
$$

where $N_T = \int_0^t n \mathrm{dt}'$ and $n$ is the publication rate at each time stamp, $t$ denotes the career stage of a scientist, $t'$ is the time when each paper was published, and $C(t, t')$ is the total paper citation that follows Eq. S12. For simplicity we assume in Eq. S13 that each scientist has constant $\mu$ and $\sigma$ for all her publications. Hence, $g_0(t)$ can be expressed as

$$
\begin{aligned}
g_0(t) &= \frac{\mathrm{d}}{\mathrm{d}t} G_0(t) \\
&= n \frac{\mathrm{d}}{\mathrm{d}t} \int_0^t m \left( e^{\Gamma_0 \Phi\left(\frac{\ln(t-t')-\mu}{\sigma}\right)} - 1 \right) \mathrm{dt}' \\
&= nm \left( e^{\Gamma_0 \Phi\left(\frac{\ln(t)-\mu}{\sigma}\right)} - 1 \right).
\end{aligned}
\tag{S14}
$$

$g_0(t)$ follows an S curve continuously increasing over time (Fig. 3d, Fig. S28) that is unable to capture a variety of $g(t)$ observed in real careers (Fig. 3d). Eq. S14 suggests $g_0(t)$ is captured by $\Gamma_0$ and two aging effect parameters $\mu$ and $\sigma$. Comparing Eq. S12 with Eq. S14 we find $g_0(t)$ is the cumulative citation of papers published at $t = 0$.

**S5.4** $g(t)$ **under hot-streak model** Next we calculate $g(t)$ under the hot-streak model by incorporating a time series $\Gamma(t)$ following Eq. 1 that

$$
\begin{aligned}
g(t) &= n\frac{\mathrm{d}}{\mathrm{d}t}G(t) \\
&= n\frac{\mathrm{d}}{\mathrm{d}t}\int_0^t m\left(e^{\Gamma(t')\Phi\left(\frac{\ln(t-t')-\mu}{\sigma}\right)} - 1\right)\mathrm{d}t' \\
&= n\int_0^t \frac{\partial}{\partial t} m e^{\Gamma(t')\Phi\left(\frac{\ln(t-t')-\mu}{\sigma}\right)}\mathrm{d}t' \\
&= nm\left[\int_0^t e^{\Gamma(t')\Phi\left(\frac{\ln(t-t')-\mu}{\sigma}\right)}\Phi\left(\frac{\ln(t-t')-\mu}{\sigma}\right)\frac{\mathrm{d}}{\mathrm{d}t'}\Gamma(t')\mathrm{d}t' - \int_0^t \frac{\partial}{\partial t'}e^{\Gamma(t')\Phi\left(\frac{\ln(t-t')-\mu}{\sigma}\right)}\mathrm{d}t'\right] \\
&= nm\int_0^t e^{\Gamma(t')\Phi\left(\frac{\ln(t-t')-\mu}{\sigma}\right)}\Phi\left(\frac{\ln(t-t')-\mu}{\sigma}\right)\frac{\mathrm{d}}{\mathrm{d}t'}\Gamma(t')\mathrm{d}t' + nm\left(e^{\Gamma_0\Phi\left(\frac{\ln(t)-\mu}{\sigma}\right)} - 1\right).
\end{aligned}
\tag{S15}
$$

Similarly, we assume a constant $\mu$ and $\sigma$ for each scientist. Eq. S15 assumes each individual has one hot streak. The second term in Eq. S15 is $g_0(t)$ under the null model prediction. The derivatives of Eq. 1 can be expressed as

$$
\begin{aligned}
\frac{\mathrm{d}}{\mathrm{d}t'}\Gamma(t') &= \frac{\mathrm{d}}{\mathrm{d}t'}[\Gamma_0 + (\Gamma_H - \Gamma_0)[\mathcal{H}(t' - t_\uparrow) - \mathcal{H}(t' - t_\downarrow)] \\
&= (\Gamma_H - \Gamma_0)[\delta(t' - t_\uparrow) - \delta(t' - t_\downarrow))].
\end{aligned}
\tag{S16}
$$

By replacing $\Gamma(t')$ in Eq. S15 with Eq. 1 and incorporating Eq. S16, we can get the expression of the first term in Eq. S15, denoted as $\Delta g(t)$, that follows

$$\Delta g(t) = \begin{cases} 0 & t < t_{\uparrow} \\ nm(\Gamma_H - \Gamma_0)\Phi\left(\frac{\ln(t-t_{\uparrow})-\mu}{\sigma}\right)e^{\Gamma_H \Phi\left(\frac{\ln(t-t_{\uparrow})-\mu}{\sigma}\right)} & t_{\uparrow} \le t < t_{\downarrow} \\ nm(\Gamma_H - \Gamma_0)\left(\Phi\left(\frac{\ln(t-t_{\uparrow})-\mu}{\sigma}\right)e^{\Gamma_H \Phi\left(\frac{\ln(t-t_{\uparrow})-\mu}{\sigma}\right)} - \Phi\left(\frac{\ln(t-t_{\downarrow})-\mu}{\sigma}\right)e^{\Gamma_H \Phi\left(\frac{\ln(t-t_{\downarrow})-\mu}{\sigma}\right)}\right) & t \ge t_{\downarrow} \end{cases} \tag{S17}$$

Hence, $g(t)$ under the hot-streak model can be expressed as

$$g(t) = \underbrace{nm(e^{\Gamma_0 \Phi\left(\frac{\ln(t)-\mu}{\sigma}\right)} - 1)}_{g_0(t)} + \underbrace{\begin{cases} 0 & t < t_{\uparrow} \\ nm(\Gamma_H - \Gamma_0)\Phi\left(\frac{\ln(t-t_{\uparrow})-\mu}{\sigma}\right)e^{\Gamma_H \Phi\left(\frac{\ln(t-t_{\uparrow})-\mu}{\sigma}\right)} & t_{\uparrow} \le t < t_{\downarrow} \\ nm(\Gamma_H - \Gamma_0)\Big(\Phi\left(\frac{\ln(t-t_{\uparrow})-\mu}{\sigma}\right)e^{\Gamma_H \Phi\left(\frac{\ln(t-t_{\uparrow})-\mu}{\sigma}\right)} & \\ \qquad -\Phi\left(\frac{\ln(t-t_{\downarrow})-\mu}{\sigma}\right)e^{\Gamma_H \Phi\left(\frac{\ln(t-t_{\downarrow})-\mu}{\sigma}\right)}\Big) & t \ge t_{\downarrow} \end{cases}}_{\Delta g(t)}, \tag{S18}$$

Eq. S18 is composed of two parts: $g_0(t)$ captures the cumulative citation of papers published when $\Gamma(t_i) = \Gamma_0$, and $\Delta g(t)$ to captures the citations contributed by papers published during hot streaks when $\Gamma(t_i) = \Gamma_H$. Fig. 4d illustrates the influence of the hot streak on individual citation: Before the hot streak, $\Delta g(t)$ has no impact on individual citation and $g(t) = g_0(t)$. During the hot streak, $g(t)$ experiences rapid growth due to $\Gamma_H$, and the trend continues for a while to reach its peak even after the end of the hot streak. Eq. S18 allows us to model a wide range of $g(t)$ that cannot be expected by the null model (Fig. 3d, Fig. 3h, Fig. S29, Fig. S30). $g(t)$ is determined by the timing of the hot streak ($t_{\uparrow}$ and $t_{\downarrow}$), the level of performance ($\Gamma_0$ and $\Gamma_H$), and the aging effect parameters

($\mu$ and $\sigma$).

**S5.5 Parameter estimation** In order to test how well Eq. S18 matches empirical data, we need to estimate the parameters $\Gamma_0$ and $\Gamma_H$, $t_\uparrow$ and $t_\downarrow$, and the aging effect parameters ($\mu$ and $\sigma$) for each scientist given her $g(t)$. Here we assume each individual has only one hot streak, and relax all the six individual-specific parameters. For each individual in our dataset, we used ordinary least square (OLS) and fitted to Eq. S18 the $g(t)$ measured from real career by minimizing the mean squared error (MSE) between data and the model result. To make sure the fitted parameters are meaningful, we also set constraints when optimizing the error function that $t_\uparrow > 0$, $t_\downarrow > 0$, $0 < \Gamma_0 \leq \Gamma_H \leq 10$, $0 \leq \mu \leq 8.5$ and $\sigma > 0$. We assume $\Gamma_H$ cannot exceed 10, otherwise the average citation per paper during hot streaks is $\langle C_{10} \rangle \approx 22026$.

**S5.6 Model validation** To systematically evaluate the accuracy of our model, we generate an uncertainty envelop of $g(t)$ for each individual for both the hot-streak model and the null model. Specifically, we simulate $g(t)$ by assigning a Gaussian noise $\mathcal{N}(0, \sigma^2)$ to the fitted $\Gamma$. The standard deviation $\sigma$ is calculated from data (Fig. S18), representing the inherent noise of the $\Gamma(N)$ sequence. Hence, for each paper $i$ we randomly generate a $\Gamma_i$ parameter from a normal distribution with mean as $\Gamma_0$ if the paper is published during the normal performance or $\Gamma_H$ during hot streaks. We assume $\mu$ and $\sigma$ to be the same as the fitting result. We simulated each individual for 1000 realizations, allowing us to get a distribution of $g(t)$ at each time step. The envelop at each time $t$ is defined as one standard deviation of all the values of $g(t)$ at $t$ . Then we measure the model performance by calculating the fraction of $g(t)$ falling within the envelope. We repeated the procedures for the null model to calculate the uncertainty envelope as well. Our hot-streak model outperforms the null model as more data fall within the envelop (Fig. 3f). We also compare the distribution of the Mean Absolute Percentage Error (*MAPE*) between data and model predictions, finding our hot-streak model has systematically smaller error compared with the null model (Fig. S31).

## S6 Testing Alternative Hypotheses

Our hot-streak model is the simplest function to approximate impacts of works within a career. The accuracy of the hot-streak model prompts us to ask whether it is unique in its ability to capture the impacts of individual careers across different domains. We therefore seek other models corresponding to different dynamics of hot streaks. In this section we test the validity of four alternative hypotheses (A–D) proposed in the main text (Figs. S34a–f), by comparing each model prediction on the relative timing of the top six hit works within a carer with empirical observations. Indeed, the symmetric patterns of normalized joint probability $\Phi$ and $R(\Delta N/N)$ for any pair of the top three hits, suggests one hit is equal likely to appear before or after another hit among the top three highest impact works within a career across three domains. We find such randomness on the relative position among hits is not just limited to the top three. We measured the position of the top three hits $\tilde{N}$ relative to the top six hits of the career. For each of the three hits, we compute $P(\tilde{N})$ for artists, directors, and scientists, finding a lack of predictive patterns for $P(\tilde{N})$ across three domains, suggesting the relative order among the top six hits in real careers are random (Figs. S34gnu).

To test if hypotheses A–D can reproduce the randomness among top six hits within a career, we then compare $P(\tilde{N})$ predicted by each hypothesis with the corresponding distributions measured from real careers. For our hot-streak model and each alternative hypothesis, we simulate a sequence of work impact for each individual. We use real productivity $N$, and the mean impact of each individual as input, and assumes for each individual a period of outstanding performance that randomly appears within a career. We assume $\Gamma_0$ for each individual as the mean impact within a career measured from data, and $\Gamma_H = \Gamma_0 + 1.0$. Since the average length of hot streaks usually accounts for 20% of total $N$ works, and the alternative model only changes how the hot streak emerges and diminishes, here we assume for each individual a fixed length of $\Gamma_H$ period $(0.2N)$ for right trapezoid, left trapezoid, and isosceles trapezoid model. We also assume the hypotenuses of each trapezoid model to be $0.2N$, ensuring the impact can gradually change between $\Gamma_0$ and $\Gamma_H$ under different hypotheses. To keep the duration of hot streaks to be the same for different hypotheses, we assume a $0.4N$ length of hot streaks for hypothesis C and our hot-streak model, respectively. We assume $\Gamma_H$ as the peak value for the quadratic and tent function of each individual. The impact of each work is then generated from a normal distribution $\mathcal{N}(\Gamma(N), \sigma)$, where the mean value $\Gamma(N)$ is defined by each model, and the standard deviation $\sigma$ is measured from data (Fig. S18).

We calculated $P(\tilde{N})$ of each hypothesis, finding $P(\tilde{N})$ for hypotheses A–D shows predictive trends on the relative position of top six hits (Figs. S34h–m). For example, the top two hits of isosceles trapezoid, quadratic, and tent model always appear in the middle of the top six hits, and their probability decreases when deviating from the middle (Figs. S34j–l). For the right trapezoid model, the top three hits always appear before the other hits (Fig. S34i); whereas the case reverses for the left trapezoid model (Fig. S34m). Among these hypotheses, only our hot-streak model can reproduce the randomness of the top six hits (Fig. S34h). To quantify the different between data and each model prediction, we performed the Kolmogorov-Smirnov test to measure whether we can reject the hypothesis that predictions and data are drawn from the same distribution. We used the R package 'dgof' to conduct the discrete Goodness-of-Fit test, finding $P(\tilde{N})$ for hypotheses A–D is statistically different from empirical observations (Figs. S34i–m). Our hot-streak model is the only one that captures the randomness of the top six hits within a career (Fig. S34h). The results reported in Figs. S34h–m are based on the simulations using artists' profiles as input. To make sure the conclusion is robust to the other two domains, we repeated the procedure using profiles of directors (Figs. S34o–t) and scientists (Figs. S34v–aa) as input, finding the conclusion remains the same for other domains.

## Table S1: Empirical evidence of whether hot hand exists

| Year | Paper | Field | Hot hand | Data |
|---|---|---|---|---|
| 1985 | Gilovich et al. (1985)[1] | Basketball | No | Field goal data for 9 NBA player during the 1980 –1981 season, free-throw data for 9 NBA players during 1980–1982 seasons, a controlled shooting experiment with 26 Cornell students |
| 1989 | Larkey et al. (1989)[53] | Basketball | Yes | 18 players in 39 NBA games in 1987–1988 season |
| 1993 | Albright (1993)[54] | Baseball | No | 40 MLB players in 1987–1990 seasons |
| 1993 | Albert (1993)[55] | Baseball | Yes | Same as Albright (1993)[54] |
| 1994 | Frohlich (1994)[56] | Baseball | No | MLB No-hit games in 1989–1993 |
| 1995 | Wardrop (1995)[57] | Basketball | Inconclusive | Free throw data for 9 members of the Boston Celtics during 1980 - 1982 seasons |
| 1999 | Wardrop (1999)[58] | Basketball | Yes | A shooting experiment for a member on the UW-Madison women's varsity basketball team |
| 2000 | Vergin (2000)[59] | Basketball, Baseball | No | 29 NBA team wins and losses in 1996–1998 seasons, 28 MLB team wins and losses in 1996 season |
| 2001 | Albert and Williamson (2001)[60] | Basketball, Baseball | Yes | Same as Wardrop (1999)[58], hitting data for Javy Lopez for the 1998 MLB season |
| 2001 | Klaassen and Magnus (2001)[61] | Tennis | Yes | Point-to-point data from Wimbledon singles in 1992–1995 |
| 2003 | Koehler and Conley (2003)[62] | Basketball | No | NBA long distance shootout contest in 1994–1997 |
| 2003 | Smith (2003)[63] | Horse Pitching | Yes | 62 pitchers in 2000–2001 World Championships |
| 2004 | Dorsey-Palmateer and Smith (2004)[64] | Bowling | Yes | 43 players from PBA in 2002–2003 season |
| 2005 | Clark (2005)[65] | Golf | Yes | Hole-to-hole scores from PGA tour in 1997–1998 |
| 2008 | Bocskocsky et al. (2008)[66] | Basketball | Yes | 83,000 shots from the 2012–2013 NBA season |
| 2008 | Albert (2008)[67] | Baseball | Yes | All regular baseball players during the 2005 MBL season |
| 2010 | Arkes (2010)[68] | Basketball | Yes | Free throw data during the 2005–06 NBA season |
| 2011 | Yaari and Eisenmann (2011)[69] | Basketball | Yes | Free throws of five NBA regular seasons from 2005 to 2009 |
| 2012 | Yaari and David (2012)[70] | Bowling | Yes | Frame by frame games for 100 top players in PBA from 2002 to 2011 |
| 2012 | Aharoni and Sarig (2012)[71] | Basketball | Yes | 1218 games in 2004–2005 NBA season |
| 2012 | Raab and Gula (2012)[72] | Volleyball | Yes | 26 top players' offensive performance in Germany s first-division volleyball league |
| 2012 | Gabel and Redner (2012)[73] | Basketball | No | 6087 games from the 2006/07 – 2009/10 seasons of NBA |
| 2014 | Miller and Sanjurjo (2014)[74] | Basketball | Yes | Controlled shooting experiment for players from the semi-professional basketball team Santo Domingo de Betanzos |
| 2014 | Csapo and Raab (2014)[75] | Basketball | Yes | 666 NBA games from the 2011-12 to 2013-14 seasons |
| 2015 | Miller and Sanjurjo (2015)[76] | Basketball | Yes | NBA Three-Point Contest |
| 2015 | Rosenqvist and Skans (2015)[77] | Golf | Yes | Male European PGA tournaments in 2000–2012 |
| 2015 | Parsons and Rohde (2015)[78] | Football/Soccer | Yes | 20 teams from EPL in 2010–2013 seasons |
| 2015 | Miller and Sanjurjo (2015)[76] | Baseball | Yes | NBA Three point shooting contest in 1986–2005 |
| 2016 | Miller and Sanjurjo (2016)[3] | Basketball | Yes | Same as Miller and Sanjurjo (2014)[74] |
| 2017 | Green and Zwiebel (2017)[79] | Baseball | Yes | MLB players' batter |
| 1993 | Hendricks (1993)[8] | Mutual funds | Yes | Risk-adjusted mutual-fund returns in 1975-1988 |
| 1995 | Malkiel (1995)[80] | Equity mutual funds | Inconclusive | All risk-adjusted mutual-fund returns in 1971–1991 |
| 1997 | Carhart (1997)[81] | Equity mutual funds | No | All risk-adjusted equity mutual-fund returns in 1962–1993 |
| 2008 | Huij et al. (2008)[82] | Bond funds | Yes | Risk-adjusted bond fund returns in 1990–2003 |
| 2010 | Jagannathan et al. (2010)[9] | Hedge funds | Yes | Risk-adjusted hedge fund returns in 1996–2005 |

Table S2: Statistical methods to detect hot hand

| Method | Paper |
|---|---|
| Conditional probablity | Gilovich et al. (1985)[1], Larkey et al. (1989)[53], Wardrop (1995)[57], Koehler and Conley (2003), Smith (2003)[63], Koehler and Conley (2003)[62], Dorsey-Palmateer and Smith (2004)[64], Clark (2005)[65], Albert (2008)[67], Yaari and Eisenmann (2011)[69], Aharoni and Sarig (2012)[71], Raab and Gula (2012)[72], Csapo and Raab (2014)[75], Miller and Sanjurjo (2016)[3], Malkiel (1995)[80] |
| Serial correlation | Gilovich et al. (1985)[1], Larkey et al. (1989)[53], Wardrop (1999)[58], Klaassen and Magnus (2001)[61], Smith (2003)[63], Raab and Gula (2012)[72], Gabel and Redner (2012)[73], Green and Zwiebel (2017)[79], Hendricks (1993)[8] |
| Number of runs | Gilovich et al. (1985)[1], Wardrop (1999)[58], Albright (1993)[54], Vergin (2000)[59], Koehler and Conley (2003)[62], Smith (2003)[63], Albert (2008)[67], Raab and Gula (2012)[72], Csapo and Raab (2014)[75], Miller and Sanjurjo (2014)[74], Parsons and Rohde (2015)[78], Miller and Sanjurjo (2015)[76] |
| Stationary of successful rate | Gilovich et al. (1985)[1], Frohlich (1994)[56], Wardrop (1999)[58], Albert (2008)[67], Yaari and Eisenmann (2011)[69], Yaari and David (2012)[70], Gabel and Redner (2012)[73] |
| hidden Markov model | Albright (1993)[54], Albert (1993)[55], Albert and Williamson (2001)[60], Albert (2008)[67], Wetzels *et al.*(2016)[83] |
| Long streaks | Larkey et al. (1989)[53], Vergin (2000)[59], Frohlich (1994)[56],Dorsey-Palmateer and Smith (2004)[64], Albert (2008)[67], Aharoni and Sarig (2012), Csapo and Raab (2014)[75], Miller and Sanjurjo (2014)[74], Miller and Sanjurjo (2015)[76] |
| Regression | Albright (1993)[54], Bocskocsky et al. (2008)[66], Arkes (2010)[68], Aharoni and Sarig (2012)[71], Csapo and Raab (2014)[75], Parsons and Rohde (2015)[78], Hendricks (1993)[8], Carhart (1997)[81], Huij et al. (2008)[82], Jagannathan et al. (2010)[9] |
| Time interval between scoring event | Gabel and Redner (2012)[73] |
| Regression discontinuity design | Rosenqvist and Skans (2015)[77] |

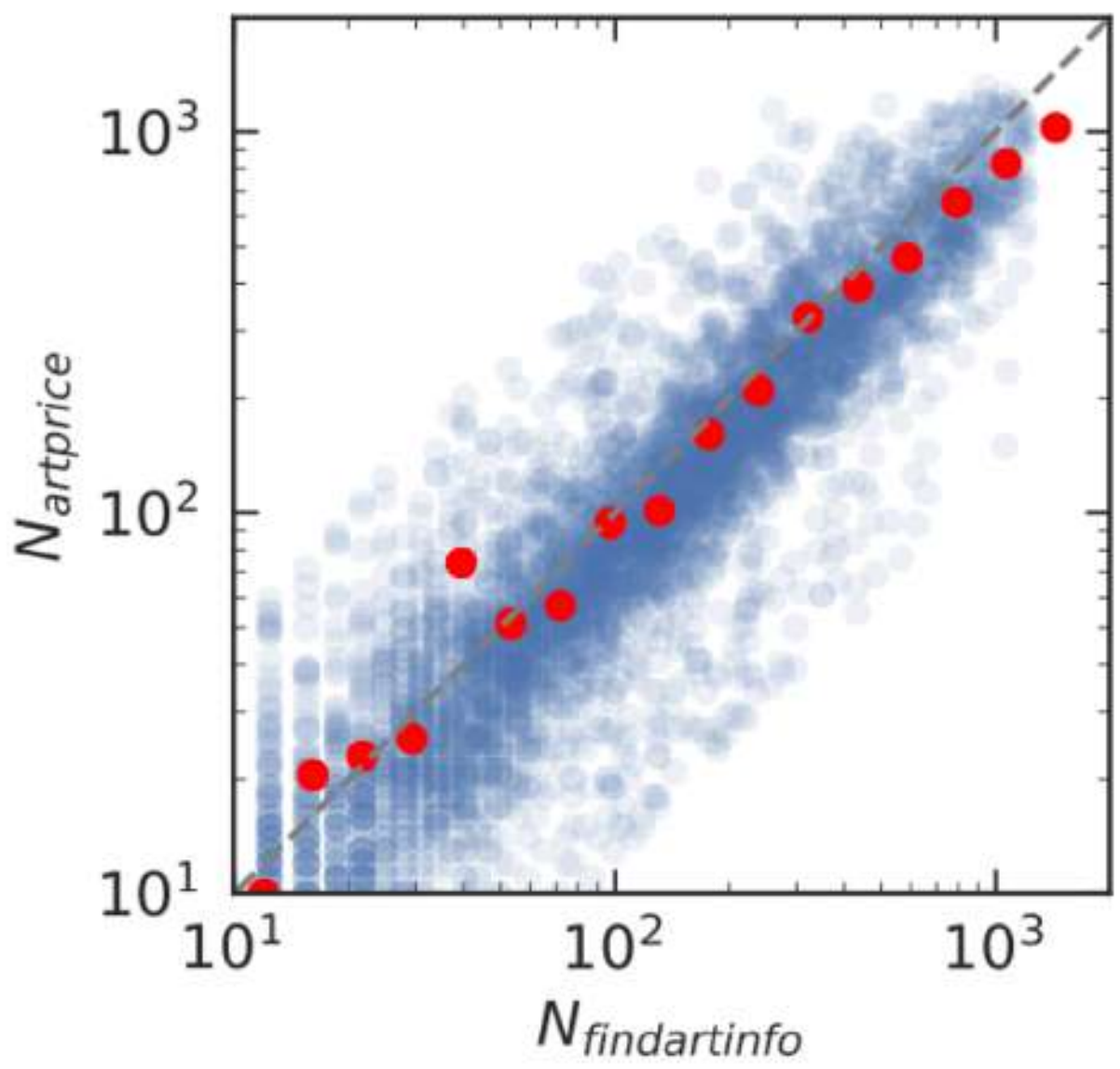

Figure S1: **Entity linking result of two auction databases.** The relationship of the number of auction records in Findartinfo and Artprice for matched artists in log-log scale, where $N_{findartinfo}$ is the number of records in Findartinfo and $N_{artprice}$ is the number of records in Artprice. Blue dots denote data, red dots denote the logarithmic binning of the scattered data, and the dashed grey line indicates $y = x$.

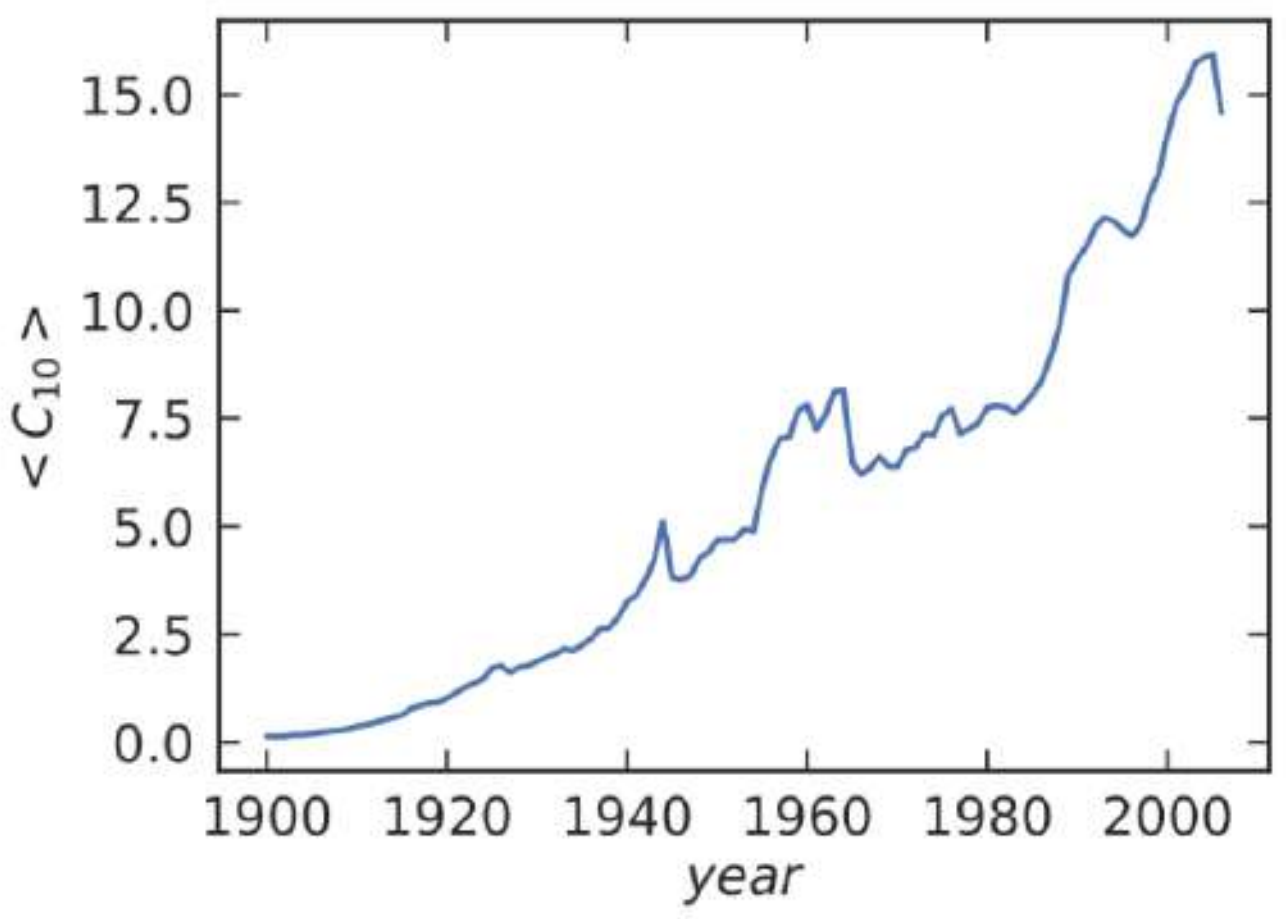

Figure S2: **Citation inflation in WoS.** The average impact of papers published in the same year, quantified with number of citations 10 years after publication, steadily grows as a function of the publication year ranging from 1900 to 2004.

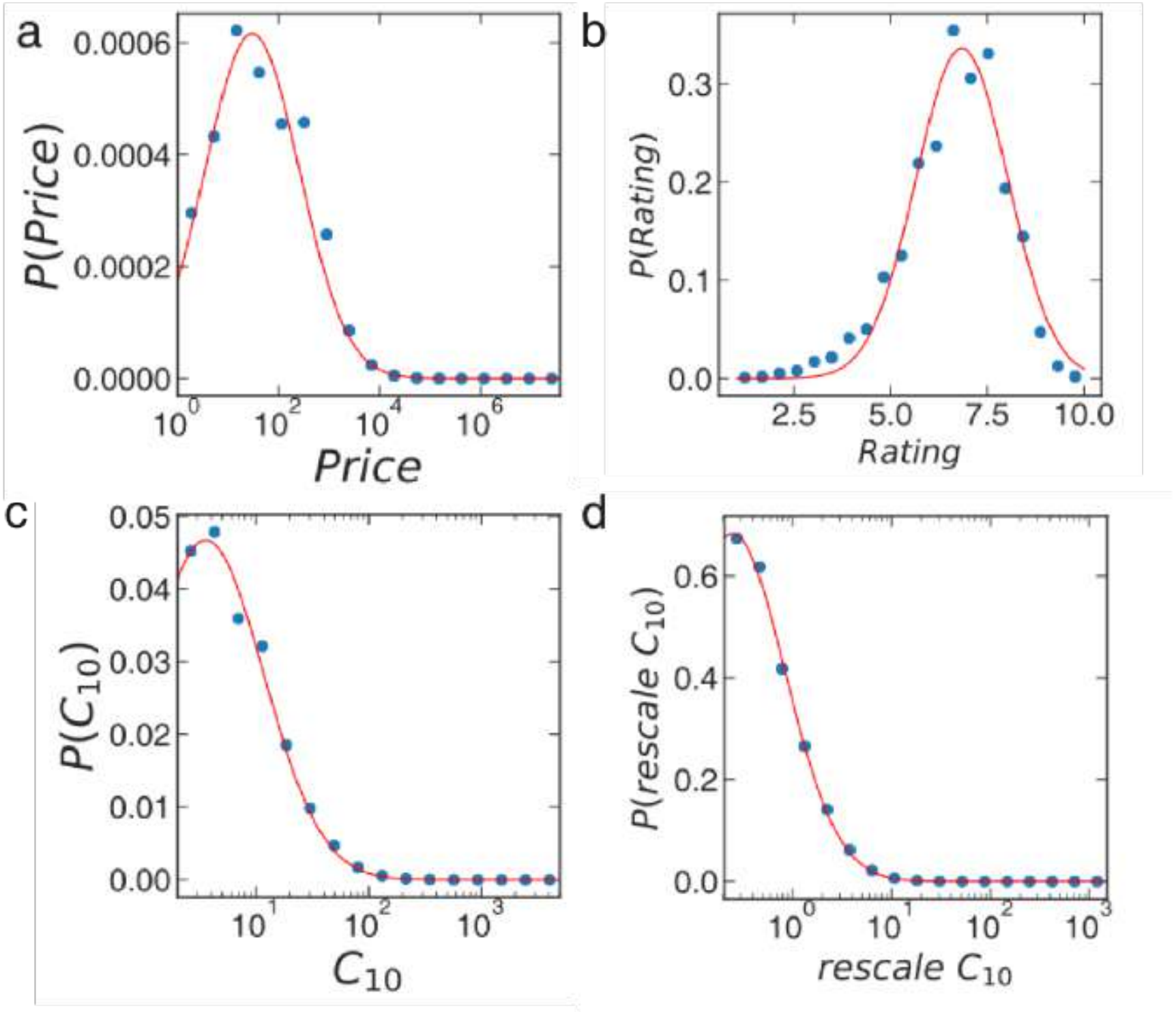

Figure S3: **Impact distribution.** (a) The distribution of auction price $P(\text{Price})$ for artists. Blue dots denote data, and the red line is a log-normal distribution with average $\mu = 8.0$ and standard deviation $\sigma = 2.15$. (b) The distribution of movie rating $P(\text{Rating})$ for directors. The red line is a normal distribution with average $\mu = 6.8$ and standard deviation $\sigma = 1.19$. (c–d), The distribution of (c) raw and (d) rescaled $C_{10}$ for scientists. The red line is a log-normal distribution, with $\mu = 2.7$ and $\sigma = 1.18$ for (c) and $\mu = 0.1$ and $\sigma = 1.42$ for (d).

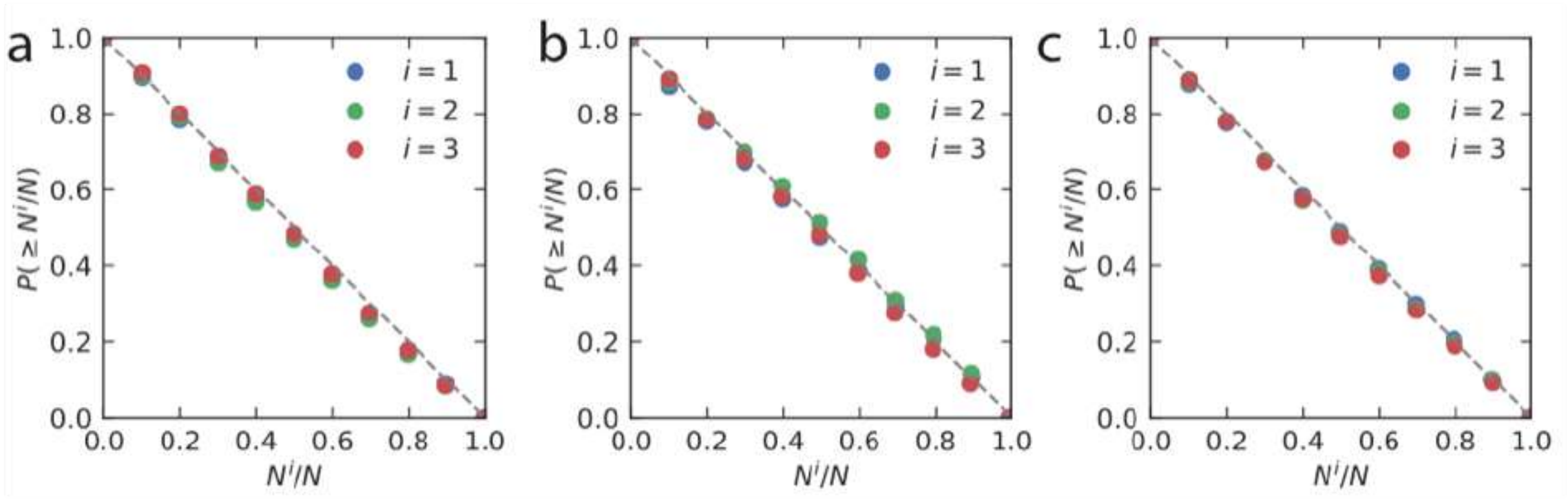

Figure S4: **Random impact rule.** $P(\geq N^i/N)$ of the top three hit works within a career for (a) artists, (b) directors, and (c) scientists. $N^i$ denotes the order of the $i^{th}$ hit work within a career. The color denotes different hit works, and the dashed grey line denotes $P(\geq N^i/N)$ for a uniform distribution.

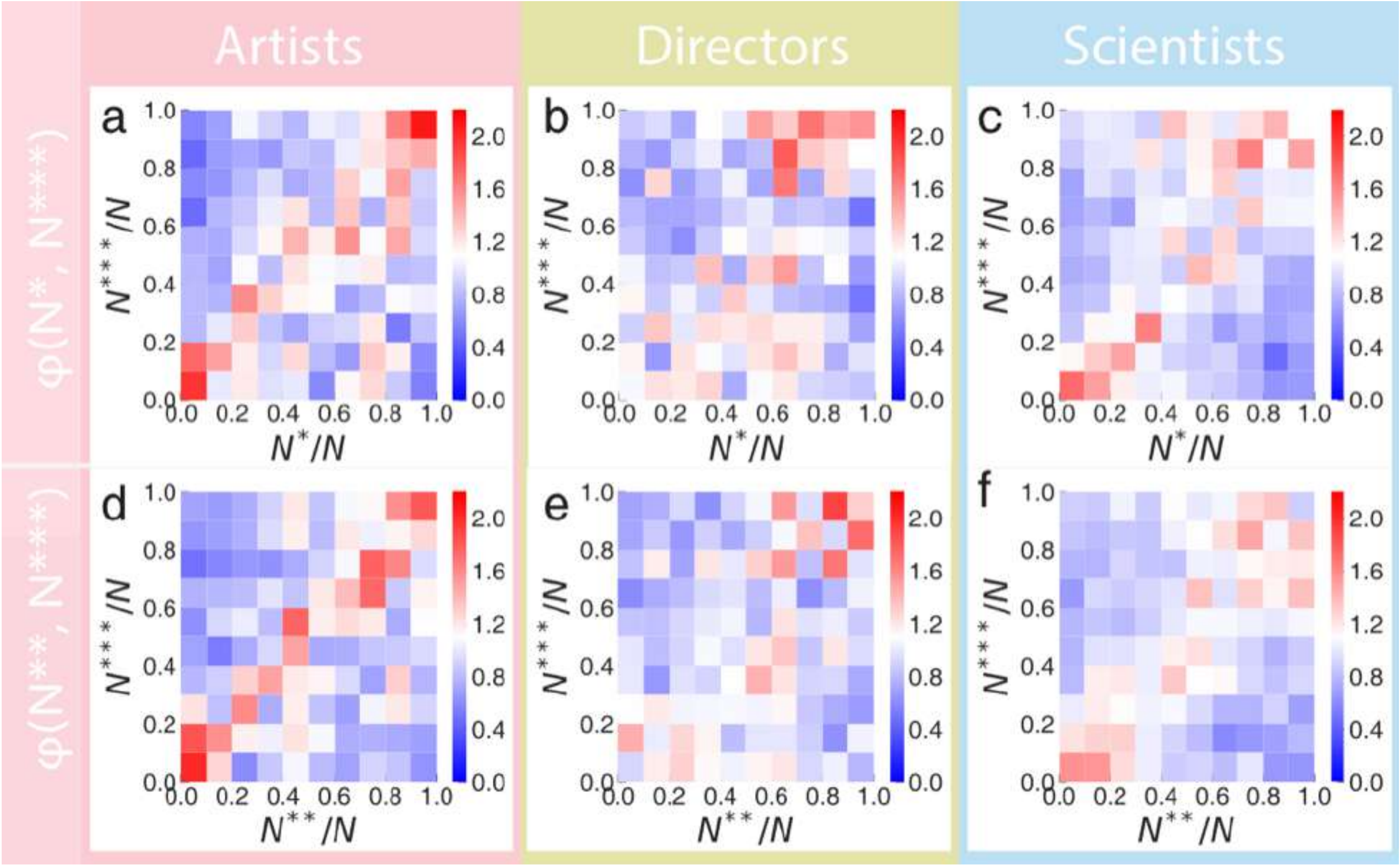

Figure S5: $\Phi$ **for other pairs of hit works.** (a–c), $\Phi(N^{*}, N^{***})$ of real careers for (a) artists, (b) directors, and (c) scientists. (d–f), $\Phi(N^{**}, N^{***})$ of real careers for (d) artists, (e) directors, and (f) scientists. We find the same patterns along the diagonal for $\Phi(N^{*}, N^{***})$ and $\Phi(N^{**}, N^{***})$.

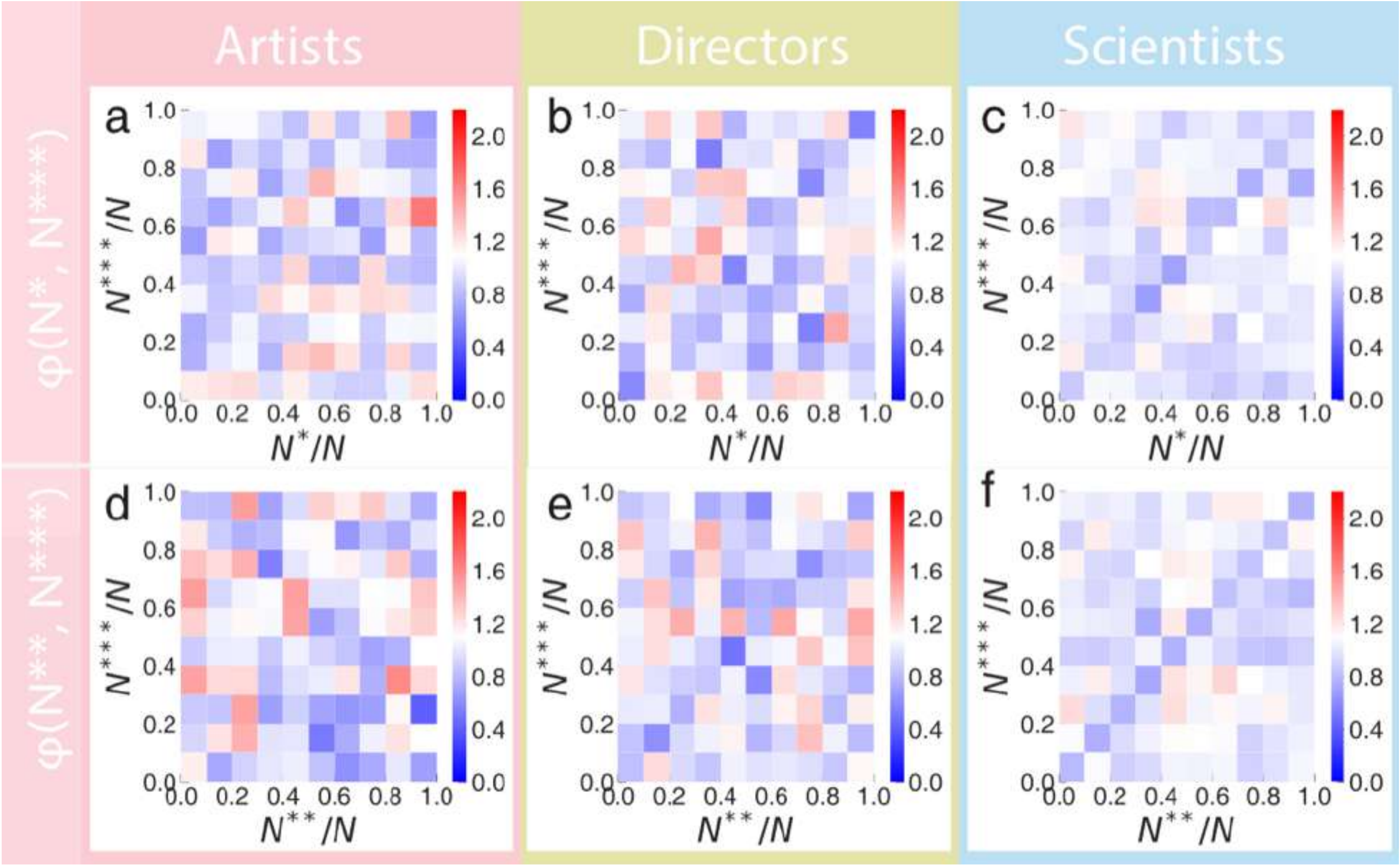

Figure S6: $\Phi$ **for shuffled careers.** (a–c), $\Phi(N^{*}, N^{***})$ of shuffled careers for (a) artists, (b) directors, and (c) scientists. (d–f), $\Phi(N^{**}, N^{***})$ of shuffled careers for (d) artists, (e) directors, and (f) scientists. We find for shuffled careers the patterns along the diagonal for $\Phi(N^{*}, N^{***})$ and $\Phi(N^{**}, N^{***})$ disappear.

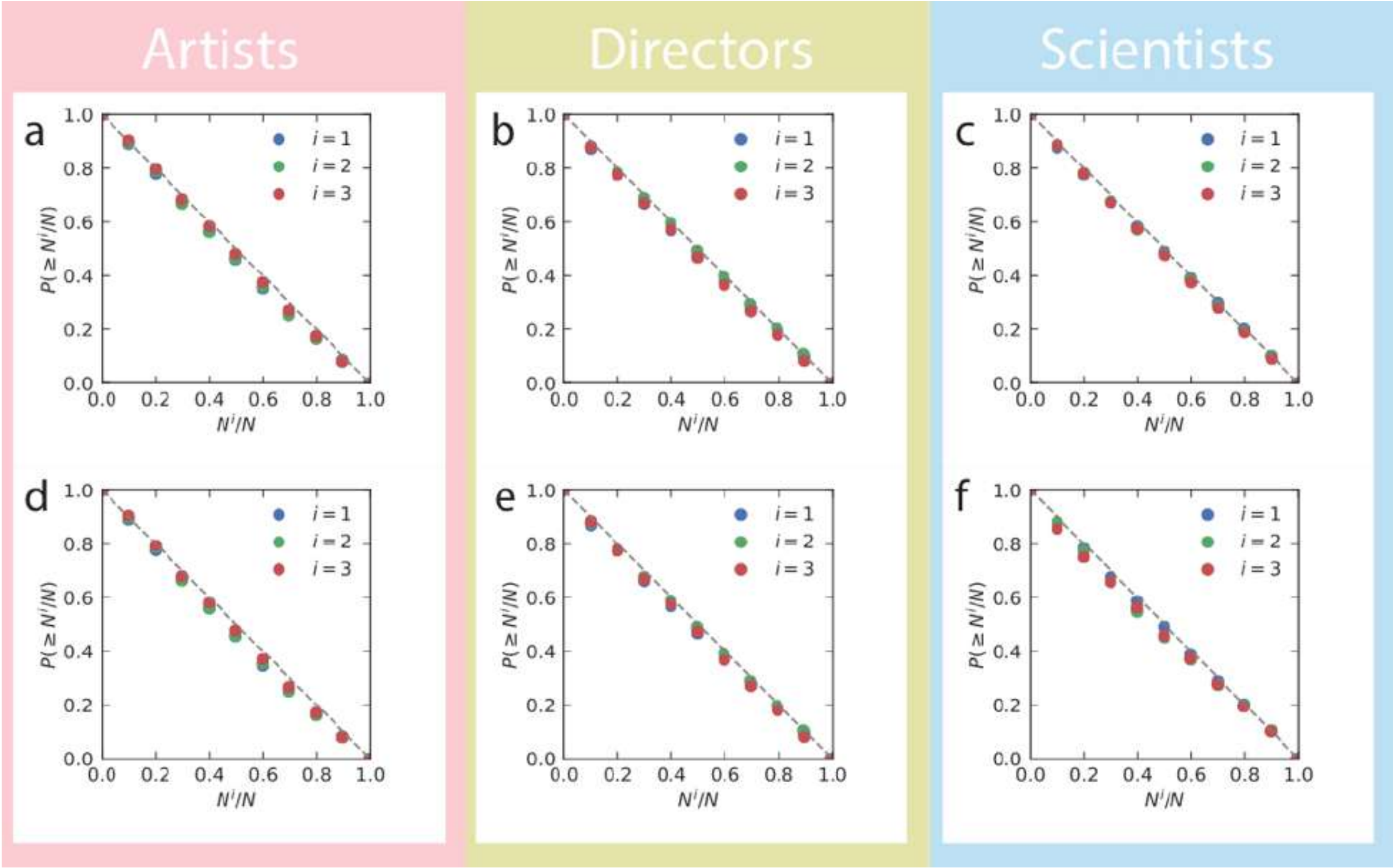

Figure S7: **Random impact rule under different career length.** $P(\geq N^i/N)$ for individuals under different career length: Artists with at least (a) 20 years and (d) 30 years of career length, directors with at least (b) 20 years and (e) 30 years of career length, and scientists with at least (c) 30 years and (f) 40 years of career length. $N^i$ denotes the order of the $i^{th}$ hit work within a career. The color denotes different hit works, and the dashed grey line denotes $P(\geq N^i/N)$ for a uniform distribution.

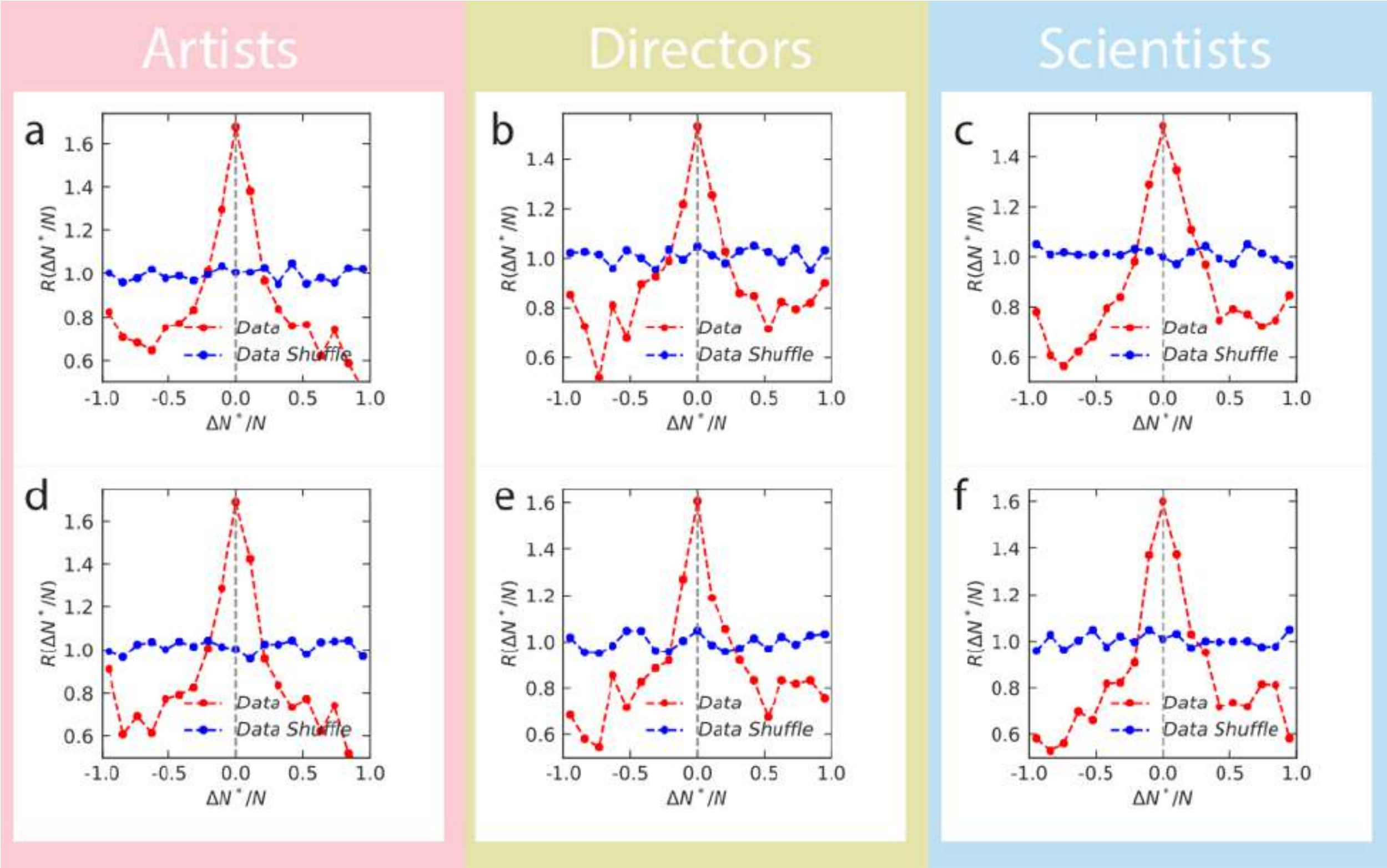

Figure S8: $R(\Delta N^*/N)$ **under different career length.** $R(\Delta N^*/N)$ for individuals under different career length: Artists with at least (a) 20 years and (d) 30 years of career length, directors with at least (b) 20 years and (e) 30 years of career length, and scientists with at least (c) 30 years and (f) 40 years of career length, where $\Delta N^* = N^* - N^{**}$. Red dots denote data and blue dots denote shuffled careers. We find the $R(\Delta N^*/N)$ still peaks around zero for individuals with different career length.

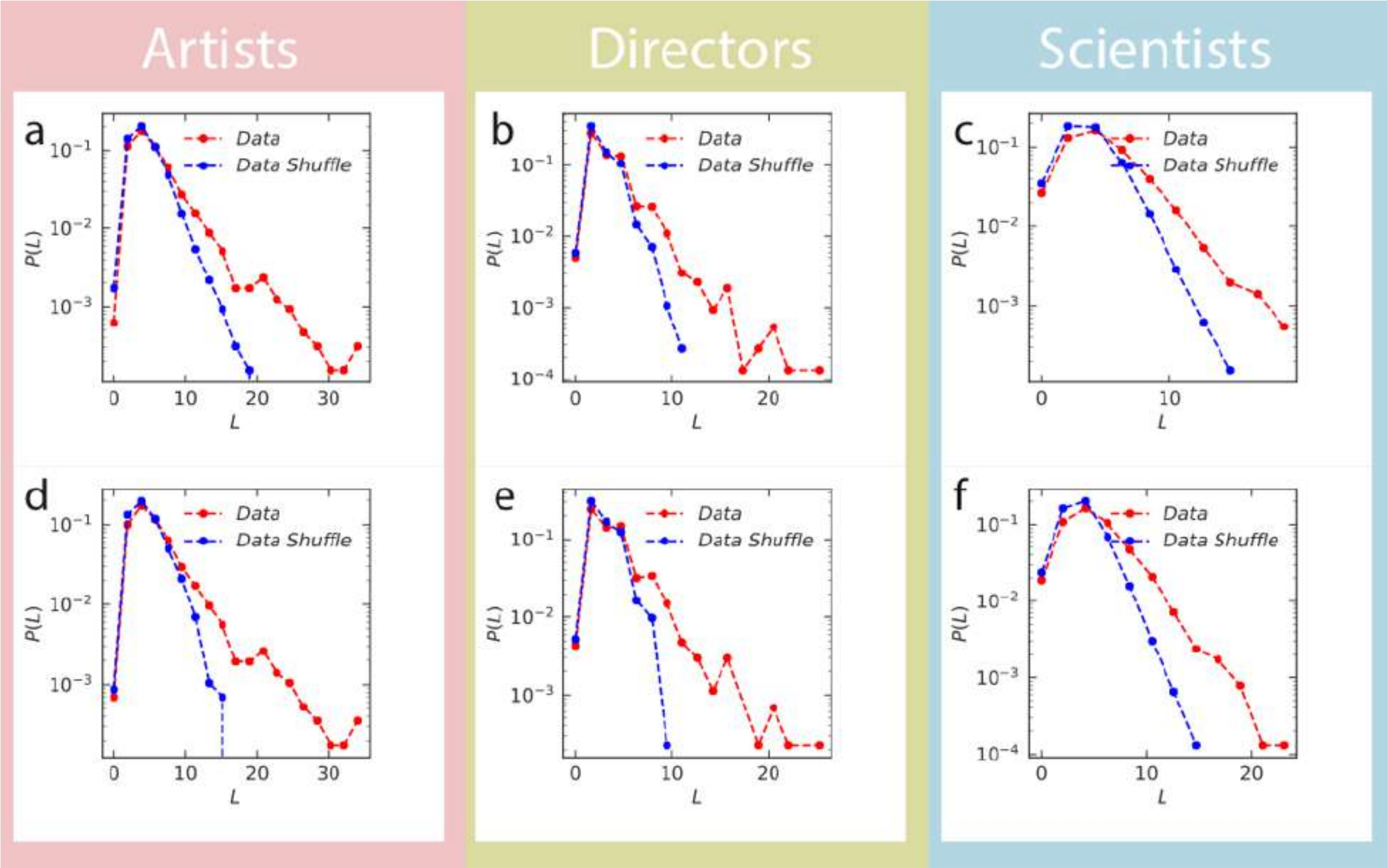

Figure S9: **Streak length under the different career length.** $P(L)$ for individuals under different career length: Artists with at least (a) 20 years and (d) 30 years of career length, directors with at least (b) 20 years and (e) 30 years of career length, and scientists with at least (c) 30 years and (f) 40 years of career length. Red dots denote data and blue dots denotes shuffled careers. We find the $P(L)$ of data has exponential tail that is wider than the shuffled careers for individuals with different career length.

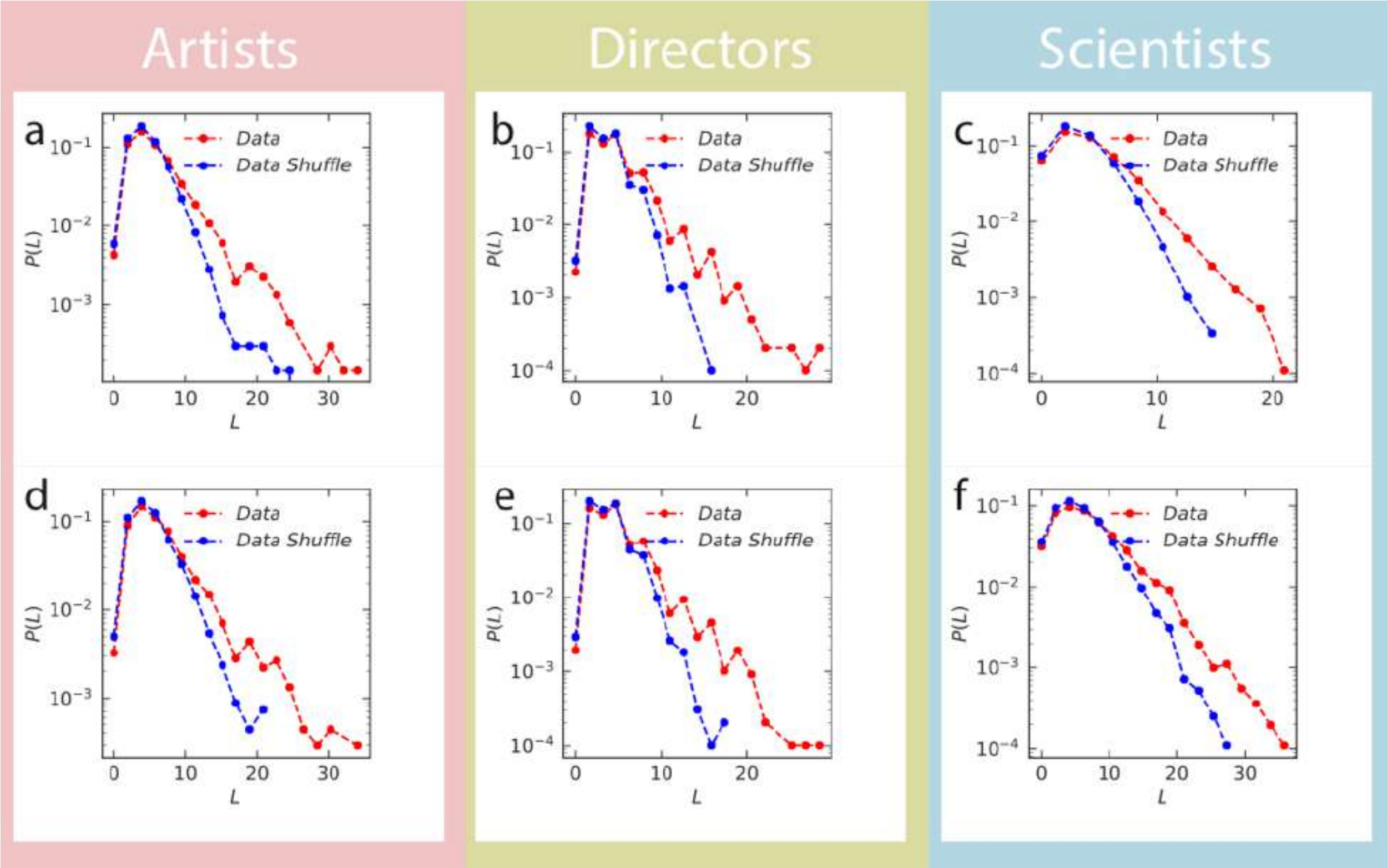

Figure S10: **Streak length under different thresholds**. (a–c), $P(L)$ and $P(L_S)$ when the threshold is the mean of impact within a career for (a) artists, (b) directors and (c) scientists. (d–f) $P(L)$ and $P(L_S)$ when the threshold is the geometric mean of impact within a career for individuals across three domains.

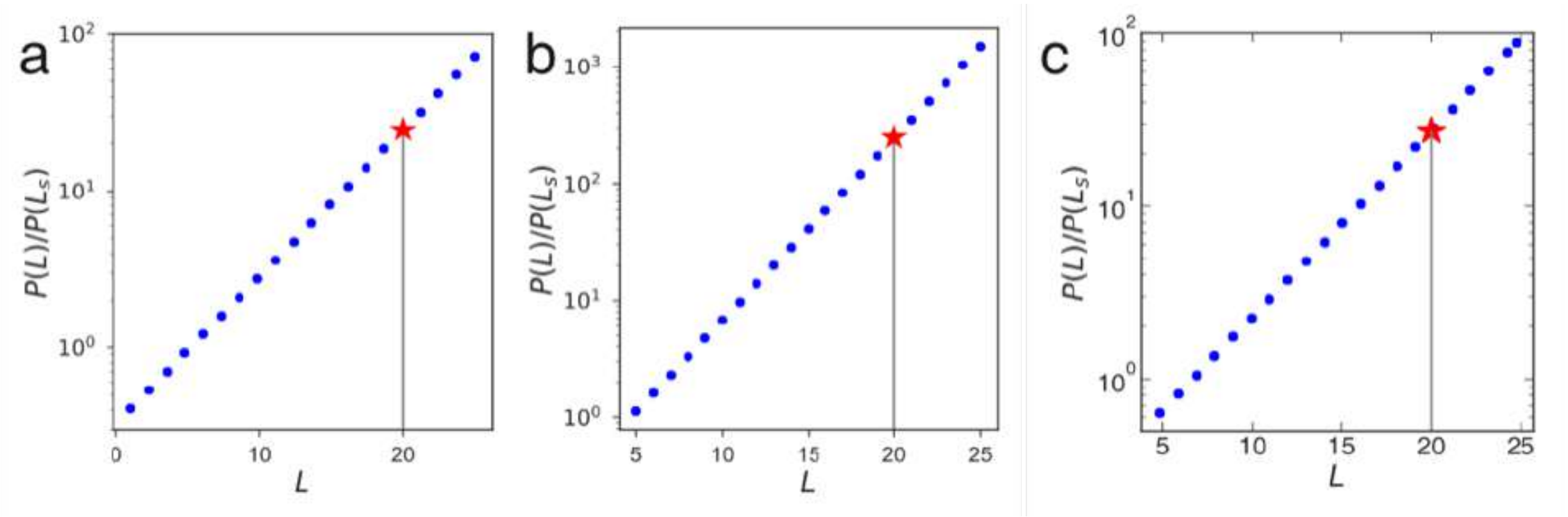

Figure S11: **The difference of $P(L)$ and $P(L_S)$.** (a–c), $P(L)/P(L_S)$ for (a) artists, (b) director, and (c) scientists in semi-log scale. The vertical line denotes $L = 20$, and the red star denotes the probability difference when $L = 20$.

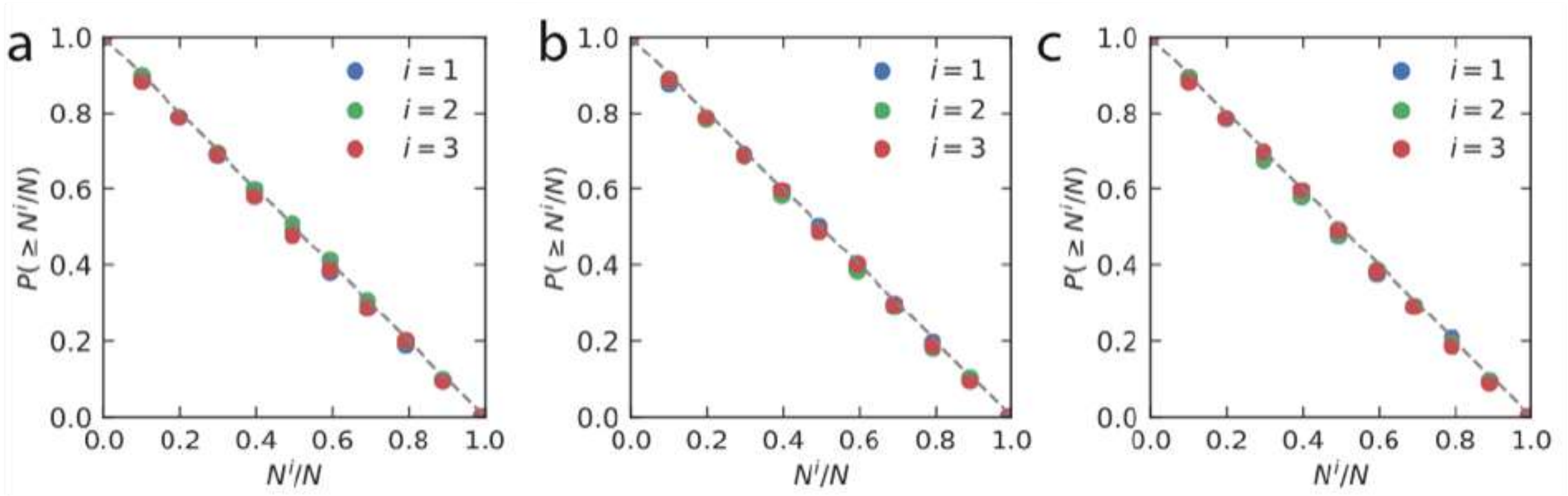

Figure S12: **Random impact rule in the null model.** $P(\geq N^i/N)$ under the null model prediction for (a) artists, (b) directors, and (c) scientists. The null model can reproduce the randomness of the top three highest impact works in a career.

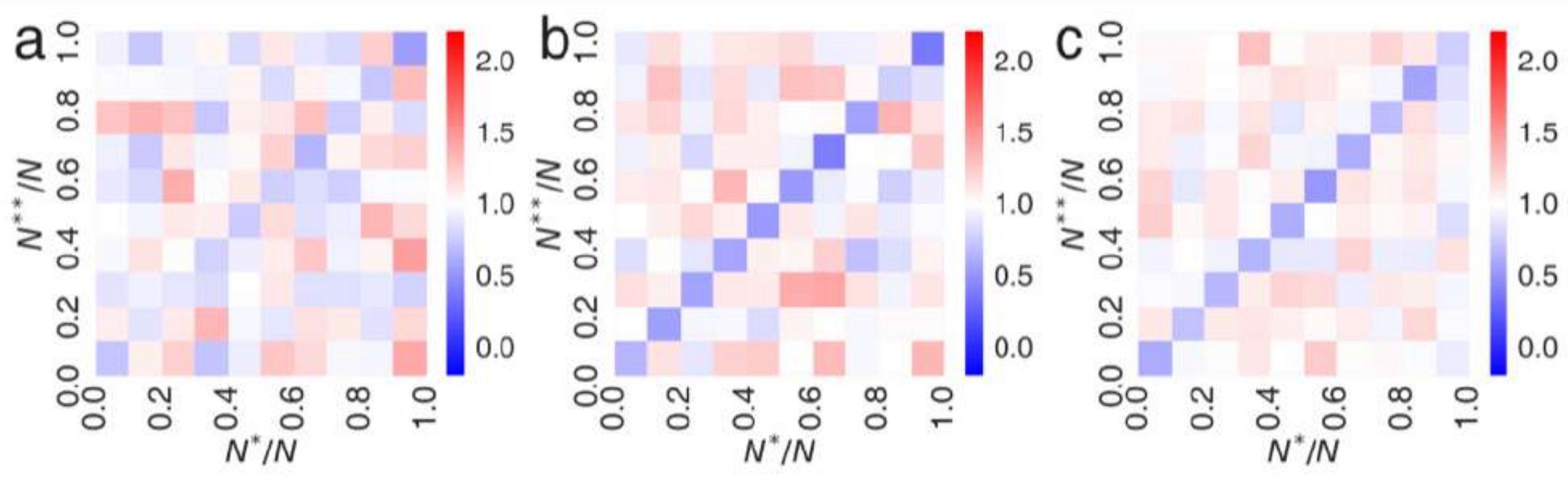

Figure S13: $\Phi(N^{*}, N^{**})$ **predicted by the null model.** The normalized joint probability $\Phi(N^{*}, N^{**})$ for the top two hit works within a career under the null model prediction for (a) artists, (b) directors, and (c) scientists. The null model cannot reproduce the temporal colocation of the two highest impact works within a career.

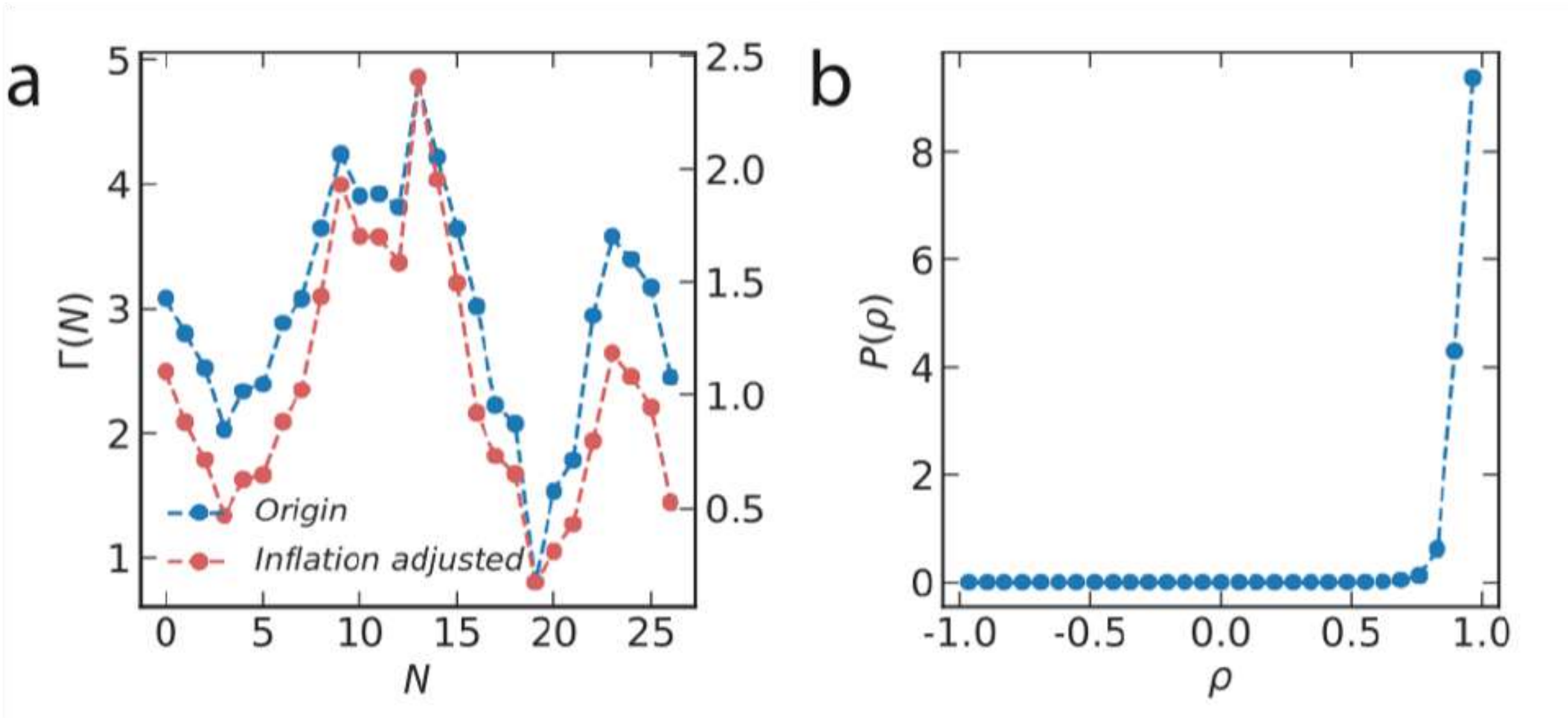

Figure S14: **Comparison of $\Gamma(N)$ calculated from raw and rescale $\log C_{10}$.** (a) The $\Gamma(N)$ sequence for a scientist in our dataset calculated from raw $C_{10}$ (blue dots) and rescaled $C_{10}$ (red dots). (b) The distribution of Pearson correlation $P(\rho)$ for $\Gamma(N)$ sequence calculated from the raw $C_{10}$ and rescaled $C_{10}$. $P(\rho)$ peaks around 1.0, with mean value 0.93.

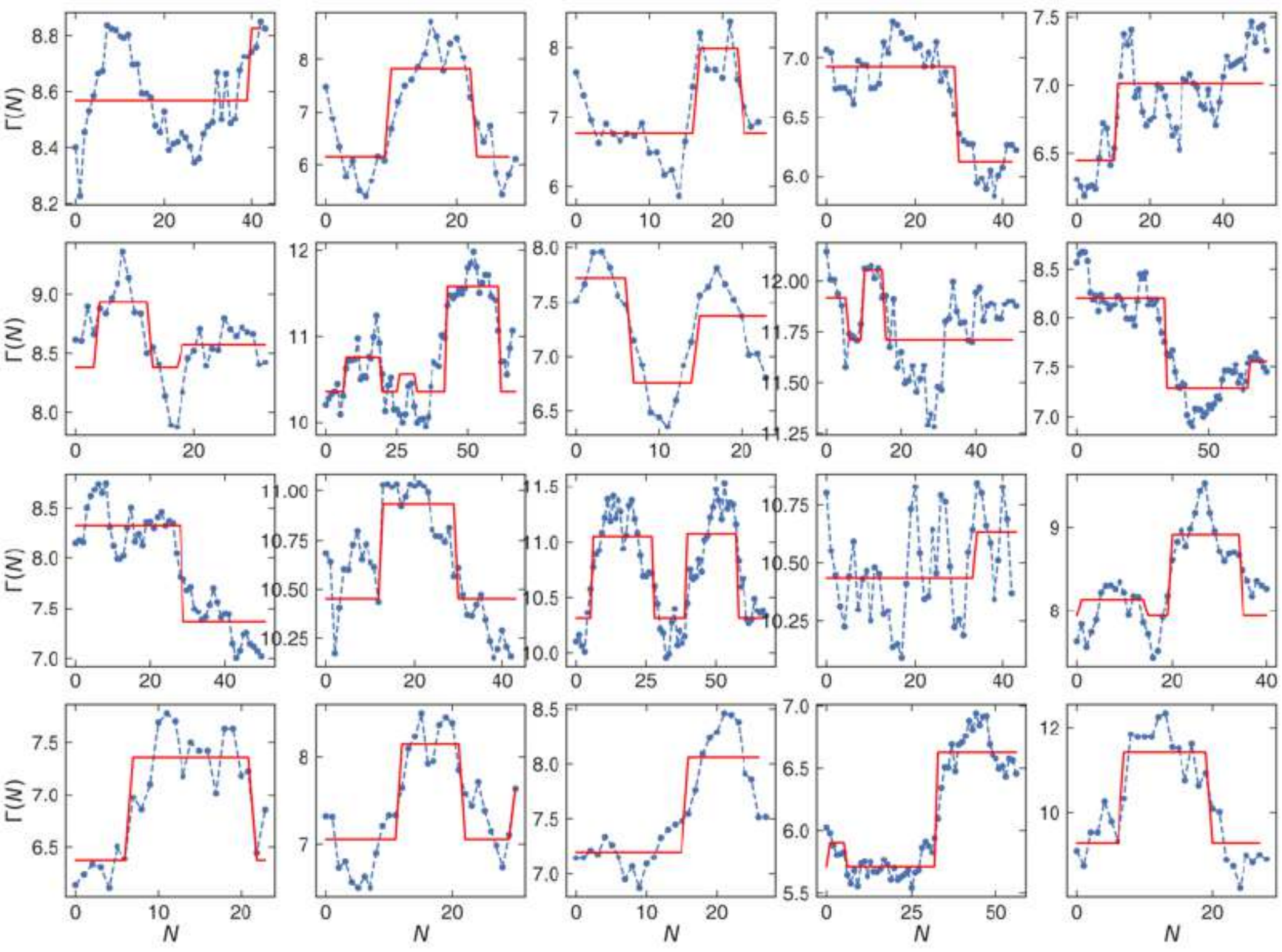

Figure S15: **The fitting results of 20 randomly selected artists.** Each subplot denotes the $\Gamma(N)$ sequence of an individual in our dataset, where blue dots denote data and red lines denote the best fitting result of each individual.

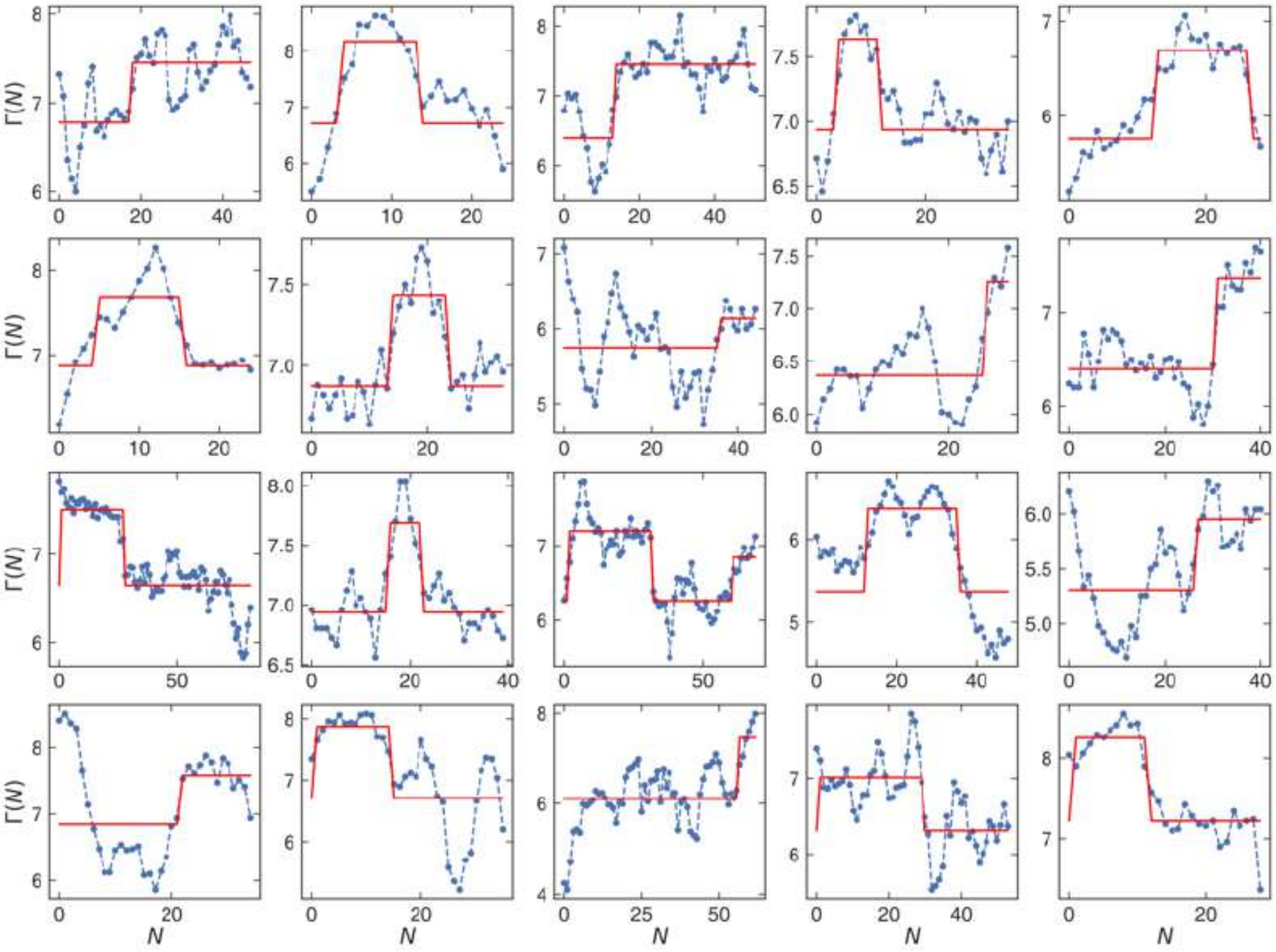

Figure S16: **The fitting results for 20 randomly selected directors.** Each subplot denotes the $\Gamma(N)$ sequence of an individual in our dataset, where blue dots denote data and red lines denote the best fitting result of each individual.

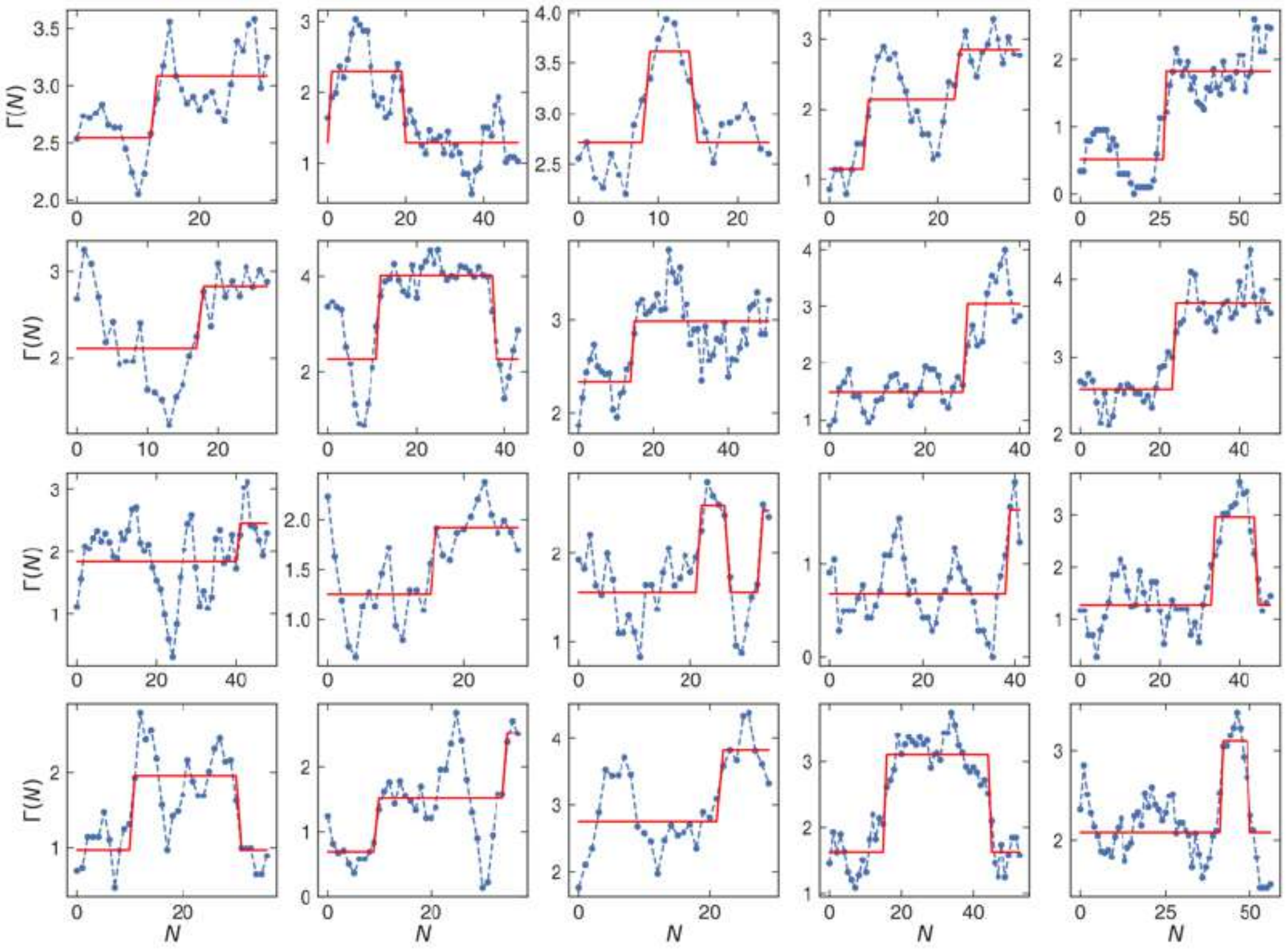

Figure S17: **The fitting results for 20 randomly selected scientists.** Each subplot denotes the $\Gamma(N)$ sequence of an individual in our dataset, where blue dots denote data and red lines denote the best fitting result of each individual.

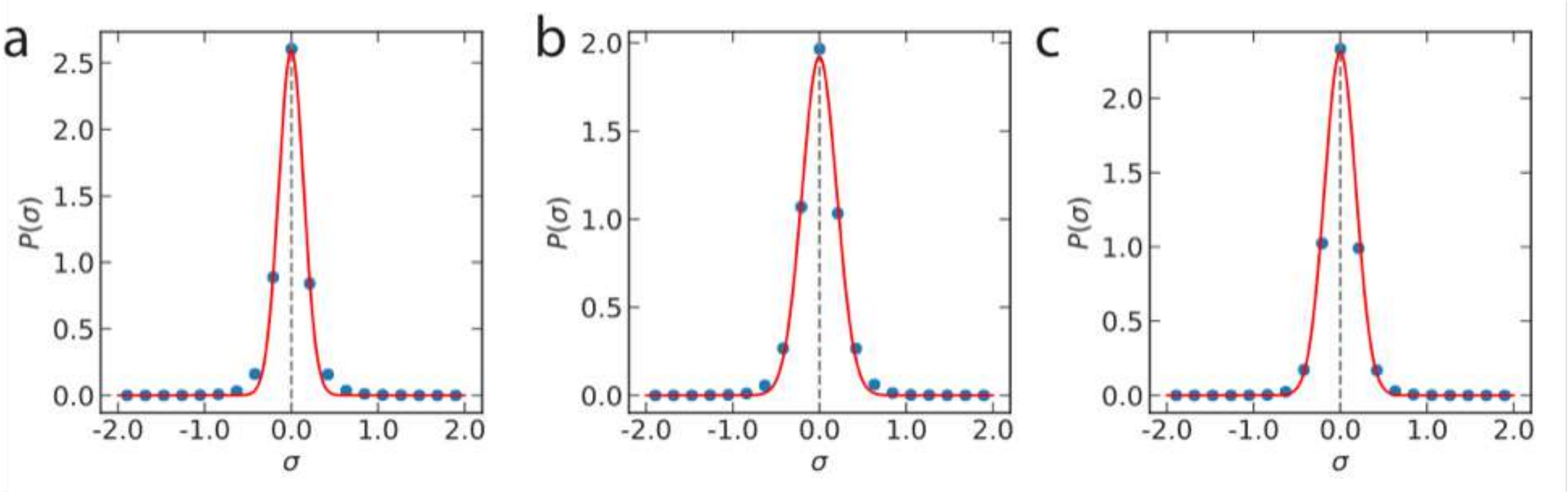

Figure S18: **Measuring the noise of** $\Gamma(N)$**.** The distribution $P(\sigma)$ for (a) artists, (b) directors, and (c) scientists, where $\sigma$ is the difference between real and fitted $\Gamma(N)$ sequence for all points in a domain. The blue dots denote data and the red line is a normal distribution. $P(\sigma)$ peaks around zero for the three domains. The standard deviation for the normal distribution is $0.186$ for artists, 0.229 for directors, and 0.189 for scientists, respectively.

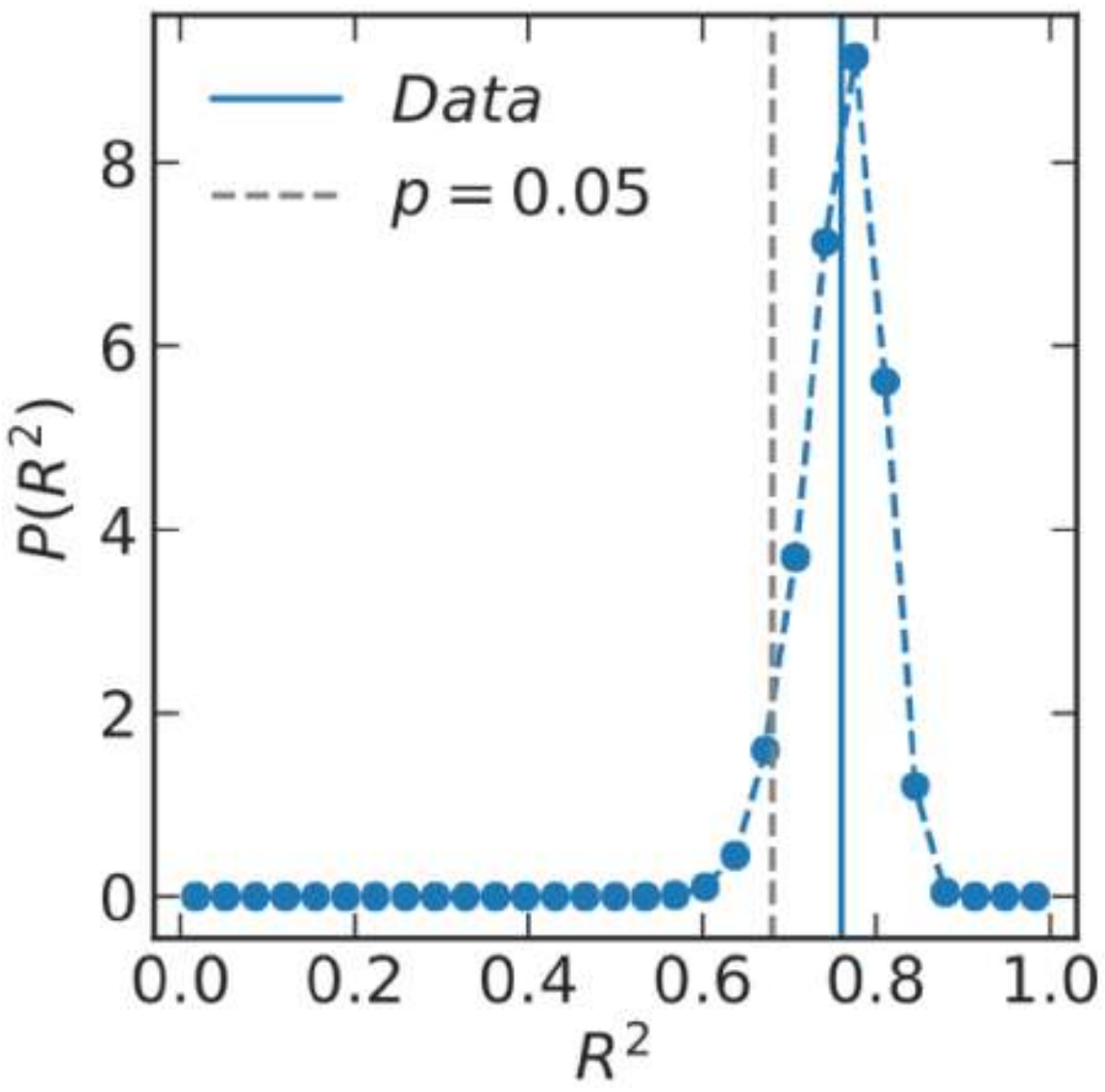

Figure S19: **An illustration of $R^2$ distribution.** The distribution $P(R^2)$ of simulated careers for an individual in our dataset. The blue dots denote $P(R^2)$ calculated from 1000 simulated careers, the vertical blue line denotes $R^2$ between data and fitted $\Gamma(N)$, and the dashed grey line is the $R^2$ when p-value is 0.05.

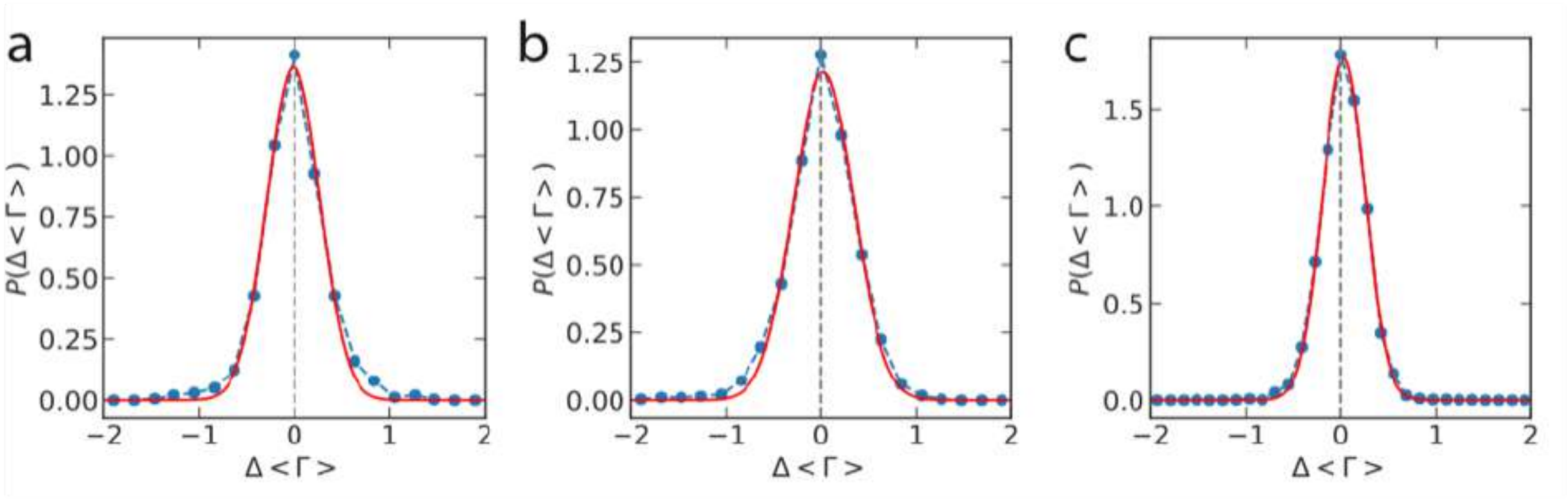

Figure S20: **Impact before and after hot hand.** The distribution of $P(\Delta\langle\Gamma\rangle)$ for (a) artists, (b) directors, and (c) scientists, where $\Delta\langle\Gamma\rangle$ measures the average impact difference before and after a hot streak. The blue dots denote data, and the red line is a normal distribution. The mean value of the normal distribution is 0.01 for artists, -0.02 for directors, and 0.02 for scientists, respectively.

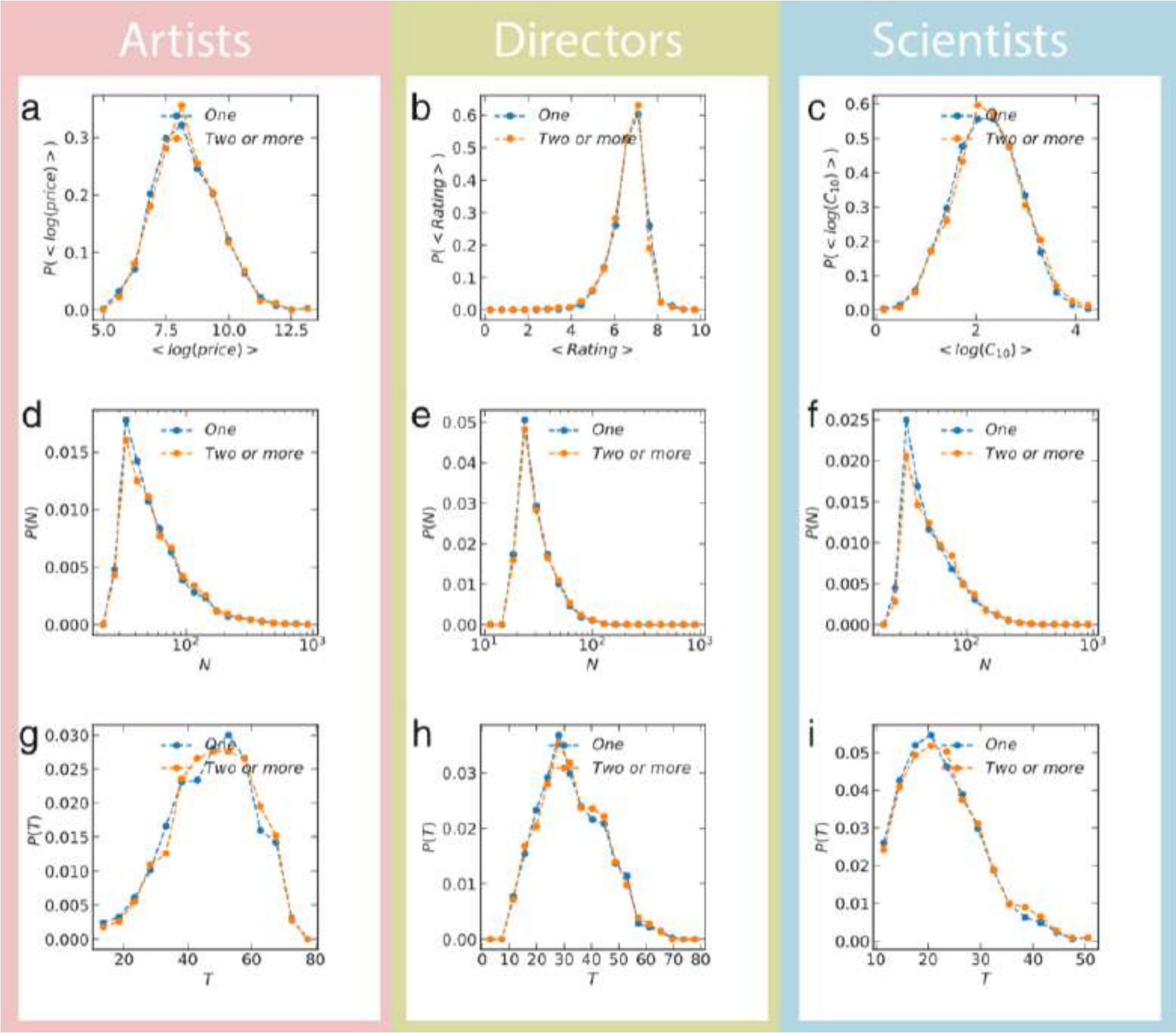

Figure S21: **Individuals with one and more than one hot streaks.** (a–c) The distribution of average impact for individuals with one, and more than one hot streaks for (a) artists, (b) directors, and (c) scientist. (d–f), The distribution of number of works $P(N)$ within a career for individuals with one and more than one hot streaks across three domains. (g–i), The distribution of career length $P(\tau)$ for individuals with one and more than one hot streaks across three domains. The blue dot denotes individuals with one hot streak, and the orange dot denote the ones with at least two hot streaks.

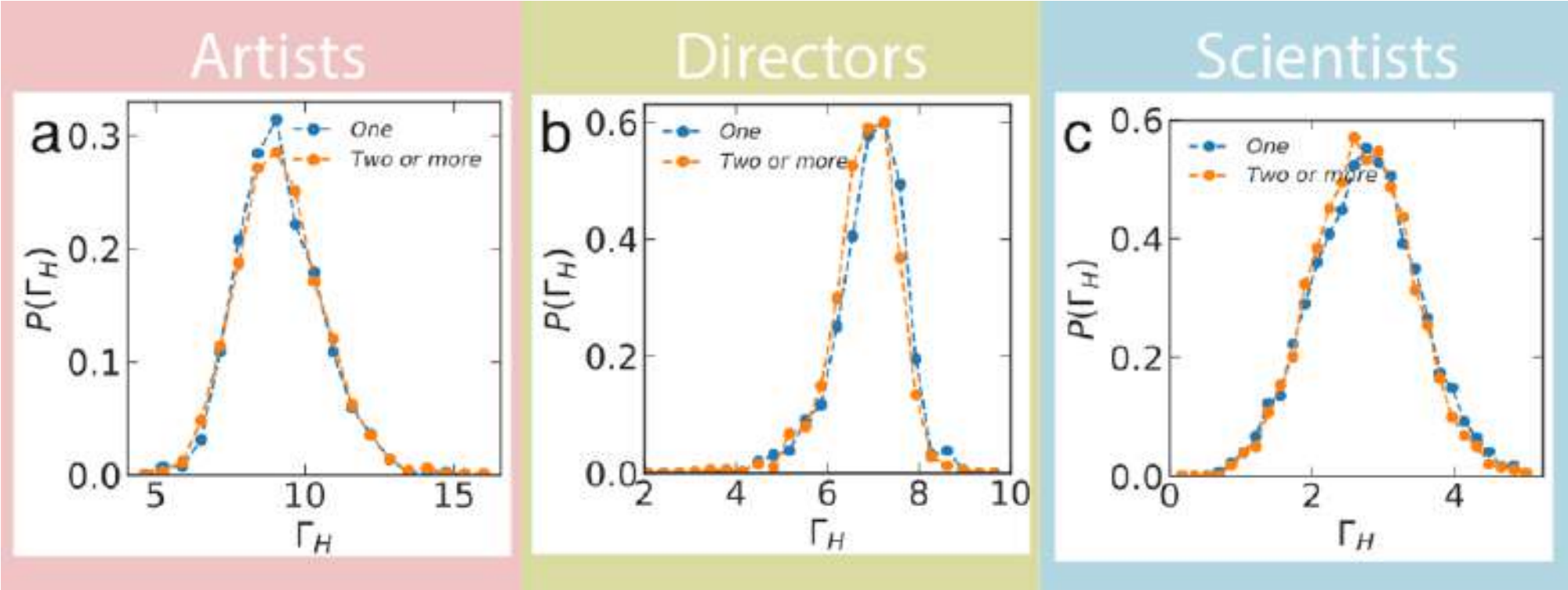

Figure S22: $\Gamma$ **for individuals with different number of hot streaks.** (a–c), $P(\Gamma_0)$ of individuals with ($\geq 1$) and without hot streaks. $P(\Gamma_0)$ for artists is the same for the two population, while $P(\Gamma_0)$ is smaller for individuals with hot streaks for directors and scientists. (d–f), $P(\Gamma_H)$ of individuals with one and more than one hot streak. $P(\Gamma_H)$ for artists is the same for the two population. While for directors and scientists, individuals with more than one hot streaks have smaller $P(\Gamma_H)$.

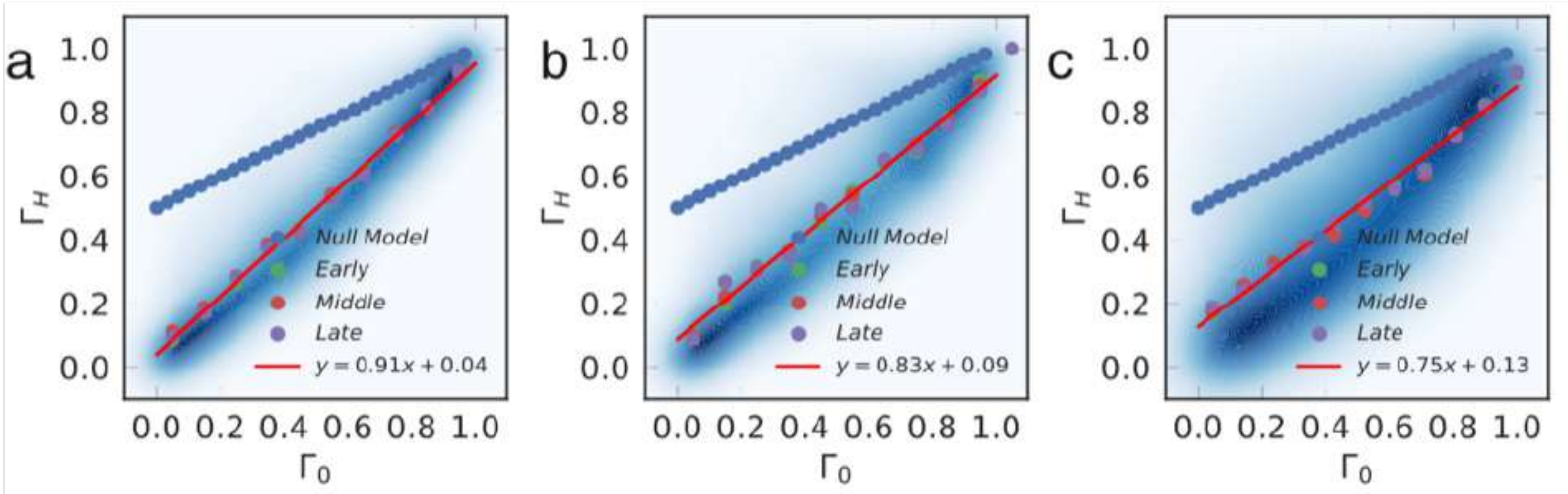

Figure S23: $\Gamma_0$ **and** $\Gamma_H$ **in percentile.** (a–c), The relationship between $\Gamma_0$ and $\Gamma_H$ in percentile for (a) artists, (b) directors, and (c) scientists. The blue area denotes the kernal dentisy of data. The colored dots are the logarithmic binning of the data, where the color corresponds to different timing of hot streaks. The red line is a linear curve. The blue dot denotes the null model prediction. $\Gamma_0$ and $\Gamma_H$ follows a linear relationship measured in percentile.

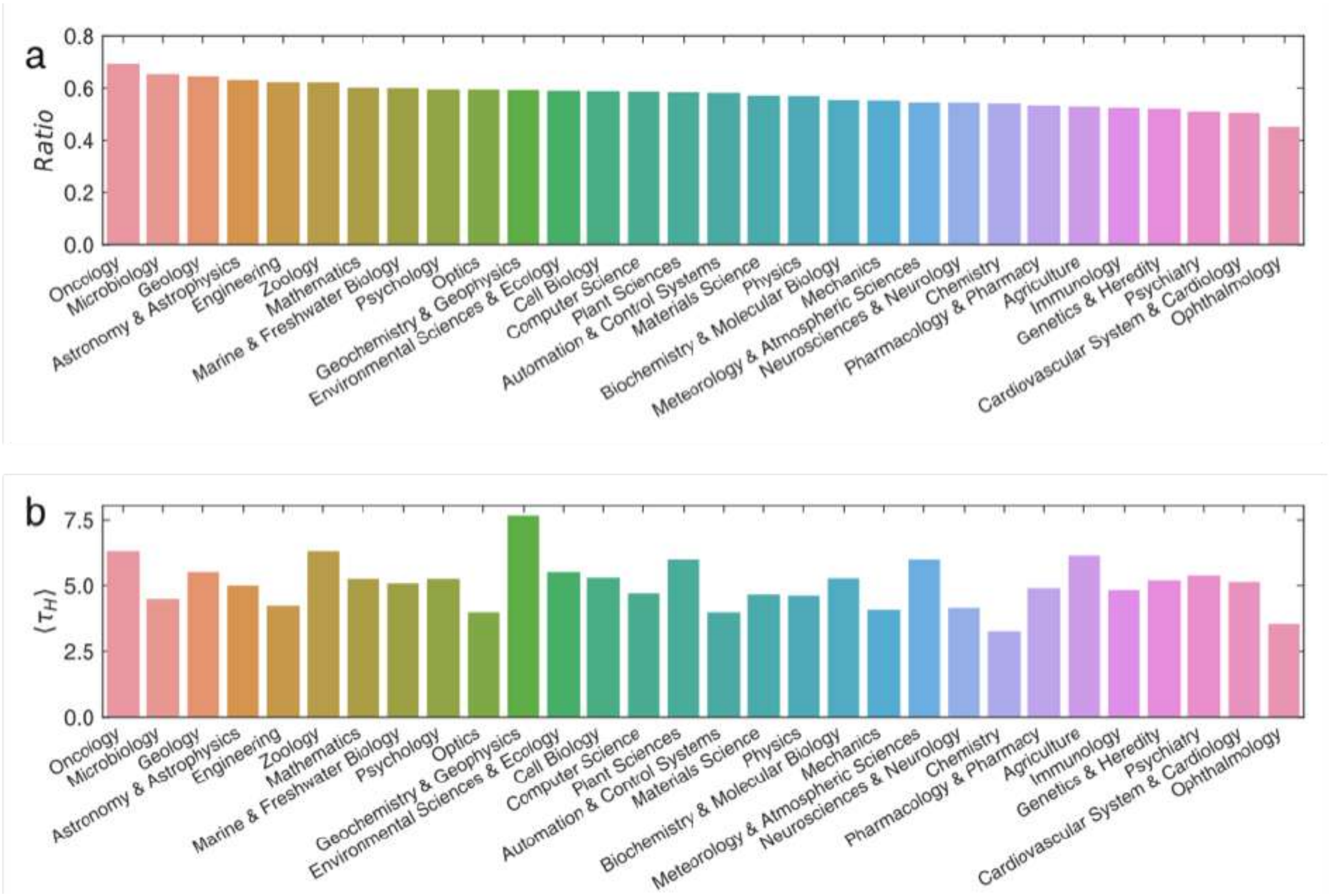

Figure S24: **Hot streaks for different scientific disciplines.** (a) The proportion of scientists with one hot streak in each discipline. (b) The average duration $\langle \tau_H \rangle$ for scientists within each discipline.

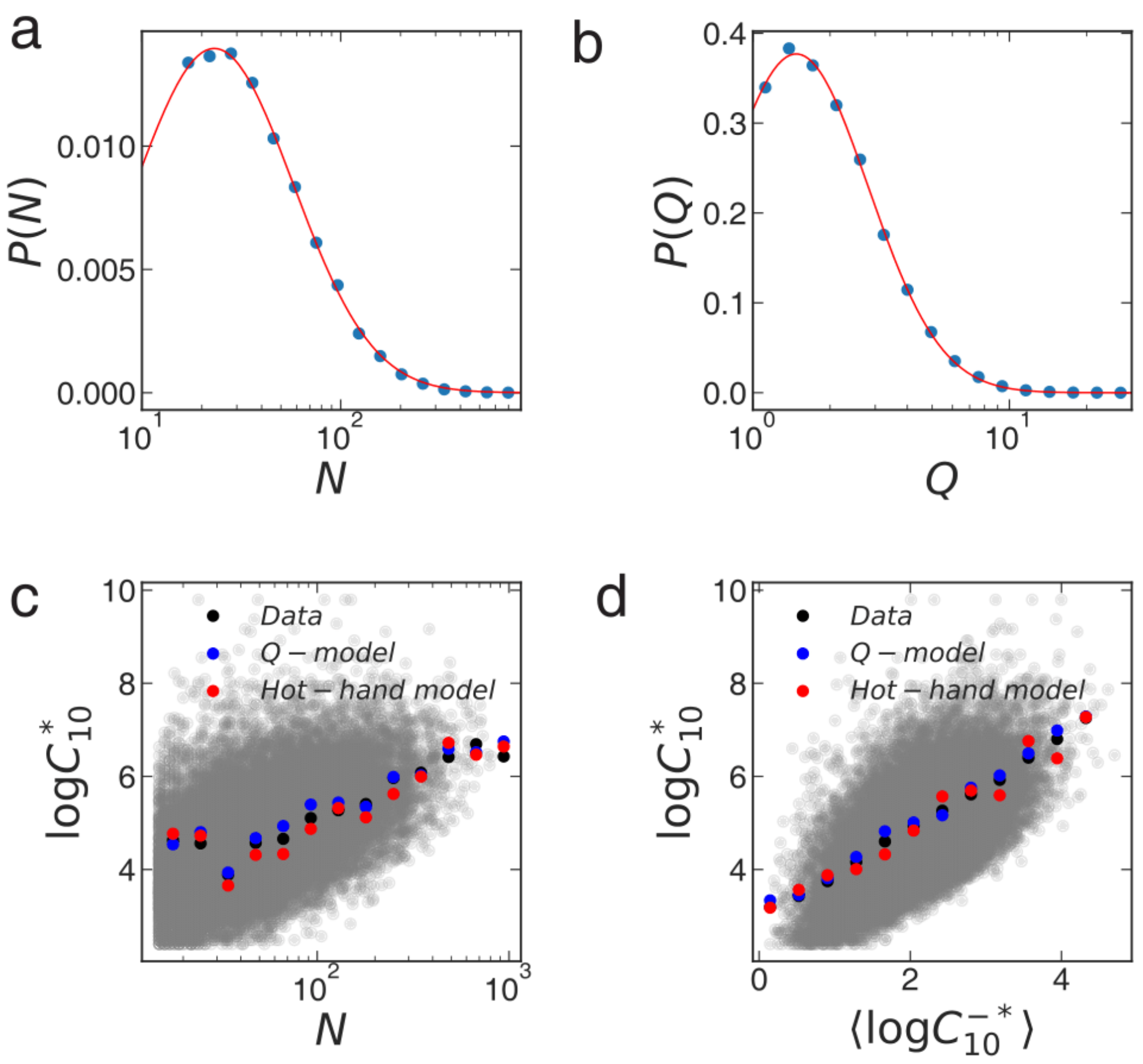

Figure S25: **Q model validation for scientists.** (a) The distribution $P(N)$ for scientists, where dots denote data, and the red line corresponds to a log-normal function with average $\mu = 3.9$ and standard deviation $\sigma = 0.84$. (b) The distribution $P(Q)$ for scientists. The red line corresponds to a log-normal function with $\mu = 0.8$ and $\sigma = 0.58$ (c) The highest impact $\log C^{*}_{10}$ versus the number of works $N$ within a career. Each grey dot corresponds to an artist. The black circles are the logarithmic binning of the scattered data. The blue and red circles represent the prediction of the $Q$-model and the hot-streak model, respectively. (d) $\log C^{*}_{10}$ versus $\langle \log C^{-*}_{10} \rangle$. Each grey dot corresponds to an artist, where $\langle \log C^{-*}_{10} \rangle$ is the average impact within a career without $\log C^{*}_{10}$.

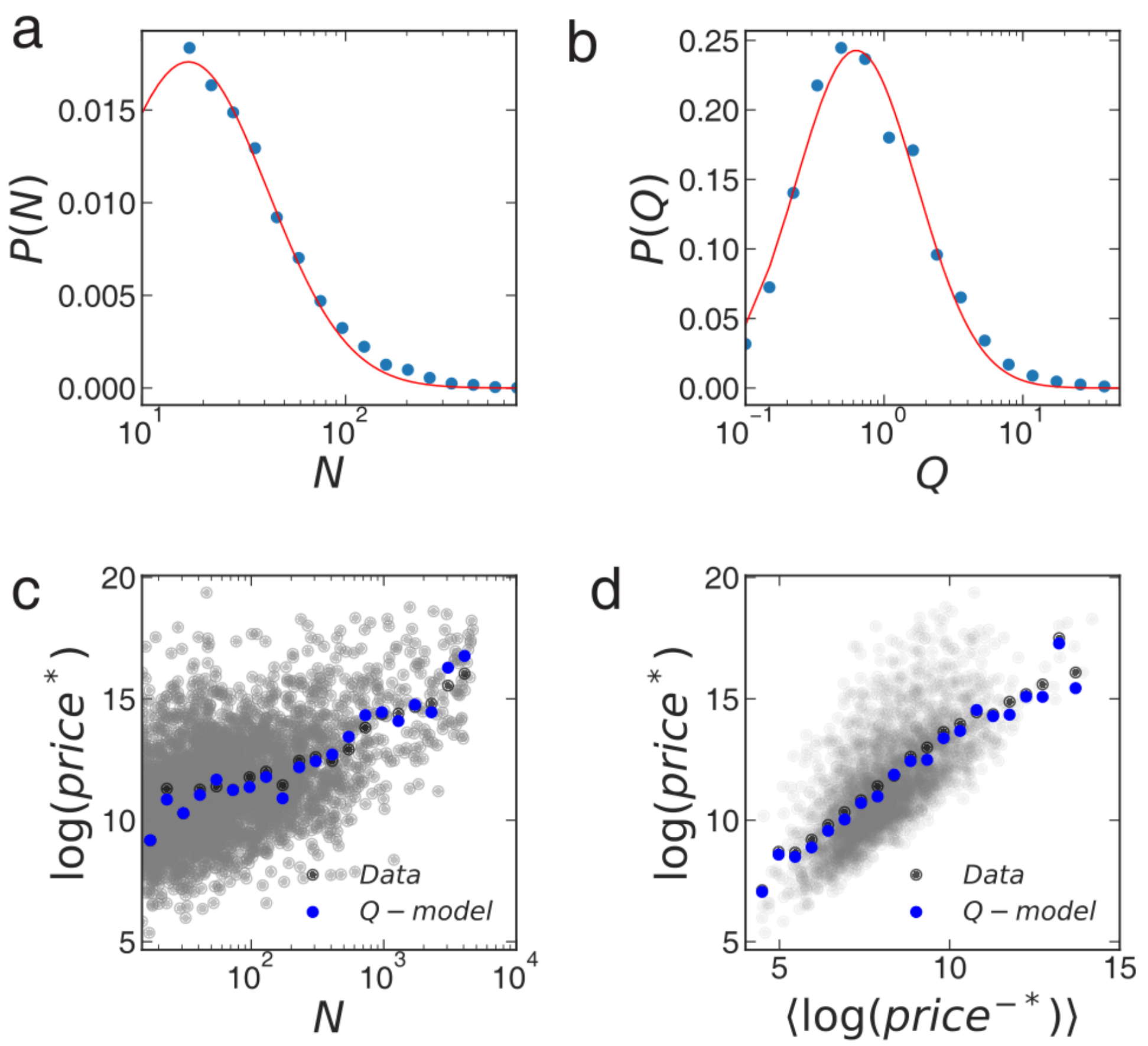

Figure S26: **Q model validation for artists.** (a) The distribution $P(N)$ for artists, where dots denote data, and the red line corresponds to a log-normal function with average $\mu = 3.6$ and standard deviation $\sigma = 0.90$. (b) The distribution $P(Q)$ for artists. The red line corresponds to a log-normal function with $\mu = 1.1$ and $\sigma = 1.21$ (c) The highest impact $\log(\text{price}^*)$ versus the number of works $N$ within a career. Each grey dot corresponds to an artist. The black circles are the logarithmic binning of the scattered data. The blue circles represent the prediction of the $Q$-model. (d) $\log(\text{price}^*)$ versus $\langle\log(\text{price}^{-*})\rangle$. Each grey dot corresponds to an artist, where $\langle\log(\text{price}^{-*})\rangle$ is the average impact within a career without $\log(\text{price}^*)$.

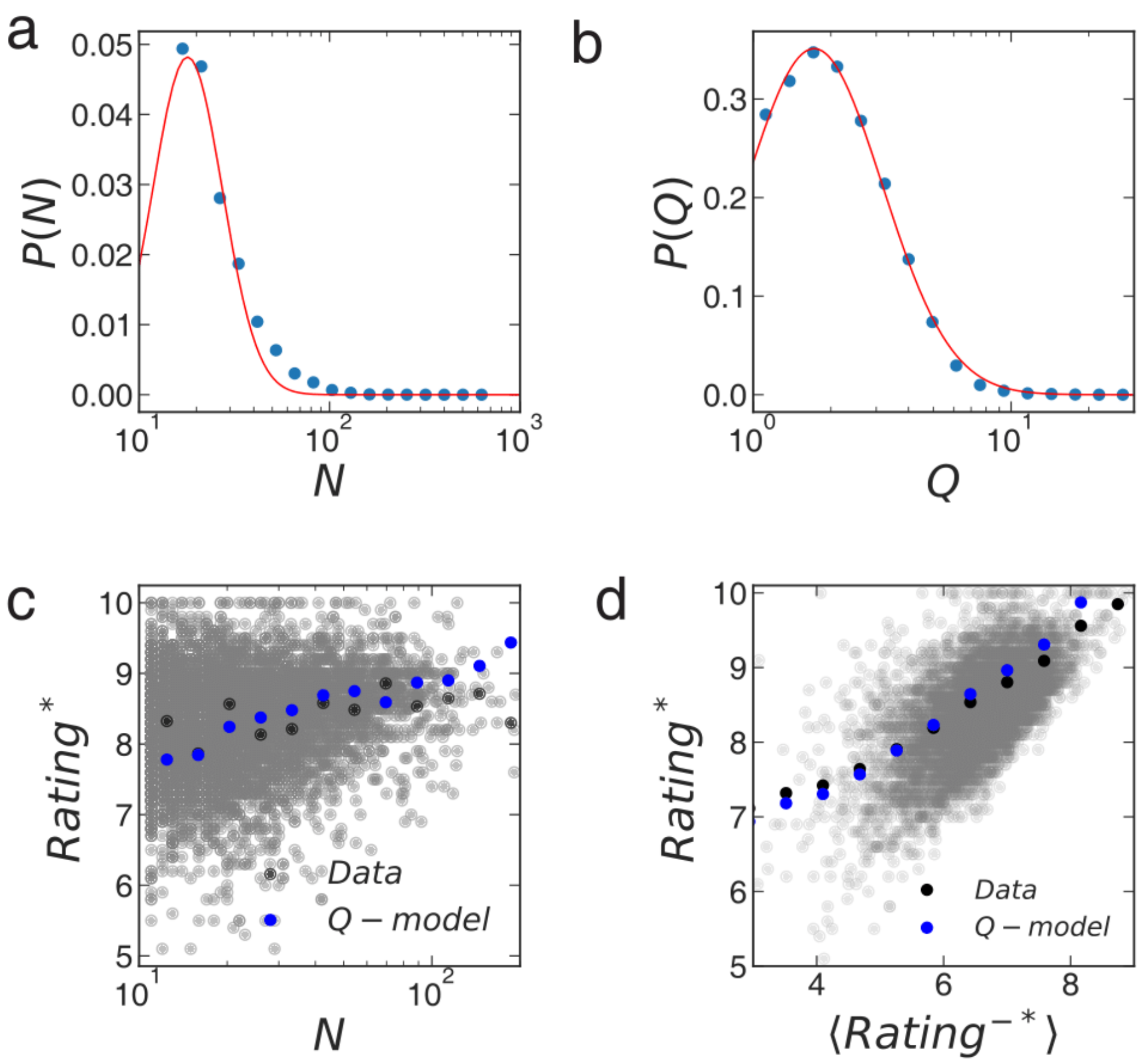

Figure S27: **Q model validation for directors.** (a) The distribution $P(N)$ for directors, where dots denote data, and the red line corresponds to a log-normal function with average $\mu = 3.1$ and standard deviation $\sigma = 0.42$. (b) The distribution $P(Q)$ for directors. The red line corresponds to a log-normal function with $\mu = 0.9$ and $\sigma = 0.56$ (c) The highest rating Rating$^*$ versus the number of works $N$ within a career. Each grey dot corresponds to a director. The black circles are the logarithmic binning of the scattered data. The blue circles represent the prediction of the $Q$-model. (d) Rating$^*$ versus $\langle \text{Rating}^{-*} \rangle$. Each grey dot corresponds to a director, where $\langle \text{Rating}^{-*} \rangle$ is the average ratings within a career without Rating$^*$.

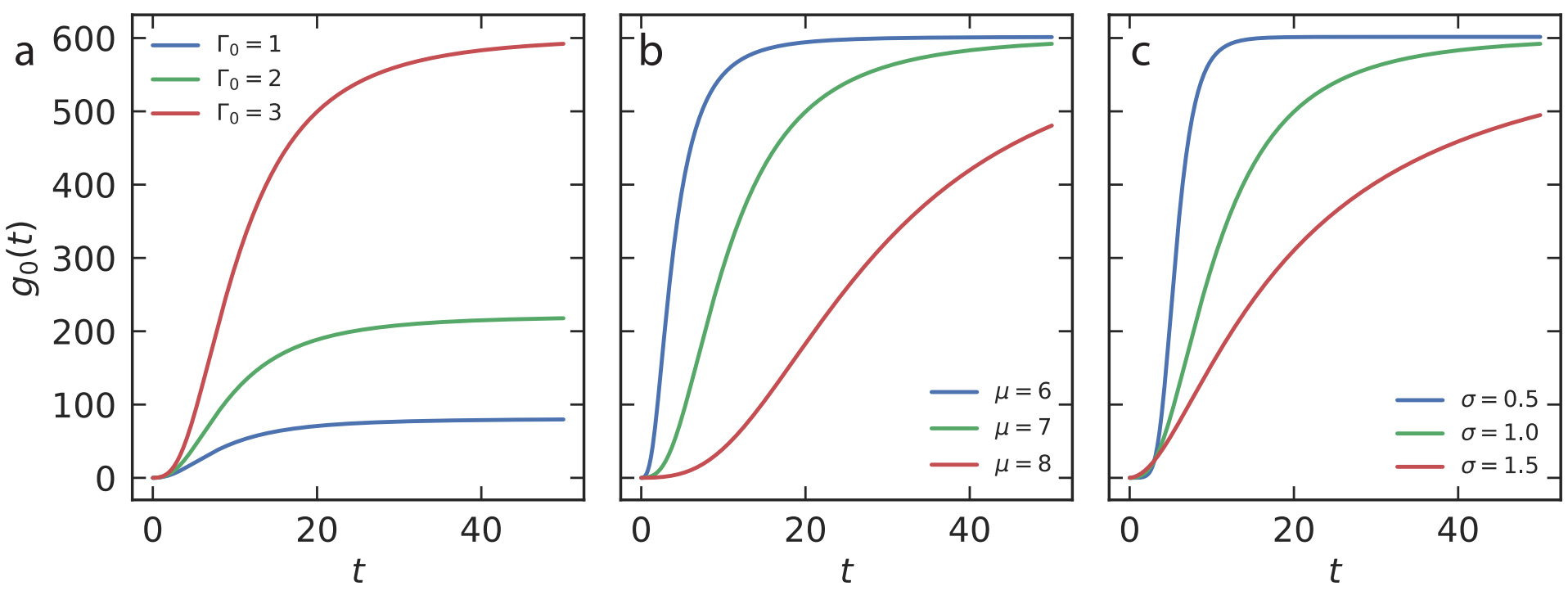

Figure S28: $g(t)$ **under the null model.** (a–c), The behaviour of $g_0(t)$ under the null model prediction with different (a) $\Gamma_0$, (b) $\mu$, and (c) $\sigma$ parameters. We use $\Gamma_0 = 3.0$, $\mu = 7.0$, and $\sigma = 1.0$ as input.

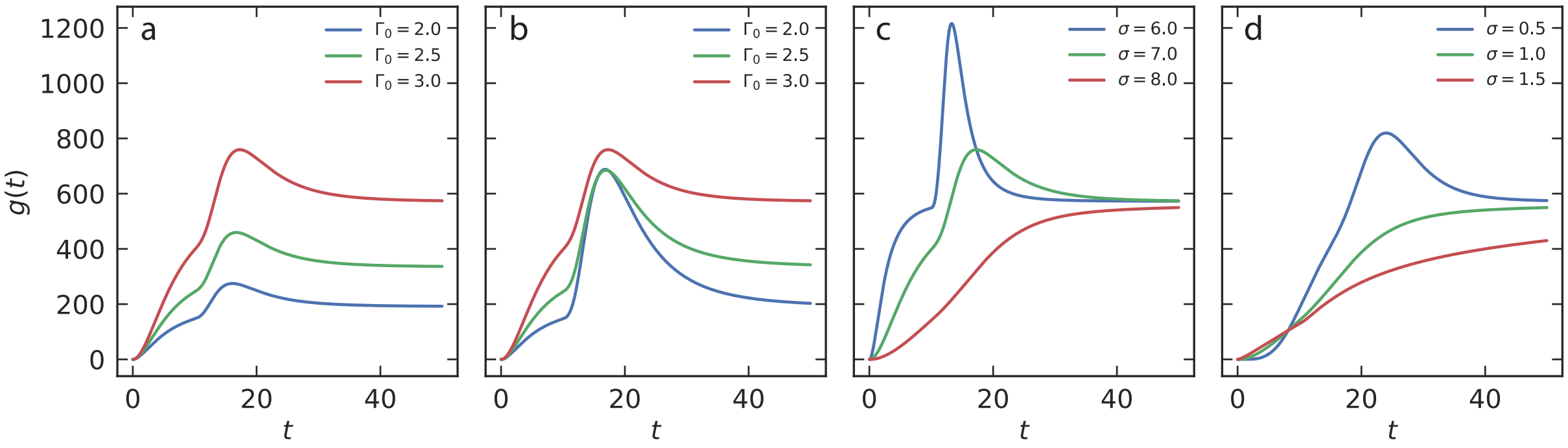

Figure S29: $g(t)$ **under the hot-streak model.** (a–d), $g(t)$ under the hot-streak model prediction with different (a) $\Gamma_0$, (b) $\Delta\Gamma$, (c) $\mu$, and (d) $\sigma$ parameters. For (a–c) we use $\Gamma_H = 4.0$, $\Gamma_0 = 3.0$, $\mu = 7.0$, $\sigma = 1.0$, $t_\uparrow = 10$ years and $t_\downarrow = 12$ years as input. We use $\Gamma_H = \Gamma_0 + 1.0$ for (a).

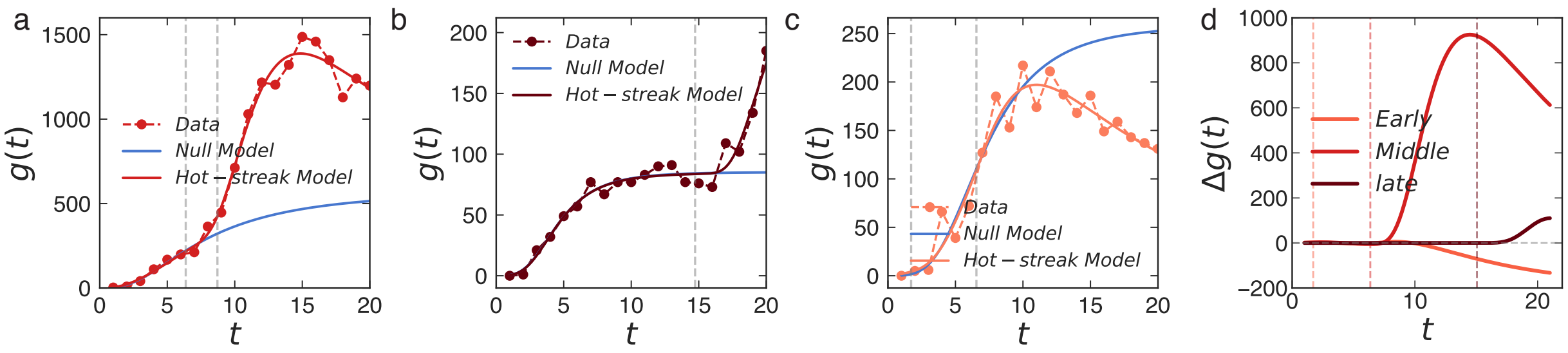

Figure S30: **Policy implications of** $g(t)$. (a–c) Individuals in our dataset with (a) mid, (b) late, and (c) early onset of the hot streak. Red dots denote data, the blue line is the null model's prediction based on early performance, and the red line captures the predictions from the hot-streak model, with dashed grey lines denoting the start and end of hot streaks. (d) The difference $\Delta g(t)$ between our hot-streak model and the null model for each individual, where dashed lines with corresponding color denotes the start of a hot streak. (d) illustrates the systematic discrepancies in predicting individual's future impact, if we ignore the uncovered hot streak phenomenon.

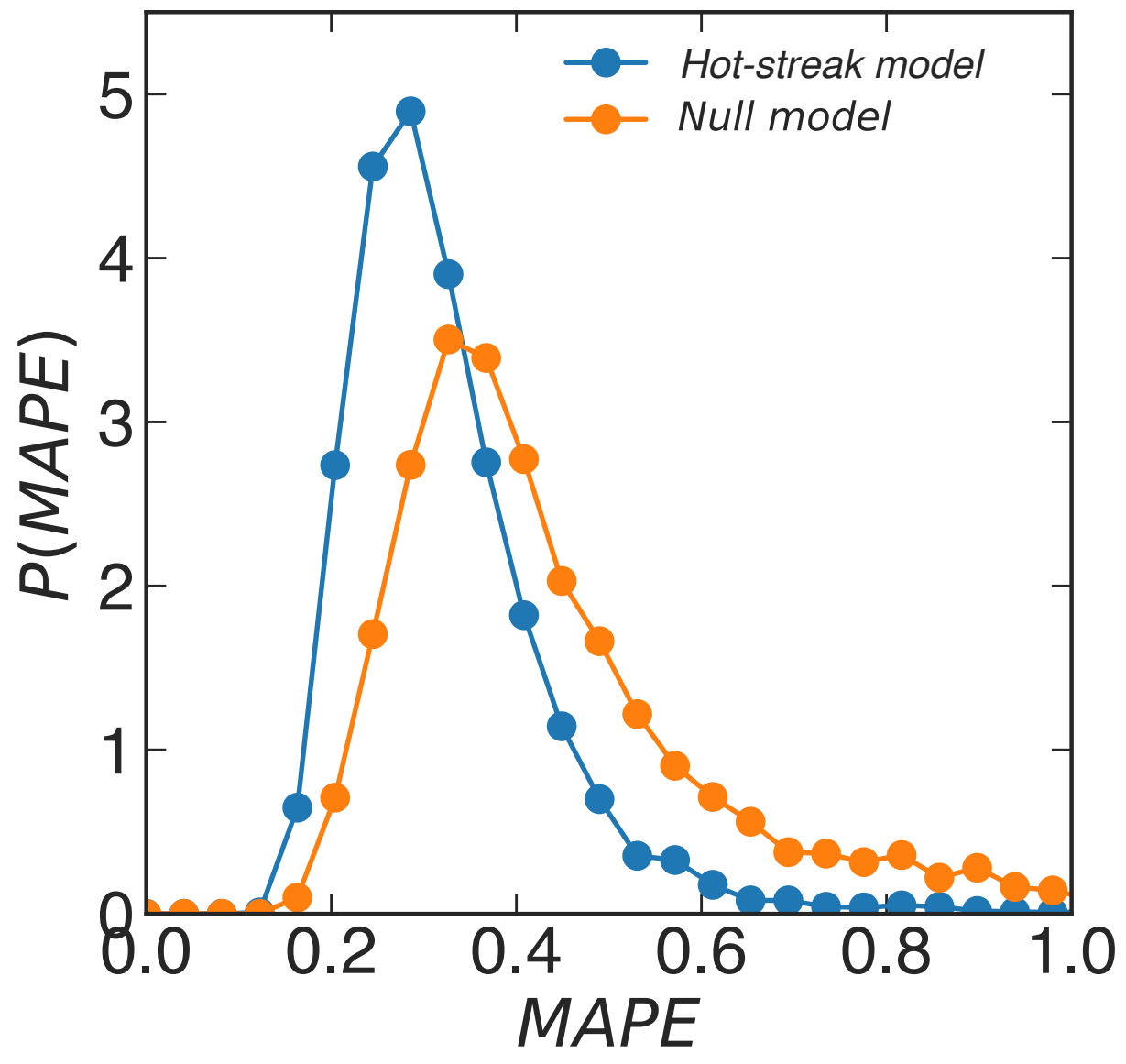

Figure S31: **The performance of different models.** The distribution of $P(MAPE)$ for the null model (orange dots) and our hot-streak model (blue dots). Compared with the null model, our hot-streak model has systematically smaller error.

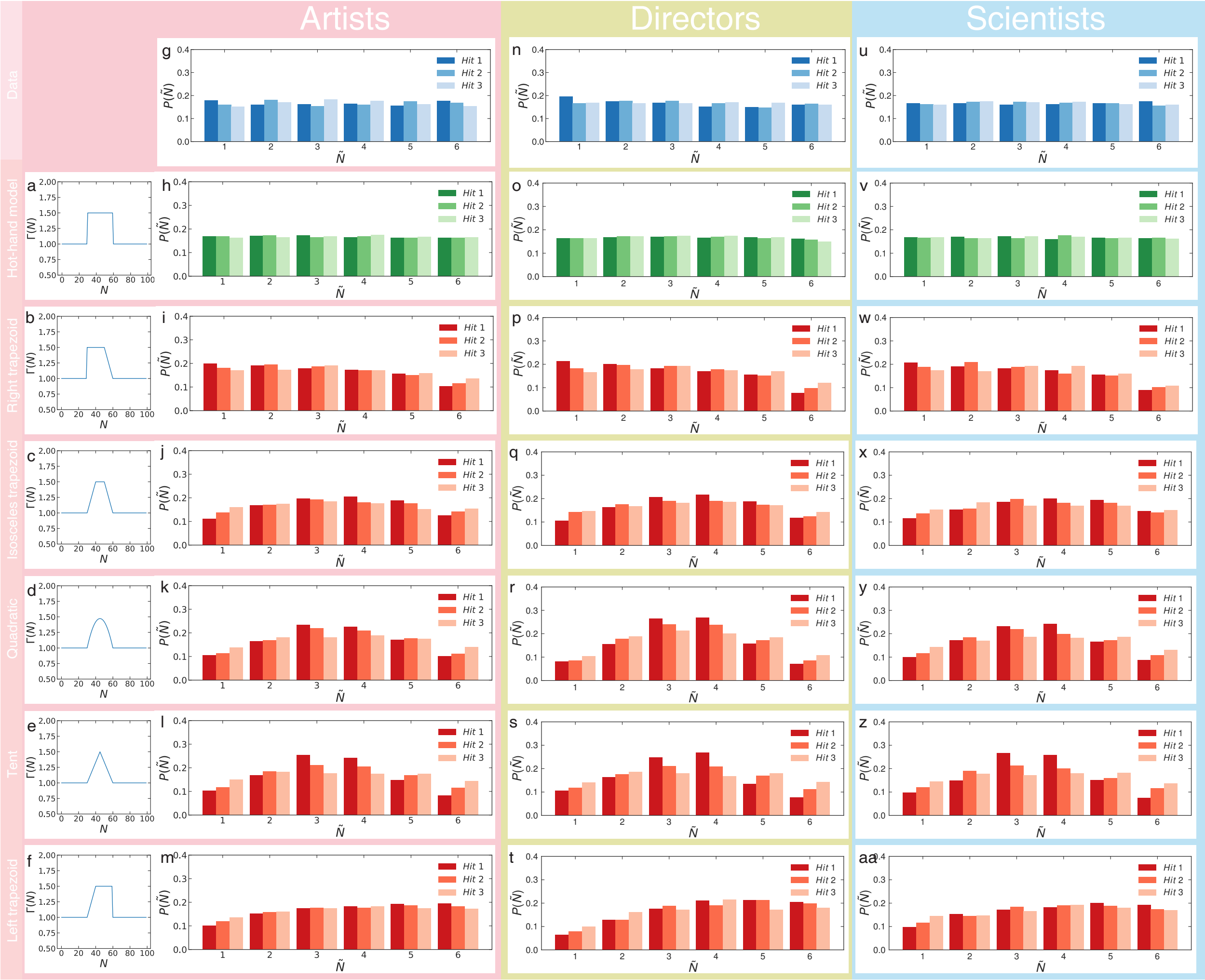

Figure S32: **Alternative models of hot streaks.** (a–f), An illustration of $\Gamma(N)$ for the (a) hot-streak model, (b) right trapezoid function, (c) isosceles trapezoid function, (d) quadratic function, (e) tent function, and (f) left trapezoid function. (g), The distribution of the relative position $P(\tilde{N})$ of the top three hit works among the top six hits within a career for artists. The shades of color correspond to different hits.The relative position $\tilde{N} = 1$ means the hit paper appears first among the top six hit, whereas $\tilde{N} = 6$ means the last one to appear. The relative order among the top six hits are random. The conclusion remains the same for (n) directors and (u) scientists. (h-m), $P(\tilde{N})$ predicted by corresponding model shown in (a–f) using artists' profiles as input. We again use shades to color different hits. We measure the statistical difference between data and the model's predictions, using the p-value of KS test for discrete distributions. We color the bars green if we cannot reject the hypothesis that data and model's prediction come from the same distributions (p-value $\geq 0.05$), and color them red otherwise. Among all alternative models considered, the hot-streak model turns out to be the only model that successfully reproduces the randomness observed in the top six hit works in (a). While the alternative functions show different trend of probability, contradicting to the randomness measured from data. We repeated the analyses using (o–t) directors' profiles and (v–aa) scientists' profiles as input, finding our results are robust to different domains.